\documentclass[twocolumn]{aastex631}
\usepackage[caption=false]{subfig}
\usepackage{multirow}
\usepackage{amsmath}

\newcommand{\HI}{H$\,${\sc i}}

\DeclareMathSymbol{:}{\mathord}{operators}{"3A}
\newcommand{\Msun}{M$_{\hbox{$\odot$}}$}  
\newcommand{\kms}{km\,s$^{-1}$}           

\accepted{April 14, 2024}

\submitjournal{AJ}

\shorttitle{WLM Early Star Formation}
\shortauthors{Archer et al.}

\begin{document}

\title{Probing the Relationship Between Early Star Formation and CO in the Dwarf Irregular Galaxy WLM with JWST}

\correspondingauthor{Haylee N. Archer}
\email{harcher@lowell.edu}

\author[0000-0002-8449-4815]{Haylee N. Archer}
\affiliation{School of Earth and Space Exploration, Arizona State University, Tempe, AZ 85287, USA}
\affiliation{Lowell Observatory, 1400 W Mars Hill Rd, Flagstaff, AZ 86001, USA}

\author[0000-0002-3322-9798]{Deidre A. Hunter}
\affiliation{Lowell Observatory, 1400 W Mars Hill Rd, Flagstaff, AZ 86001, USA}

\author[0000-0002-1723-6330]{Bruce G.\ Elmegreen}
\affiliation{IBM T.\ J.\ Watson Research Center, 1101 Kitchawan Road, Yorktown Heights, NY 10598, USA}

\author[0000-0002-5307-5941]{Monica Rubio}
\affiliation{Departamento de Astronom\'{i}a, Universidad de Chile, Casilla 36-D, Santiago, 8320000, Chile}

\author[0000-0002-8736-2463]{Phil Cigan}
\affiliation{United States Naval Observatory, 3450 Massachusetts Ave NW, Washington, DC 20392, USA}

\author[0000-0001-8156-6281]{Rogier A. Windhorst}
\affiliation{School of Earth and Space Exploration, Arizona State University, Tempe, AZ 85287, USA}

\author[0000-0001-9998-9741]{Juan R. Cort{\'e}s}
\affiliation{Joint ALMA Observatory, Alonso de C{\'o}rdova 3107, Vitacura, Santiago, 7630355, Chile}
\affiliation{National Radio Astronomy Observatory, Avenida Nueva Costanera 4091, Vitacura, Santiago, 7630197, Chile}

\author[0000-0003-1268-5230]{Rolf A. Jansen}
\affiliation{School of Earth and Space Exploration, Arizona State University, Tempe, AZ 85287, USA}



\begin{abstract}
Wolf-Lundmark-Melotte (WLM) is a Local Group dwarf irregular (dIrr) galaxy with a metallicity 13\% of solar. At 1 Mpc, the relative isolation of WLM provides a unique opportunity to investigate the internal mechanisms of star formation at low metallicities. The earliest stages of star formation in larger spirals occur in embedded clusters within molecular clouds, but dIrrs lack the dust, heavy metals, and organized structure of spirals believed necessary to collapse the molecular clouds into stars. Despite actively forming stars, the early stages of star formation in dIrrs is not well understood. We examine the relationship between early star formation and molecular clouds at low metallicities. We utilize ALMA-detected CO cores, \textit{JWST} near-infrared (NIR) images (F090W, F150W, F250M, and F430M), and \textit{GALEX} far-ultraviolet (FUV) images of WLM to trace molecular clouds, early star formation, and longer star formation timescales respectively. We compare clumps of NIR-bright sources (referred to as objects) categorized into three types based on their proximity to FUV sources and CO cores. We find objects, independent of their location, have similar colors and magnitudes and no discernible difference in temperature. However, we find that objects near CO have higher masses than objects away from CO, independent of proximity to FUV. Additionally, objects near CO are coincident with Spitzer 8 $\mu$m sources at a higher frequency than objects elsewhere in WLM. This suggests objects near CO may be embedded star clusters at an earlier stage of star formation, but accurate age estimates for all objects are required for confirmation.
\end{abstract}



\section{Introduction} \label{sec:intro}
Wolf-Lundmark-Melotte (WLM) is a nearby dwarf irregular galaxy (dIrr) in the Local Group at a distance of approximately 980 kpc \citep{Leaman_2012,Lee_2021}. WLM is a relatively isolated galaxy, with distances far enough from both the Milky Way and M31 to indicate a low probability of previously interacting with either \citep{Teyssier_2012,Albers_2019}. However, recent observations of \HI\ morphology and kinematics in WLM find evidence of ram-pressure stripping, suggesting the galaxy may be less isolated than previously thought \citep{Yang_2022}. Studying star formation in WLM's potentially less perturbed environment can provide insights into the internal mechanisms of star-forming regions and the impact of the surrounding environment on the process. Like other nearby dIrrs, the proximity and low metal content of WLM at $12 + \textnormal{log(O/H)} = 7.8$ \citep{Lee_2005}, which is $\sim$13\% of solar metallicity, allow us to probe the details of star formation at low metallicity.

Star formation occurs in molecular clouds that are primarily composed of H$_2$, which is typically traced with low rotational transitions of CO \citep[e.g.][]{Kennicutt_1998,Bigiel_2011,Glover_2011,Bolatto_2013}. In larger spirals like the Milky Way, Giant Molecular Clouds (GMCs) account for nearly all star formation, but in dwarf galaxies there is little CO to trace the molecular clouds despite ongoing star formation. In environments with low metal content, molecular clouds have altered properties compared to those in metal-rich environments \citep[e.g.][]{Elmegreen_1989,Hunter_1998,Brosch_1998,Bolatto_1999,Leroy_2008,Chevance_2020}. For instance, scarcity of heavy elements limits the formation of dust grains, which play a crucial role in shielding and cooling molecular clouds \citep[e.g.][]{Draine_2007,Fukui_2010,Wakelam_2017,Osman_2020}. Additionally, lower amounts of carbon and dust necessary for cooling result in larger photodissociation regions, longer CO formation time, and smaller CO clouds that are typically used to trace the molecular clouds \citep{Elmegreen_1980,Isreal_1986,Taylor_1998,Leroy_2009,Schruba_2012}.

The earliest stages of star formation in larger spiral galaxies occur within embedded clusters in GMCs \citep[e.g.][]{Lada_2003,Bastian_2006,Dale_2015}. \citet{Elmegreen_2018} find embedded star clusters occurring at regular intervals within the main spiral arms of nearby spiral galaxies, suggesting that star formation likely originated from the collapse of dense gas, which was then compressed into narrow dust lanes due to the impact of stellar spiral arm shocks \citep{Elmegreen_2019,Elmegreen_2020}. Without the organized structures of spiral arms, however, star formation in dwarf galaxies tends to be more irregular and sporadic, and the early stages of star formation in these low metallicity environments are not well understood.

Following the discovery of CO(3–2) emission in two star-forming regions of WLM by \citet{Elmegreen_2013} using the APEX telescope, \citet{Rubio_2015} targeted these regions with CO(1–0) and mapped 10 CO cores with the Atacama Large Millimeter Array (ALMA). \citet{archer_2022} then examined the relation between these 10 -- plus an additional 35 (Rubio et al.\ in preparation) -- CO cores in WLM and FUV used to define star-forming regions, along with the \HI\ reservoir out of which the star-forming clouds formed, and found no obvious characteristics driving the formation of the CO cores. Recently, the \textit{JWST} Resolved Stellar Populations ERS program (PID 1334) imaged a large portion of the star-forming area of WLM in the near infrared (NIR) at 0.9 $\mu$m, 1.5 $\mu$m, 2.5 $\mu$m, and 4.3 $\mu$m with the Near Infrared Camera (NIRCam) \citep{Weisz_2023}. The NIR images now provide a clearer view of the young, embedded star-forming regions to explore the role of the CO cores in these regions and better understand early star formation at low metallicities.

\begin{figure*}[bht!]
\epsscale{1.1}\plotone{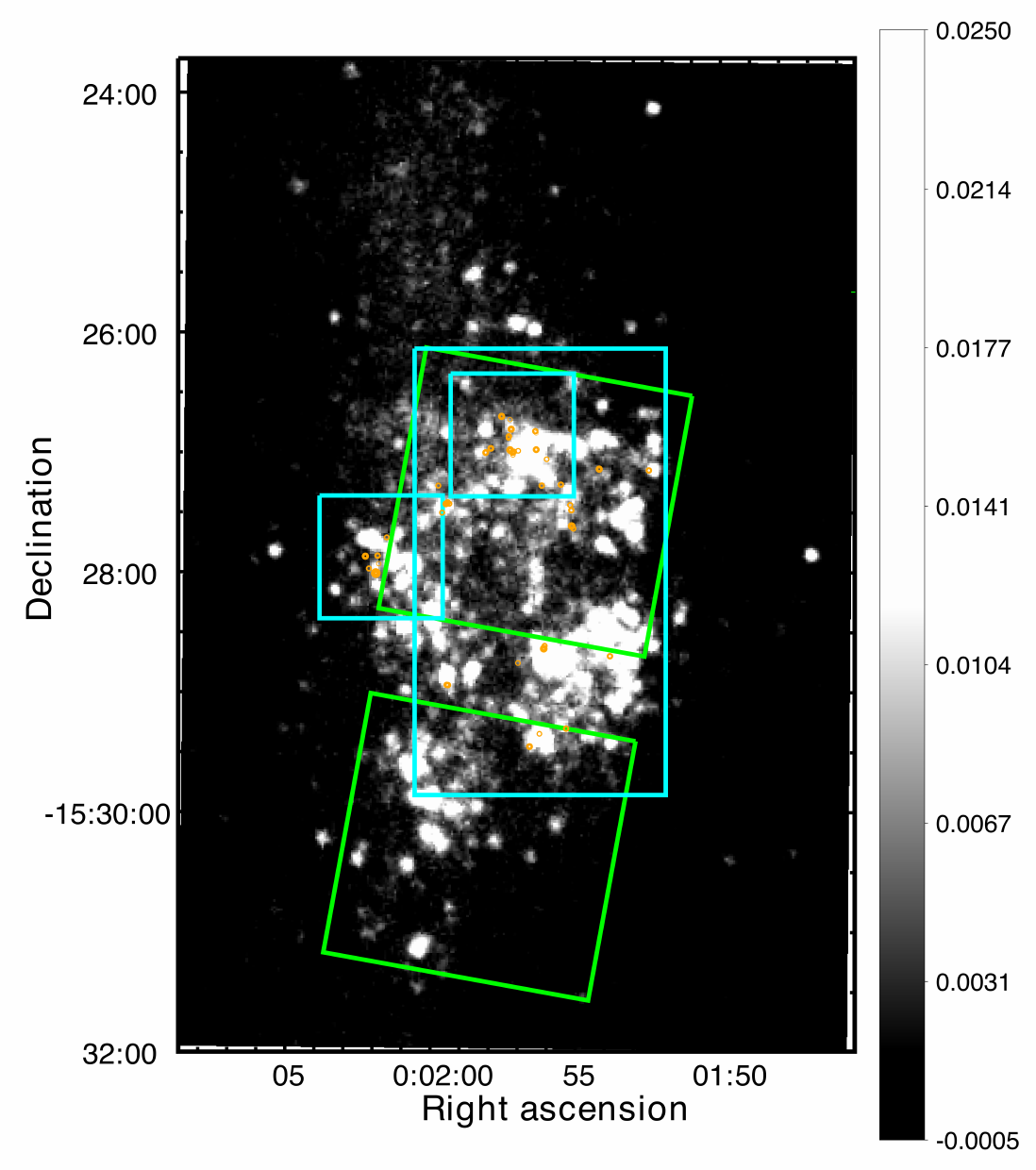}
\caption{$FUV$ image of WLM showing the field of view from ALMA Cycle 1 \citep[smaller cyan squares,][]{Rubio_2015} and Cycle 6 (larger cyan rectangle, Rubio et al.\ in preparation) and the 45 detected CO cores (tiny orange circles), along with the field of view from \textit{JWST} ERS PID 1334 \citep[green squares,][]{Weisz_2023}. The ERS fields will be referred to as the North field and the South field. The orientation of the image is such that North is up and East is to the left. Colorbar values can be converted from counts/second to calibrated AB magnitudes with the equation: FUV$_{A\!B} = -2.5$log$_{10}($FUV$_{counts/s}) + 18.82$.
\label{fig:fuv_fov}}
\end{figure*}

Gaining insights into the role of CO in the initial stages of star formation in dwarf galaxies not only sheds light on how stars form at low metallicities but also provides insights into the fundamental physical processes of star formation and the interplay between stars and their environments.
In this paper we investigate young star-forming regions in WLM to determine (1) if early star-forming regions have the same properties in different local environments and (2) the relationship between early star formation and CO in low metallicity environments. We use NIR images as tracers of young, embedded star formation \citep[$<$5 Myr,][]{Lada_2003}, the CO cores to trace the molecular clouds from which stars form, and FUV to trace star formation on longer time scales \citep[up to $\sim$100 Myr,][]{Calzetti_2013}.

The paper is organized as follows. In Section \ref{sec:data} we introduce our data and describe our object selection and definitions of the object types, along with our methods for determining the blackbody temperature, luminosity, and mass of the objects. We present our results in Section \ref{sec:results} and discuss our findings in Section \ref{sec:discussion}. Finally, we summarize our conclusions in Section \ref{sec:summary}.

\begin{figure*}[bht!]
\hspace{-2.5em}\epsscale{1.05}\plotone{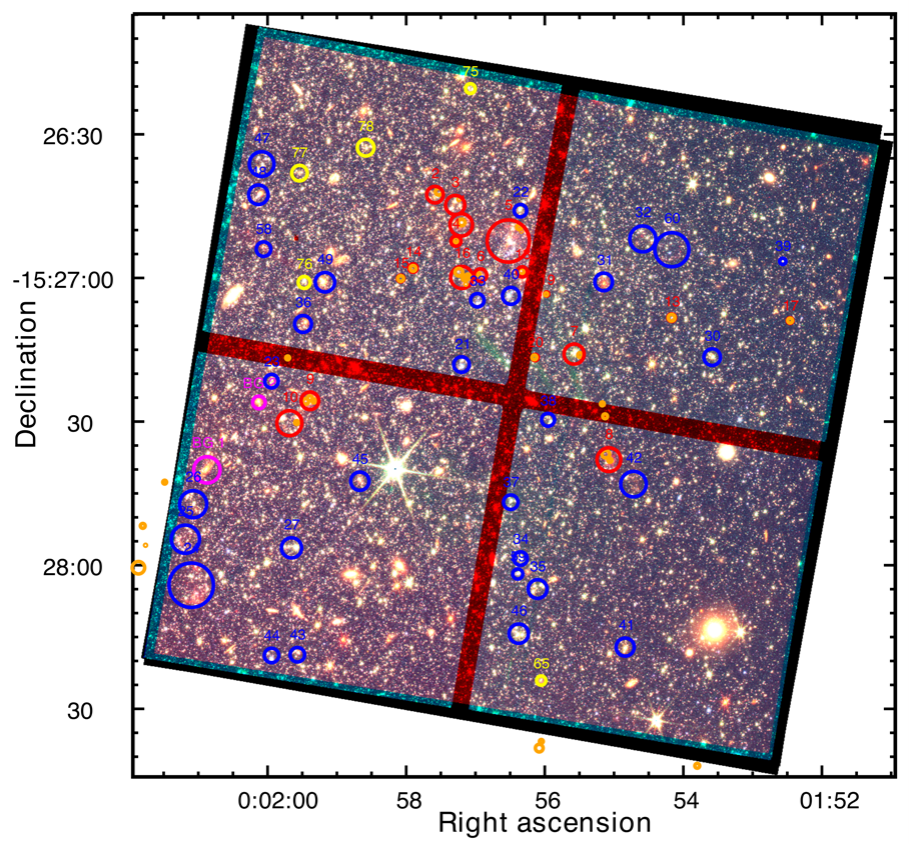}
\caption{F250M (red), F150W (green), F090W (blue) three-color image of \textit{JWST} ERS north field overlaid with type 1 objects (in FUV and near CO, red circles), type 2 objects (in FUV and away from CO, blue circles), type 3 objects (away from FUV and away from CO, yellow), resolved background galaxies (magenta circles), and CO cores (orange circles). The orientation of the image is such that North is up and East is to the left.
\label{fig:jwst_north}}
\end{figure*}

\begin{figure*}[bht!]
\epsscale{1.15}\plotone{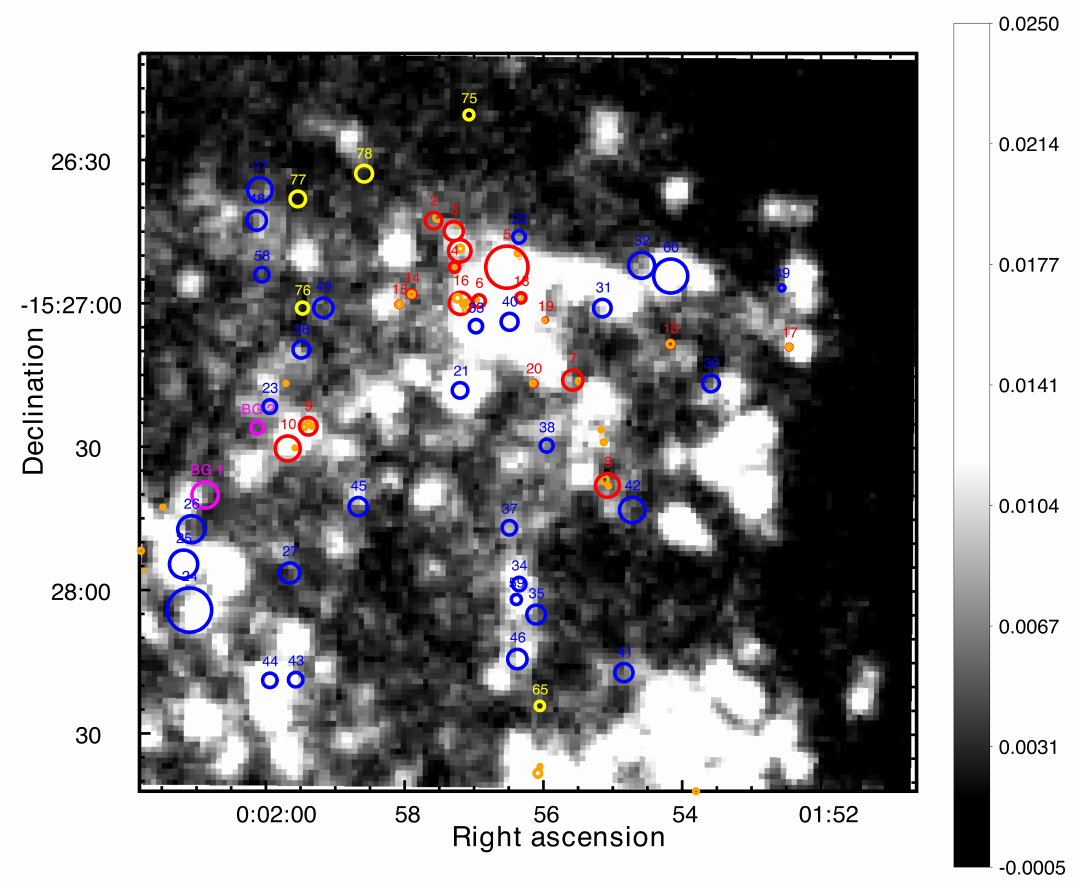}
\caption{FUV image of \textit{JWST} ERS north field overlaid with type 1 objects (in FUV and near CO, red circles), type 2 objects (in FUV and away from CO, blue circles), type 3 objects (away from FUV and away from CO, yellow), resolved background galaxies (magenta circles), and CO cores (orange circles). The orientation of the image is such that North is up and East is to the left.  Colorbar values can be converted from counts/second to calibrated AB magnitudes with the equation: FUV$_{A\!B} = -2.5$log$_{10}($FUV$_{counts/s}) + 18.82$.
\label{fig:fuv_north}}
\end{figure*}

\begin{figure*}[bht!]
\hspace{-2.5em}\epsscale{1.05}\plotone{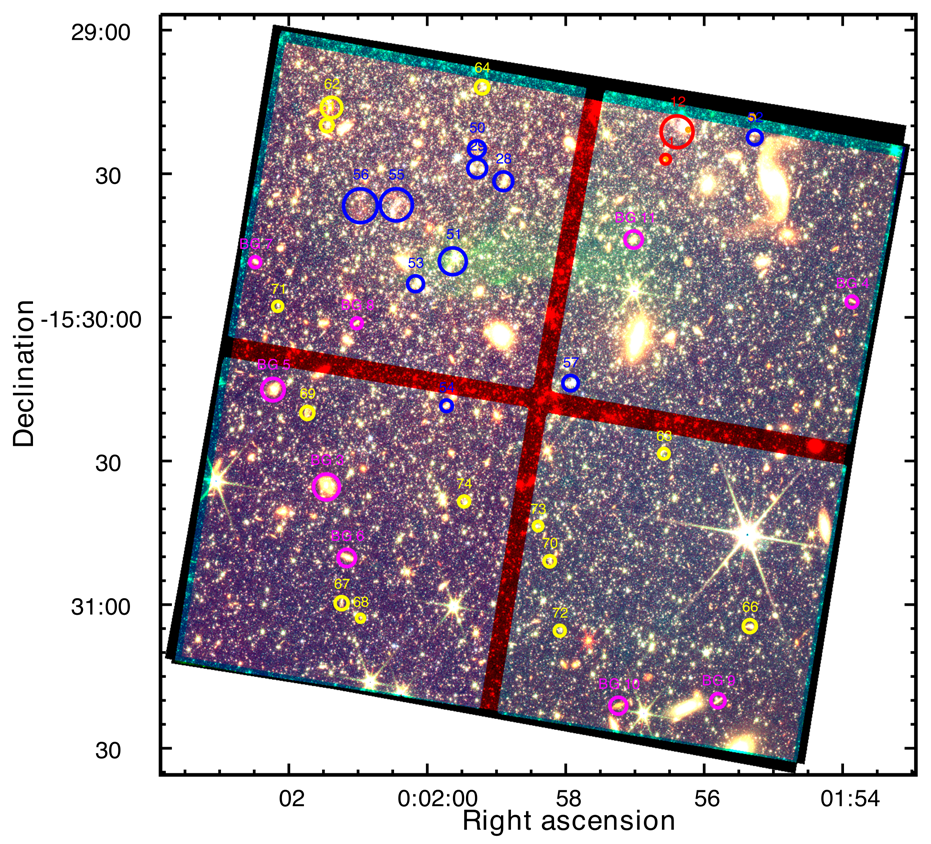}
\caption{F250M (red), F150W (green), F090W (blue) three-color image of \textit{JWST} ERS south field overlaid with type 1 objects (in FUV and near CO, red circles), type 2 objects (in FUV and away from CO, blue circles), type 3 objects (away from FUV and away from CO, yellow), resolved background galaxies (magenta circles), and CO cores (orange circles). The orientation of the image is such that North is up and East is to the left.
\label{fig:jwst_south}}
\end{figure*}

\begin{figure*}[bht!]
\epsscale{1.15}\plotone{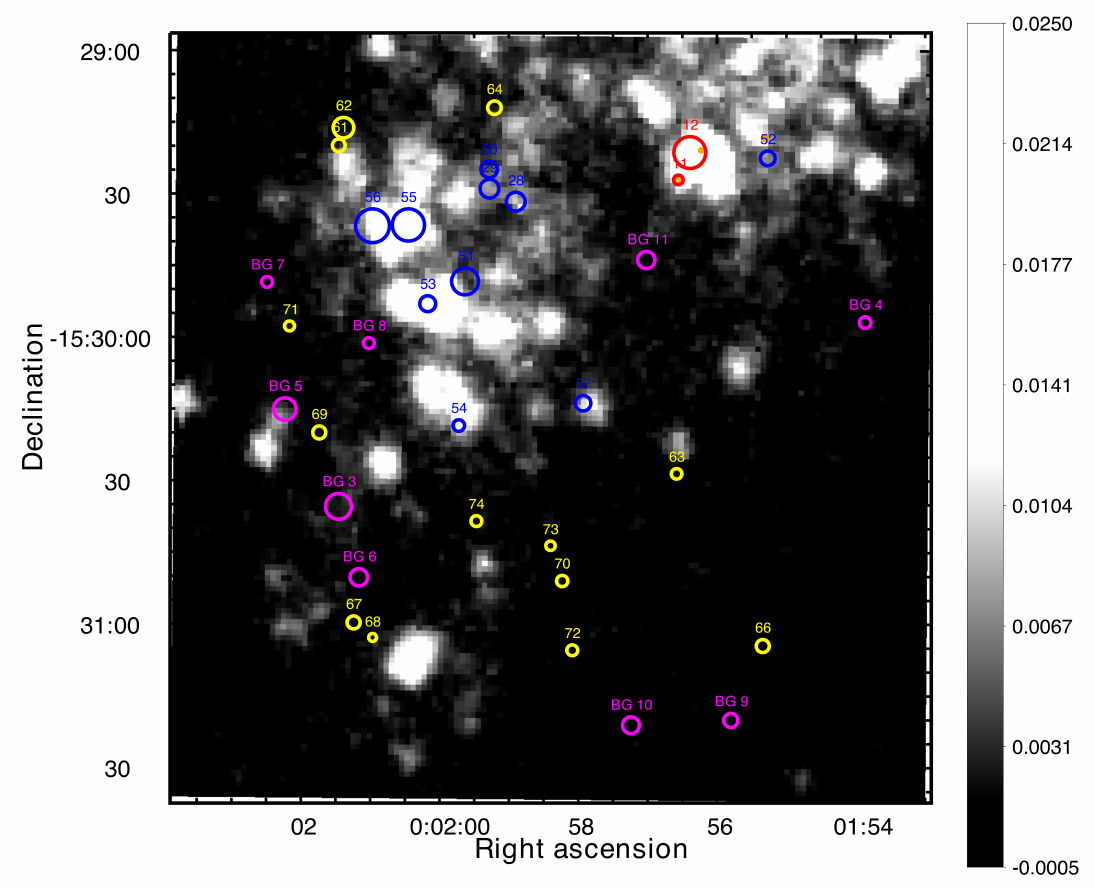}
\caption{FUV image of \textit{JWST} ERS south field overlaid with type 1 objects (in FUV and near CO, red circles), type 2 objects (in FUV and away from CO, blue circles), type 3 objects (away from FUV and away from CO, yellow), resolved background galaxies (magenta circles), and CO cores (orange circles). The orientation of the image is such that North is up and East is to the left.  Colorbar values can be converted from counts/second to calibrated AB magnitudes with the equation: FUV$_{A\!B} = -2.5$log$_{10}($FUV$_{counts/s}) + 18.82$.
\label{fig:fuv_south}}
\end{figure*}

\begin{figure*}[bht!]
\epsscale{1.15}\plotone{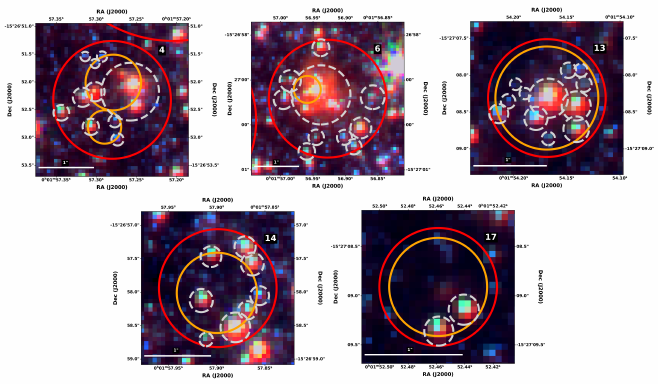}
\caption{F250M (red), F150W (green), F090W (blue) three-color image postage stamps of select type 1 objects (red circles) and the individual NIR sources within (gray, dashed circles) that we used to compare photometry between our larger objects and the smaller, individual sources within. We also overlay the CO cores (orange circles) inside the type 1 objects. The orientation of the image is such that North is up and East is to the left.
\label{fig:indSource}}
\end{figure*}

\section{Data} \label{sec:data}

\subsection{CO cores}
\citet{Rubio_2015} imaged two star-forming regions of WLM in CO(1--0) with ALMA Cycle 1 and detected 10 CO cores, with an additional 35 cores detected from CO(2--1) ALMA Cycle 6 observations (Rubio et al., in preparation). The beam size was $0\farcs9 \times 1\farcs3$ for Cycle 1 and $0\farcs6 \times 0\farcs5$ for Cycle 6, and the observations covered most of the star-forming area of WLM. Further details on the CO(1--0) detection are described by \citet{Rubio_2015}.

Rubio et al.\ (in preparation) used the CPROPS algorithm \citep{Rosolowsky_2006} in the ALMA CO(2--1) data-cube to identify the molecular clouds. They considered as true clouds those with sizes larger than the beam size spatial resolution, a velocity width (FWHM) greater than 3 channels ($> $0.5 \kms), and a noise level above 3 sigma rms. With these criteria 35 clouds were identified and their properties derived. All these clouds have signal-to-noise (S/N) ratios larger than $5$ and $V_{lsr}$ between $-150$ and $-110$ \kms. 

The properties of the CO(2--1) clouds such as the radius ($R$), velocity dispersion ($\sigma_{\upsilon}$), CO flux ($F_{\rm CO}$) were calculated by CPROPS using the moment method in the position-position-velocity data cube. For the calculation of the radius $R$, CPROPS uses the \citet{Solomon_1987} definition for spherical clouds assuming a factor of 1.91 to convert the second moments of the emission along the major and minor axes of clouds ($\sigma_{r}$) to $R$. For unresolved clouds along the minor axis, Rubio et al.\ obtained the radii following the \citet{Saldano_2023} calculation, in which the radii were recalculated assuming that the minor axis is equal to the minor beam axis size. 

They found that the WLM clouds are tiny with sizes between 0.6 pc to 3.8 pc. The velocity dispersion of the clouds is quite small ranging form 0.4 \kms\ to 1.3 \kms. The CO luminosity ranges from $2.5 \times 10^{2}$ K \kms\ to $15.2 \times10^{2}$ K \kms. The derived virial mass of the clouds ranges between $0.4 \times10^{3}$ \Msun\ to $7.6 \times10^{3}$ \Msun.

\subsection{JWST ERS \#1334}
The F090W ($\lambda = 0.9\ \mu m$), F150W ($\lambda = 1.5\ \mu m$), F250M ($\lambda = 2.5\ \mu m$), and F430M ($\lambda = 4.3\ \mu m$) NIRCam images come from the \textit{JWST} Resolved Stellar Populations ERS program (PID 1334). All the \textit{JWST} data used in this paper can be found in MAST: \dataset[https://doi.org/10.17909/wq6j-x975]{https://doi.org/10.17909/wq6j-x975}. The exposure time was 30492 seconds for each of the F090W and F430M filters and 23707 seconds for each of the F150W and F250M filters. Further information on the ERS program and acquired images can be found in \citet{Weisz_2023}.

\begin{deluxetable}{ccc}
\tabletypesize{\normalsize}
\tablecaption{Object type, definition based on location, and color used for the clumps of NIR sources (referred to as objects) examined in this work. \label{tab:obj_type}}
\tablehead{\colhead{\textbf{Object Type}} & \colhead{\textbf{Location}} & \colhead{\textbf{Color}}}
\startdata
\multirow{2}{*}{Type 1} & In FUV knots, &  \multirow{2}{*}{Red} \\[-3pt]
& near CO cores & \\
\tableline
\multirow{2}{*}{Type 2} & In FUV knots, &  \multirow{2}{*}{Blue} \\[-3pt]
& away from  CO cores & \\
\tableline
\multirow{2}{*}{Type 3} & Away from FUV knots, & \multirow{2}{*}{Yellow}\\[-3pt]
& away from CO cores & \\
\enddata
\end{deluxetable}

\begin{deluxetable*}{CCCCCCCCCCC}
\centerwidetable
\tabletypesize{\scriptsize}
\tablecaption{Object number, right ascension, declination, radius, F090W AB magnitude, F090W--F150W color, F250M AB magnitude, F250M--F430M color, blackbody temperature corresponding to the F090W--F150W and F250M--F430M colors, and total mass for type 1 objects. A full table including all objects is available in the online materials. \label{tab:type1}}
\tablehead{\colhead{} & \colhead{RA} & \colhead{Dec} & \colhead{Radius} & \colhead{F090W} & \colhead{} & \colhead{F250M} & \colhead{} & \colhead{T$_{090\!-\!150}$} & \colhead{T$_{250\!-\!430}$} & \colhead{Mass}\\[-6pt]
\colhead{Object} & \colhead{[\textit{J2000}]} & \colhead{[\textit{J2000}]} & \colhead{[\arcsec]} & \colhead{[\textit{AB mag}]} & \colhead{F090W--F150W} & \colhead{[\textit{AB mag}]} & \colhead{F250M--F430M} & \colhead{[\textit{K}]} & \colhead{[\textit{K}]} & \colhead{[$M_\odot$]}\\[-6pt]
\colhead{(1)} & \colhead{(2)} & \colhead{(3)} & \colhead{(4)} & \colhead{(5)} & \colhead{(6)} & \colhead{(7)} & \colhead{(8)} & \colhead{(9)} & \colhead{(10)} & \colhead{(11)}}
\startdata
\multicolumn{11}{c}{Object Type 1 (red)} \\
\tableline
1 & 00:01:57.20 & -15:26:48.9 & 2.34 & 19.84 \pm 0.02 & \phantom{-}0.20 \pm 0.03 & 20.10 \pm 0.03 & -0.65 \pm 0.05 & 3910 \pm \phantom{0}70 & 1420 \pm \phantom{0}310 & 1134 \pm 1118 \\
2 & 00:01:57.58 & -15:26:42.6 & 1.73 & 19.89 \pm 0.02 & \phantom{-}0.38 \pm 0.03 & 20.18 \pm 0.03 & -0.75 \pm 0.05 & 3520 \pm \phantom{0}60 & 1280 \pm \phantom{0}530 & \phantom{0}555 \pm \phantom{0}540 \\
3 & 00:01:57.29 & -15:26:44.7 & 2.00 & 19.73 \pm 0.02 & \phantom{-}0.32 \pm 0.03 & 19.87 \pm 0.03 & -0.60 \pm 0.04 & 3640 \pm \phantom{0}60 & 1320 \pm \phantom{0}230 & \phantom{0}878 \pm \phantom{0}859 \\
4 & 00:01:57.28 & -15:26:52.3 & 1.06 & 22.05 \pm 0.03 & \phantom{-}0.01 \pm 0.05 & 22.35 \pm 0.07 & -0.02 \pm 0.10 & 4450 \pm 160 & 1610 \pm \phantom{0}120 & 1961 \pm 1959 \\
5 & 00:01:56.52 & -15:26:52.2 & 4.41 & 18.15 \pm 0.02 & -0.07 \pm 0.03 & 18.92 \pm 0.02 & -0.73 \pm 0.03 & 4750 \pm 100 & 1720 \pm \phantom{0}310 & 2355 \pm 2285 \\
6 & 00:01:56.92 & -15:26:59.4 & 1.31 & 21.32 \pm 0.03 & \phantom{-}0.29 \pm 0.04 & 21.36 \pm 0.04 & -0.62 \pm 0.07 & 3710 \pm \phantom{0}80 & 1350 \pm \phantom{0}450 & \phantom{0}821 \pm \phantom{0}817 \\
7 & 00:01:55.57 & -15:27:15.9 & 2.10 & 18.91 \pm 0.02 & -0.13 \pm 0.03 & 19.44 \pm 0.02 & -0.52 \pm 0.04 & 4980 \pm 110 & 1800 \pm \phantom{0}150 & 3388 \pm 3348 \\
8 & 00:01:55.07 & -15:27:37.9 & 2.48 & 19.87 \pm 0.02 & \phantom{-}0.36 \pm 0.03 & 18.71 \pm 0.02 & \phantom{-}1.38 \pm 0.03 & 3560 \pm \phantom{0}60 & 1300 \pm \phantom{00}10 & 1012 \pm \phantom{0}941 \\
9 & 00:01:59.38 & -15:27:25.6 & 1.84 & 20.05 \pm 0.02 & -0.06 \pm 0.03 & 20.31 \pm 0.03 & -0.45 \pm 0.05 & 4680 \pm 120 & 1700 \pm \phantom{0}150 & 2644 \pm 2629 \\
10 & 00:01:59.68 & -15:27:30.3 & 2.61 & 18.82 \pm 0.02 & \phantom{-}0.28 \pm 0.03 & 19.22 \pm 0.02 & -0.87 \pm 0.04 & 3730 \pm \phantom{0}60 & 1360 \pm \phantom{0}900 & \phantom{0}589 \pm \phantom{0}553 \\
11 & 00:01:56.55 & -15:29:26.9 & 1.00 & 21.65 \pm 0.03 & \phantom{-}0.01 \pm 0.04 & 21.99 \pm 0.06 & -0.90 \pm 0.10 & 4450 \pm 130 & 1610 \pm 2990 & \phantom{0}357 \pm \phantom{0}354 \\
12 & 00:01:56.39 & -15:29:21.2 & 3.33 & 18.81 \pm 0.02 & -0.27 \pm 0.03 & 19.20 \pm 0.02 & -0.08 \pm 0.03 & 5650 \pm 150 & 2040 \pm \phantom{00}50 & 5930 \pm 5875 \\
13 & 00:01:54.16 & -15:27:08.3 & 0.80 & 22.39 \pm 0.04 & \phantom{-}0.09 \pm 0.05 & 22.83 \pm 0.08 & \phantom{-}0.24 \pm 0.11 & 4210 \pm 160 & 1530 \pm \phantom{0}100 & 1303 \pm 1301 \\
14 & 00:01:57.90 & -15:26:57.9 & 0.88 & 21.99 \pm 0.03 & \phantom{-}0.31 \pm 0.04 & 22.68 \pm 0.08 & -1.04 \pm 0.15 & 3660 \pm \phantom{0}90 & 1330 \pm 3280 & \phantom{00}92 \pm \phantom{00}91 \\
15 & 00:01:58.07 & -15:27:00.1 & 0.70 & 23.12 \pm 0.05 & -0.42 \pm 0.08 & 24.74 \pm 0.20 & -1.25 \pm 0.40 & 6640 \pm 680 & 2380 \pm 4150 & 1867 \pm 1866 \\
16 & 00:01:57.19 & -15:26:59.8 & 2.31 & 19.72 \pm 0.02 & \phantom{-}0.24 \pm 0.03 & 20.22 \pm 0.03 & -0.60 \pm 0.05 & 3830 \pm \phantom{0}70 & 1390 \pm \phantom{0}260 & \phantom{0}883 \pm \phantom{0}867 \\
17 & 00:01:52.45 & -15:27:08.8 & 0.60 & 24.10 \pm 0.07 & \phantom{-}0.12 \pm 0.10 & 25.06 \pm 0.23 & -1.28 \pm 0.47 & 4110 \pm 300 & 1490 \pm 4770 & \phantom{000}4 \pm \phantom{000}4 \\
18 & 00:01:56.32 & -15:26:58.7 & 1.07 & 21.17 \pm 0.03 & -0.07 \pm 0.04 & 21.94 \pm 0.06 & -0.85 \pm 0.10 & 4720 \pm 160 & 1710 \pm 2800 & \phantom{0}381 \pm \phantom{0}378 \\
19 & 00:01:55.97 & -15:27:03.3 & 0.44 & 22.96 \pm 0.05 & \phantom{-}0.31 \pm 0.06 & 22.62 \pm 0.08 & \phantom{-}0.15 \pm 0.10 & 3660 \pm 130 & 1330 \pm \phantom{0}100 & \phantom{0}937 \pm \phantom{0}936 \\
20 & 00:01:56.14 & -15:27:16.6 & 0.66 & 24.03 \pm 0.07 & \phantom{-}0.07 \pm 0.10 & 24.93 \pm 0.22 & -0.55 \pm 0.35 & 4280 \pm 320 & 1550 \pm 3670 & \phantom{000}1 \pm \phantom{000}1 \\
\enddata
\tablecomments{Table 2 is published in its entirety in the supplemental materials in machine-readable format.}
\end{deluxetable*}
\begin{deluxetable}{LCCCC}
\centerwidetable
\tabletypesize{\normalsize}
\tablecaption{CO core number, right ascension, declination, and radius for the 28 CO cores used in this work. \label{tab:CO_core}}
\tablehead{\colhead{} & \colhead{RA} & \colhead{Dec} & \colhead{Radius} & \colhead{$M_{vir}$}\\[-6pt]
\colhead{Object} & \colhead{[\textit{J2000}]} & \colhead{[\textit{J2000}]} & \colhead{[\arcsec]} & \colhead{[$M_\odot$]}\\[-6pt]
\colhead{(1)} & \colhead{(2)} & \colhead{(3)} & \colhead{(4)} & \colhead{(5)}}
\startdata
1 & 00:01:57.29 & -15:26:52.8 & 0.31 & 1087 \pm \phantom{0}919\\
2 & 00:01:57.90 & -15:26:58.0 & 0.56 & 1561 \pm \phantom{0}985\\
3 & 00:01:58.08 & -15:27:00.1 & 0.56 & \phantom{0}894 \pm \phantom{0}637\\
4 & 00:01:52.46 & -15:27:08.9 & 0.10 & \phantom{0}600 \pm 1100\\
5 & 00:01:54.17 & -15:27:08.3 & 0.15 & 2100 \pm 1000\\
6 & 00:01:55.06 & -15:27:38.1 & 0.08 & 2400 \pm 1200\\
7 & 00:01:55.12 & -15:27:36.8 & 0.17 & 1800 \pm 1600\\
8 & 00:01:55.49 & -15:27:16.1 & 0.10 & 1000 \pm \phantom{0}700\\
9 & 00:01:55.98 & -15:27:03.4 & 0.06 & 3600 \pm 2700\\
10 & 00:01:56.15 & -15:27:16.6 & 0.10 & 3300 \pm 2400\\
11 & 00:01:56.23 & -15:29:20.8 & 0.06 & 3100 \pm 2700\\
12 & 00:01:56.34 & -15:26:50.3 & 0.02 & \phantom{0}300 \pm \phantom{0}400\\
13 & 00:01:56.34 & -15:26:58.6 & 0.13 & 2800 \pm 1600\\
14 & 00:01:56.37 & -15:26:49.4 & 0.10 & \phantom{0}900 \pm \phantom{0}600\\
15 & 00:01:56.57 & -15:29:27.2 & 0.13 & 5600 \pm 4100\\
16 & 00:01:56.96 & -15:26:59.2 & 0.06 & \phantom{0}800 \pm \phantom{0}700\\
17 & 00:01:57.11 & -15:26:59.7 & 0.10 & 1400 \pm 1100\\
18 & 00:01:57.14 & -15:27:01.3 & 0.04 & \phantom{0}400 \pm \phantom{0}300\\
19 & 00:01:57.15 & -15:27:00.0 & 0.13 & 3400 \pm 1200\\
20 & 00:01:57.20 & -15:26:48.4 & 0.15 & 7600 \pm 2300\\
21 & 00:01:57.24 & -15:26:58.8 & 0.17 & 1200 \pm 1100\\
22 & 00:01:57.24 & -15:26:43.6 & 0.04 & \phantom{0}700 \pm \phantom{0}400\\
23 & 00:01:57.28 & -15:26:52.0 & 0.10 & 3100 \pm 1200\\
24 & 00:01:57.53 & -15:26:42.0 & 0.15 & 1400 \pm 1000\\
25 & 00:01:59.33 & -15:27:25.7 & 0.08 & \phantom{0}900 \pm \phantom{0}600\\
26 & 00:01:59.39 & -15:27:25.1 & 0.10 & 3900 \pm 1300\\
27 & 00:01:59.45 & -15:27:25.8 & 0.13 & 1800 \pm 1000\\
28 & 00:01:59.58 & -15:27:30.2 & 0.08 & 1600 \pm 1200\\
\enddata
\end{deluxetable}
\subsection{FUV}
The FUV image comes from the NASA Galaxy Evolution Explorer (\textit{GALEX}\footnote[1]{\textit{GALEX} was operated for NASA by the California Institute of Technology under NASA contract NAS5-98034.}) satellite \citep{Martin_2005} GR4/5 pipeline, and were further reduced by \citet{Zhang_2012}. Figure \ref{fig:fuv_fov} shows the $FUV$ image of WLM overlaid with the detected CO cores and the ALMA field of view (FOV) for both Cycle 1 and Cycle 6, along with the two FOVs from the \textit{JWST} Resolved Stellar Populations ERS program (PID 1334). The colorbar values can be converted from counts per second to calibrated AB magnitudes with the equation: 
\begin{equation}
    FUV_{A\!B} = -2.5\log_{10}(FUV_{counts/s}) + 18.82.
\end{equation}

\begin{deluxetable*}{CCCCCCCCC}
\centerwidetable
\tabletypesize{\normalsize}
\tablecaption{Individual source number, host object number, right ascension, declination, radius, F090W AB magnitude, F090W--F150W color, F250M AB magnitude, and F250M--F430M color for 10 individual sources within Type 1 Objects. A full table including all measured individual sources is available in the online materials. \label{tab:ind_source}}
\tablehead{\colhead{Individual} & \colhead{Host} & \colhead{RA} & \colhead{Dec} & \colhead{Radius} & \colhead{F090W} & \colhead{} & \colhead{F250M} & \colhead{}\\[-6pt]
\colhead{Source} & \colhead{Object} & \colhead{[\textit{J2000}]} & \colhead{[\textit{J2000}]} & \colhead{[\arcsec]} & \colhead{[\textit{AB mag}]} & \colhead{F090W--F150W} & \colhead{[\textit{AB mag}]} & \colhead{F250M--F430M}\\[-6pt]
\colhead{(1)} & \colhead{(2)} & \colhead{(3)} & \colhead{(4)} & \colhead{(5)} & \colhead{(6)} & \colhead{(7)} & \colhead{(8)}}
\startdata
1 & 14 & 00:01:57.87 & -15:26:57.3 & 0.16 & 24.48 \pm 0.09 & \phantom{-}0.43 \pm 0.11 & 25.74 \pm 0.31 & -0.81 \pm \phantom{0}0.54 \\
2 & 14 & 00:01:57.86 & -15:26:57.5 & 0.17 & 24.65 \pm 0.09 & \phantom{-}0.40 \pm 0.12 & 25.10 \pm 0.23 & -0.93 \pm \phantom{0}0.42 \\
3 & 14 & 00:01:57.88 & -15:26:58.5 & 0.22 & 23.00 \pm 0.05 & \phantom{-}0.57 \pm 0.06 & 23.35 \pm 0.10 & -0.96 \pm \phantom{0}0.19 \\
4 & 14 & 00:01:57.86 & -15:26:58.2 & 0.14 & 25.25 \pm 0.12 & \phantom{-}0.24 \pm 0.16 & 26.17 \pm 0.38 & -1.68 \pm \phantom{0}0.91 \\
5 & 14 & 00:01:57.85 & -15:26:58.0 & 0.14 & 25.01 \pm 0.11 & -0.84 \pm 0.20 & 27.09 \pm 0.58 & -1.14 \pm \phantom{0}1.13 \\
6 & 14 & 00:01:57.91 & -15:26:58.1 & 0.17 & 25.05 \pm 0.11 & \phantom{-}0.39 \pm 0.15 & 25.90 \pm 0.33 & -0.91 \pm \phantom{0}0.61 \\
7 & 14 & 00:01:57.90 & -15:26:57.4 & 0.15 & 25.47 \pm 0.13 & \phantom{-}0.36 \pm 0.18 & 26.41 \pm 0.42 & -0.83 \pm \phantom{0}0.75 \\
8 & 14 & 00:01:57.91 & -15:26:58.6 & 0.10 & 26.83 \pm 0.25 & \phantom{-}0.27 \pm 0.33 & 28.49 \pm 1.10 & -1.26 \pm \phantom{0}2.24 \\
9 & 4 & 00:01:57.25 & -15:26:52.1 & 0.52 & 22.67 \pm 0.04 & \phantom{-}0.12 \pm 0.06 & 22.80 \pm 0.08 & -0.11 \pm \phantom{0}0.12 \\
10 & 4 & 00:01:57.30 & -15:26:52.7 & 0.15 & 25.46 \pm 0.13 & \phantom{-}0.18 \pm 0.18 & 26.45 \pm 0.43 & -1.22 \pm \phantom{0}0.86 \\
\enddata
\tablecomments{Table 4 is published in its entirety in the supplemental materials in machine-readable format.}
\end{deluxetable*}

\subsection{Object types} \label{subsec:object}
We used a three-color image combining the \textit{JWST} NIRCam F250M (red), F150W (green), and F090W (blue) images to identify and define three types of objects: those in FUV knots and near CO cores (object \textbf{type 1}), those in FUV knots and away from CO cores (object \textbf{type 2}), and those away from FUV knots and away from CO cores (object \textbf{type 3}). We focus on clumps of NIR sources in this work, which we refer to as ``objects''. Table \ref{tab:obj_type} details each object type, location, and color used in figures and plots for reference as the reader proceeds.

We refer to the CO sources as ``CO cores'' and the FUV sources as ``FUV knots''.  We consider the FUV knots to be the visible parts of star forming regions. The determination of whether a NIR object is near or away from a CO core is made visually; nearby objects encompass the CO core. Figures \ref{fig:jwst_north}, \ref{fig:fuv_north}, \ref{fig:jwst_south}, and \ref{fig:fuv_south} show the three object types overlaid on the \textit{JWST} three-color and FUV images for each of the two \textit{JWST} pointings. Table \ref{tab:type1} lists the RA, Dec, object radius in arcseconds, F090W AB magnitude, and F250M AB magnitude for type 1 objects. A full list for objects of all three types is available in the online materials. We also present the locations, radii, and masses of the 28 CO cores included in this work in Table \ref{tab:CO_core}. All of the CO cores are inside FUV knots, which is why there is no object type ``outside FUV knots and near CO cores''. Additionally, we include four type 2 objects and 12 type 3 objects further south than was mapped with ALMA. The type 2 objects south of the ALMA FOV may contain unobserved CO, and removing them from the sample changes the ratio of type 1 to type 2 objects from 50\% to 56\%. Because the CO cores are all near FUV knots, we do not expect any type 3 objects to be near unobserved CO. A \textit{JWST} three-color postage stamp for each object is included in the appendix.

We then performed aperture photometry on the objects using the Image Reduction and Analysis Facility (\textsc{IRAF}) \citep{Tody_1986} routine \textsc{Apphot} to measure the fluxes of the objects in the F090W, F150W, F250M, and F430M images. The size of the aperture for each object corresponded to the size of the object, shown in Figures \ref{fig:jwst_north}, \ref{fig:fuv_north}, \ref{fig:jwst_south}, and \ref{fig:fuv_south}. We calculated our uncertainties using photon statistics convolved with the known 4\% NIRCam zeropoint uncertainty \citep{Windhorst_2023}. We converted the fluxes of the objects in each image to AB magnitudes using the conversions:
\begin{equation}
    \textnormal{F090W}_{A\!B}\!=\!-2.5\log_{10}(\textnormal{F090W}_{M\!Jy/sr} \times 2.2\times10^{-8}) + 8.90
\end{equation}
\begin{equation}
    \textnormal{F150W}_{A\!B}\!=\!-2.5\log_{10}(\textnormal{F150W}_{M\!Jy/sr} \times 2.3\times10^{-8}) + 8.90
\end{equation}
\begin{equation}
    \textnormal{F250M}_{A\!B}\!=\!-2.5\log_{10}(\textnormal{F250M}_{M\!Jy/sr} \times 9.3\times10^{-8}) + 8.90
\end{equation}
\begin{equation}
    \textnormal{F430M}_{A\!B}\!=\!-2.5\log_{10}(\textnormal{F430M}_{M\!Jy/sr} \times 9.3\times10^{-8}) + 8.90.
\end{equation}

We chose to look at larger clumps instead of individual NIR sources to include extended sources that appeared to be part of the same system. As a test case, we also measured the colors of the individual NIR sources inside a subset of the larger type 1 objects (Figure \ref{fig:indSource}). We include the right ascension (RA), declination (Dec), radius in arcseconds, F090W AB mag, F090W--F150W color, F250M AB mag, and F250M--F430M color for 10 of the individual sources within Type 1 Objects used in our test case in Table \ref{tab:ind_source}, with the full table included in the online materials. We find that the individual sources have similar colors to the larger objects. As such, we chose to stick with the larger objects for the purpose of this paper.

To determine whether our objects are contaminated by unresolved background galaxies, we use \citet[][Section 4 and Figures 6-8]{Windhorst_2023} to estimate the surface density of faint galaxies at our AB magnitude flux limit of $\sim$25 to be approximately 0.02 background galaxies per square arcsecond in all four NIRCam filters. For a typical object with a radius of 1.5 arcseconds, this equates to about 0.14 unresolved background galaxies within each object.

\begin{deluxetable*}{LCCCCCCC}
\centerwidetable
\tabletypesize{\normalsize}
\tablecaption{Object number, right ascension, declination, radius, F090W AB magnitude, F090W--F150W color, F250M AB magnitude, and F250M--F430M color for our sample of resolved background galaxies. \label{tab:bg_gals}}
\tablehead{\colhead{} & \colhead{RA} & \colhead{Dec} & \colhead{Radius} & \colhead{F090W} & \colhead{} & \colhead{F250M} & \colhead{}\\[-6pt]
\colhead{Object} & \colhead{[\textit{J2000}]} & \colhead{[\textit{J2000}]} & \colhead{[\arcsec]} & \colhead{[\textit{AB mag}]} & \colhead{F090W--F150W} & \colhead{[\textit{AB mag}]} & \colhead{F250M--F430M}\\[-6pt]
\colhead{(1)} & \colhead{(2)} & \colhead{(3)} & \colhead{(4)} & \colhead{(5)} & \colhead{(6)} & \colhead{(7)} & \colhead{(8)}}
\startdata
BG\ 1 & 00:02:00.87 & -15:27:40.0 & 2.74 & 19.99 \pm 0.02 & 0.44 \pm 0.03 & 19.55 \pm 0.02 & -0.02 \pm 0.04 \\
BG\ 2 & 00:02:00.12 & -15:27:26.0 & 1.32 & 20.95 \pm 0.02 & 0.66 \pm 0.03 & 20.33 \pm 0.03 & \phantom{-}0.27 \pm 0.04 \\
BG\ 3 & 00:02:01.45 & -15:30:35.4 & 2.70 & 20.09 \pm 0.02 & 0.85 \pm 0.03 & 18.72 \pm 0.02 & \phantom{-}0.89 \pm 0.03 \\
BG\ 4 & 00:01:53.86 & -15:29:56.7 & 1.21 & 22.61 \pm 0.04 & 0.44 \pm 0.05 & 21.89 \pm 0.06 & -0.38 \pm 0.09 \\
BG\ 5 & 00:02:02.22 & -15:30:15.0 & 2.35 & 20.65 \pm 0.02 & 0.57 \pm 0.03 & 19.84 \pm 0.03 & \phantom{-}0.05 \pm 0.04 \\
BG\ 6 & 00:02:01.16 & -15:30:50.2 & 1.81 & 21.37 \pm 0.03 & 0.53 \pm 0.04 & 20.56 \pm 0.03 & -0.54 \pm 0.05 \\
BG\ 7 & 00:02:02.48 & -15:29:48.4 & 1.13 & 21.99 \pm 0.03 & 0.63 \pm 0.04 & 20.90 \pm 0.04 & -0.36 \pm 0.06 \\
BG\ 8 & 00:02:01.01 & -15:30:01.2 & 1.13 & 22.39 \pm 0.04 & 0.45 \pm 0.05 & 21.63 \pm 0.05 & \phantom{-}0.35 \pm 0.07 \\
BG\ 9 & 00:01:55.80 & -15:31:20.0 & 1.47 & 21.76 \pm 0.03 & 0.50 \pm 0.04 & 21.50 \pm 0.05 & -0.49 \pm 0.07 \\
BG\ 10 & 00:01:57.24 & -15:31:21.1 & 1.76 & 23.05 \pm 0.05 & 0.99 \pm 0.06 & 21.44 \pm 0.05 & \phantom{-}0.73 \pm 0.06 \\
BG\ 11 & 00:01:57.01 & -15:29:43.7 & 1.80 & 20.90 \pm 0.02 & 0.22 \pm 0.03 & 20.97 \pm 0.04 & -0.08 \pm 0.06 \\
\enddata
\end{deluxetable*}

\begin{figure}[hbt!]
\captionsetup[subfloat]{farskip=1pt}
\centering
\subfloat{\includegraphics[width=1\columnwidth]{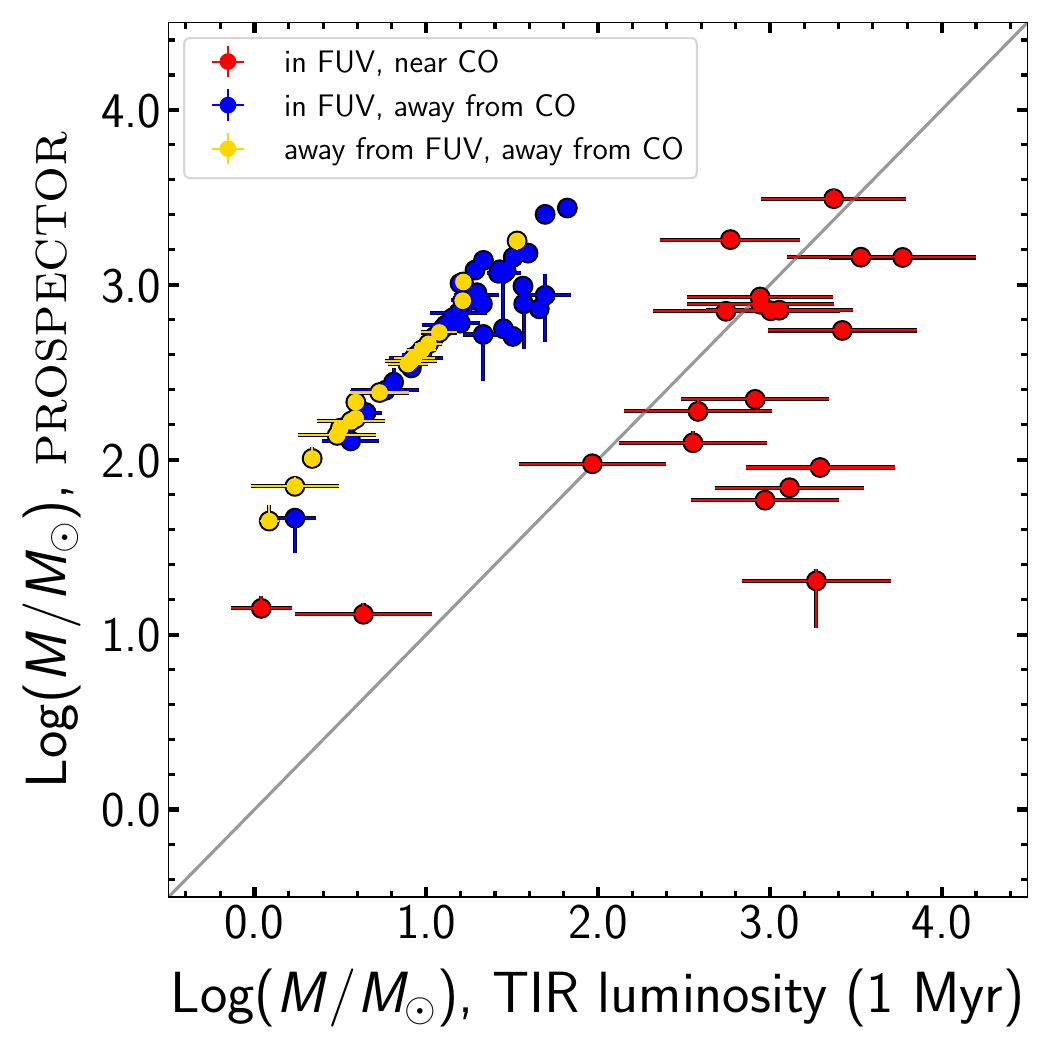}}

\subfloat{\includegraphics[width=1\columnwidth]{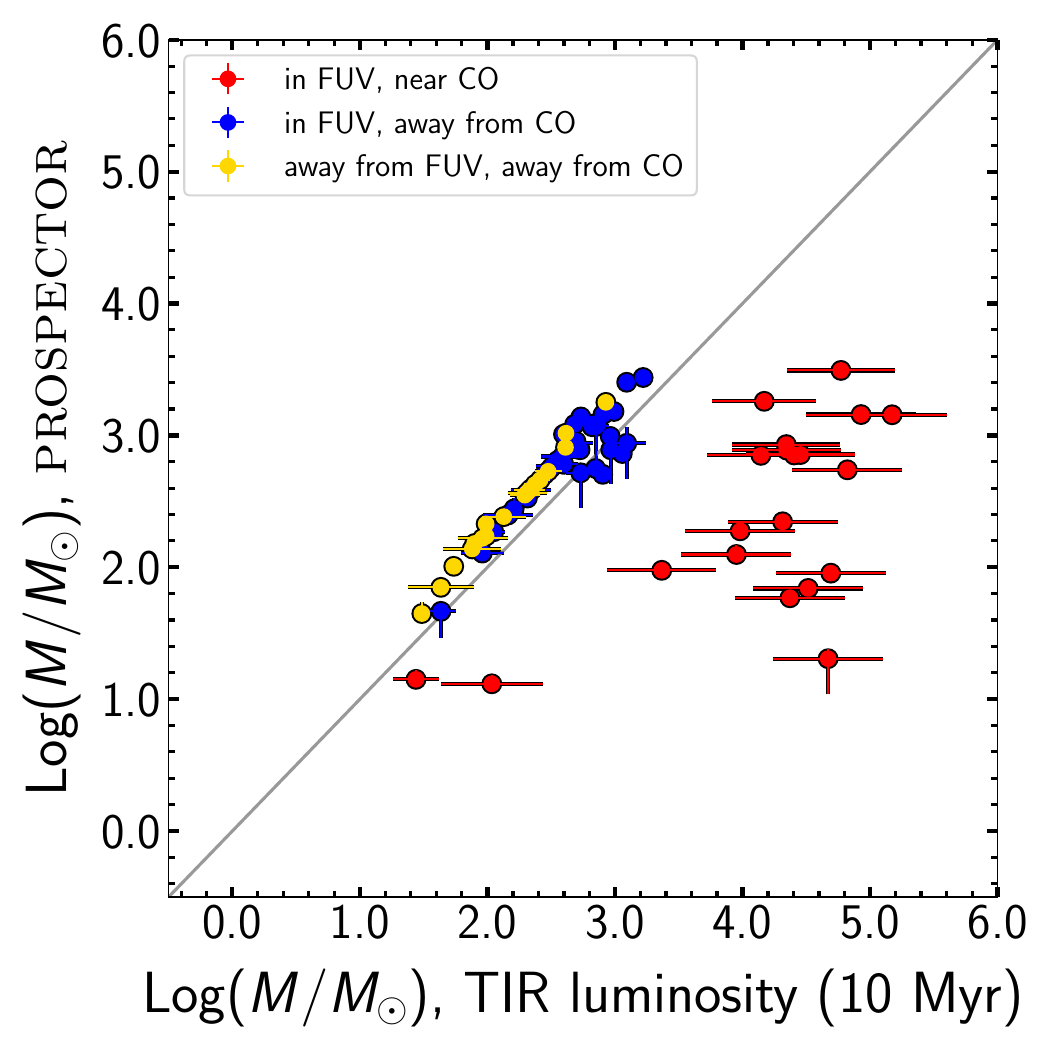}}
\caption{Comparison of object masses computed using \textsc{prospector} versus the TIR luminosity for all three object types, where the TIR luminsoity method assumes an age of 1 Myr (top) or an age of 10 Myr (bottom). The gray line shows where the masses computed using both methods are the same. 
\label{fig:mass_comp}}
\end{figure}

\begin{figure}[hbt!]
\captionsetup[subfloat]{farskip=1pt}
\centering
\subfloat{\includegraphics[width=1\columnwidth]{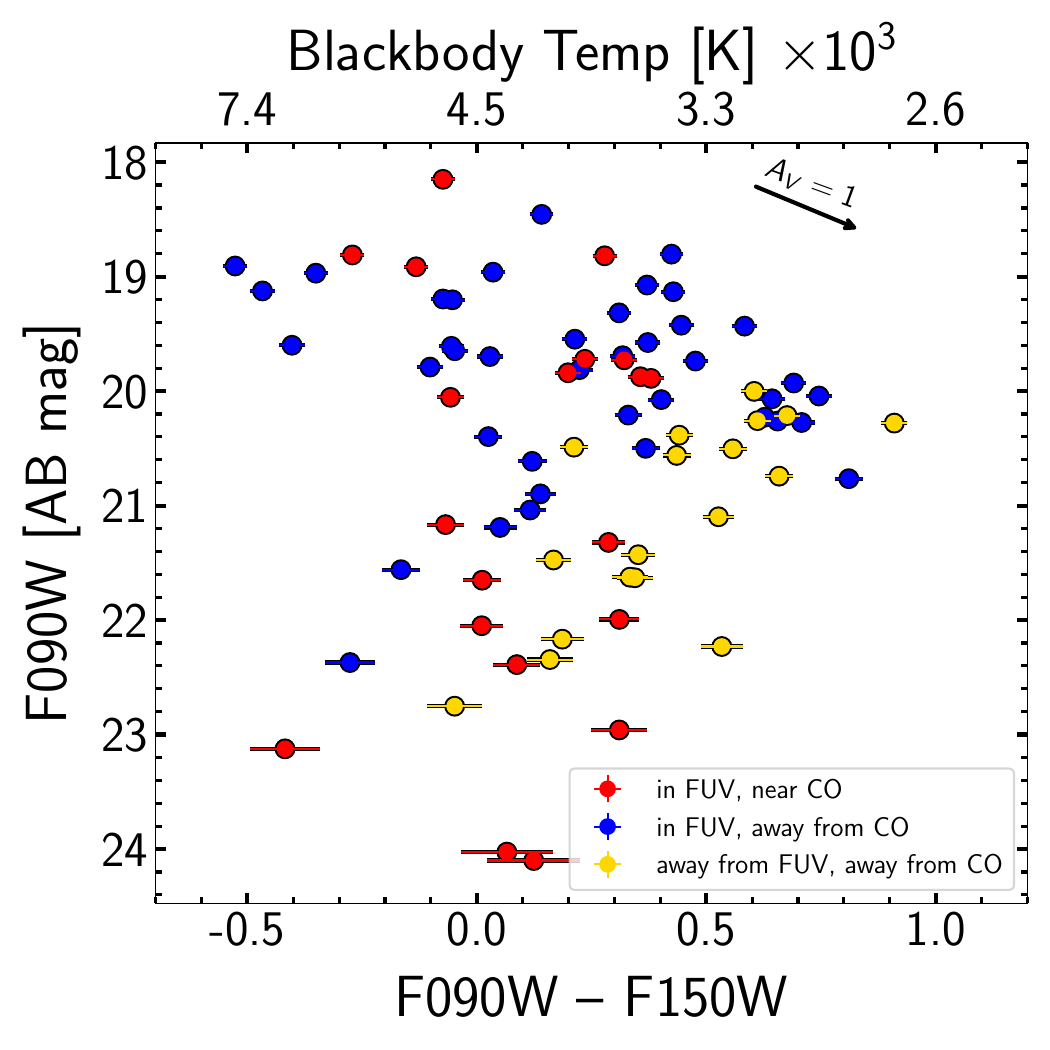}}

\subfloat{\includegraphics[width=1\columnwidth]{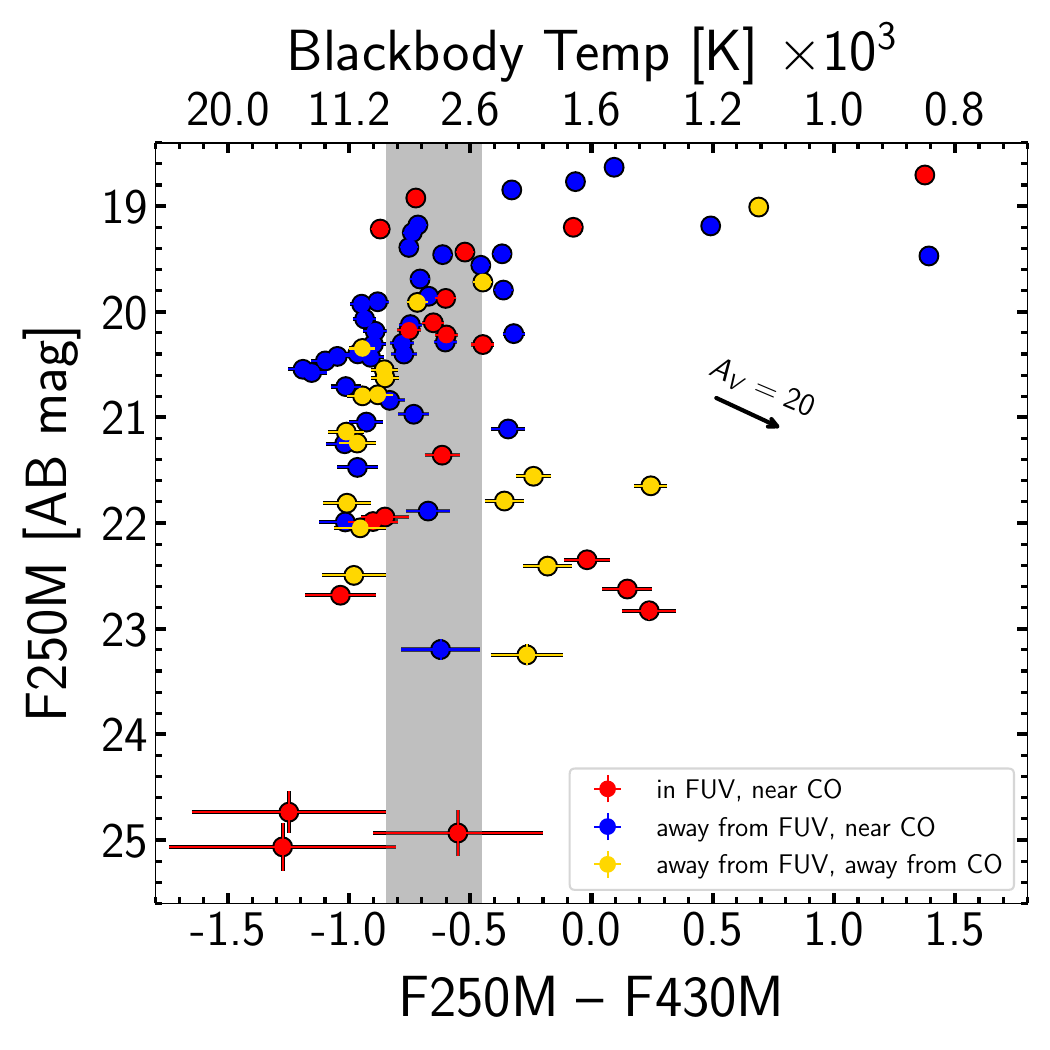}}
\caption{Top: F090W versus F090W--F150W CMD for all objects (red: in FUV knots and near CO, blue: in FUV knots and away from CO, and yellow: away from FUV knots and away from CO). The blackbody temperature corresponding to the F090W--F150W color is shown on the top x-axis, and the reddening vector $A_V=1$ assuming SMC-like extinction is included in the upper right of the plot. Bottom: F250M versus F250M--F430M CMD for all objects. The blackbody temperature corresponding to the F250M--F430M color is shown on the top x-axis and the temperature range from the F090W versus F090W--F150W CMD is shaded in gray. The reddening vector $A_V=20$ assuming SMC-like extinction is included in the upper right of the plot.
\label{fig:CMDs}}
\end{figure}
\begin{figure}[bht!]
\epsscale{1.145}\plotone{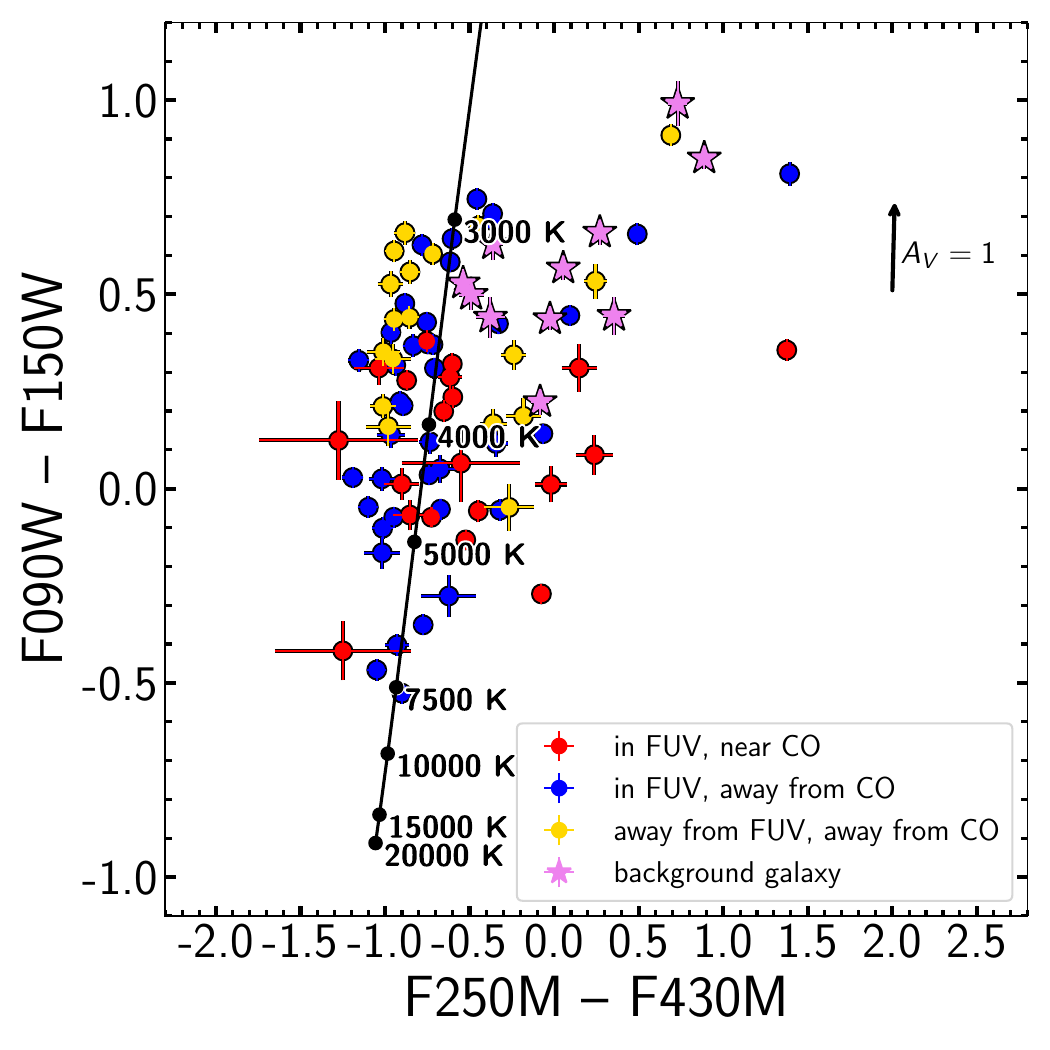}
\caption{Color-color diagram of F090W--F150W versus F250M--F430M for all objects (red circles: in FUV knots and near CO, blue circles: in FUV knots and away from CO, and yellow circles: away from FUV knots and away from CO) including seven resolved background galaxies (light purple stars). A reddening vector of $A_V=1$ assuming SMC-like extinction is included in the upper left of the plot, and the black line shows the effective temperature of a blackbody.
\label{fig:ccd}}
\end{figure}
\subsection{Blackbody Temperature, Luminosity, and Mass Estimates} \label{subsec:blackbody}
We calculated the blackbody temperatures corresponding to the color for the given filters using the Planck law and the \textsc{observate} routine from the Python package \textsc{sedpy} for spectral energy distributions (SEDs) to generate synthetic photometry through the four \textit{JWST} filters \citep{sedpy}. For each object we found the blackbody temperature corresponding to the F090W--F150W and F250M--F430M colors separately.

Assuming our objects are embedded star clusters, we found the luminosity of each object using the SEDs from Table 2 of \citet{Xu_2001}, which gives the flux density versus wavelength for six normal galaxies with $24\ \mu m$ luminosities ranging from $10^7$ to $10^{10.6}\ L_\odot$, two AGNs with $24\ \mu m$ luminosities of $10^8\ L_\odot$ and $ 10^{11}\ L_\odot$, and eight starburst galaxies with $24\ \mu m$ luminosities ranging from $10^7$ to $10^{11.5}\ L_\odot$. For the eight starburst galaxy luminosity bins, we integrated over both the entire SED and a subset of the SED the width of the four \textit{JWST} bands, and found the ratio of the full SED to the JWST-width SED for each. We converted the summed flux of our objects in the four \textit{JWST} filters to luminosity. We then took the average full SED-to-JWST SED ratio of the eight luminosity bins, which we found to be approximately 4, to convert the summed luminosities in the four \textit{JWST} bands to total IR (TIR) luminosities. We note that the higher metallicity of the \citet{Xu_2001} galaxies compared to WLM may have an effect on the luminosities.

We then converted the TIR luminosities to total mass. Taking the bolometric magnitude of a young stellar population at less than 1 Myr age to be -2.69 for Small Magellanic Cloud (SMC) metallicity of 20\% solar \citep{BruzualCharlot_2003} and 4.74 mag as the bolometric magnitude of the Sun, the bolometric luminosity per solar mass of young stars is then $3.01 \times 10^{35} \times 10^{0.4\times2.69}$ \citep{Elmegreen_2018}. We then took the TIR luminosity divided by this conversion for bolometric luminosity per solar mass of young stars at 1 Myr to get the mass of our objects.

We also computed the masses of our objects using the SED fitting toolbox \textsc{prospector} \citep{Leja_2017,Johnson_2021,python-fsps} for comparison with the masses derived from the TIR luminosities. We used dynamic nested sampling with the Python package \textsc{dynesty} \citep{Conroy_2009,Conroy_2010,Speagle_2020,dynesty} assuming a single stellar population and fixed metallicity. The mass, diffuse dust \textit{V}-band optical depth, and age were free parameters. We found that the optical depths and ages derived from this method were not reasonable for our objects, with most objects having ages in the highest age bin independent of the parameter priors used. As such, we opted to use the masses computed with the TIR method for this work. Comparing the masses between the two methods, we find that type 1 objects are similar between methods within the larger uncertainties of the TIR luminosity method, while object types 2 and 3 have systematically higher masses using \textsc{prospector} by about 1.5 dex (Figure \ref{fig:mass_comp}, top). One reason for this may be because the TIR method assumes an age of 1 Myr. One reason for this may be because the TIR method assumes an age of 1 Myr, which is reasonable for embedded star clusters. Increasing the age to 10 Myr, which changes the bolometric magnitude of the stellar population from -2.69 to +0.81, increases the mass of the object and brings the masses of object types 2 and 3 computed with the TIR luminosity method closer to those computed using \textsc{prospector} (Figure \ref{fig:mass_comp}, bottom).

\begin{figure}[hbt!]
\epsscale{1.145}\plotone{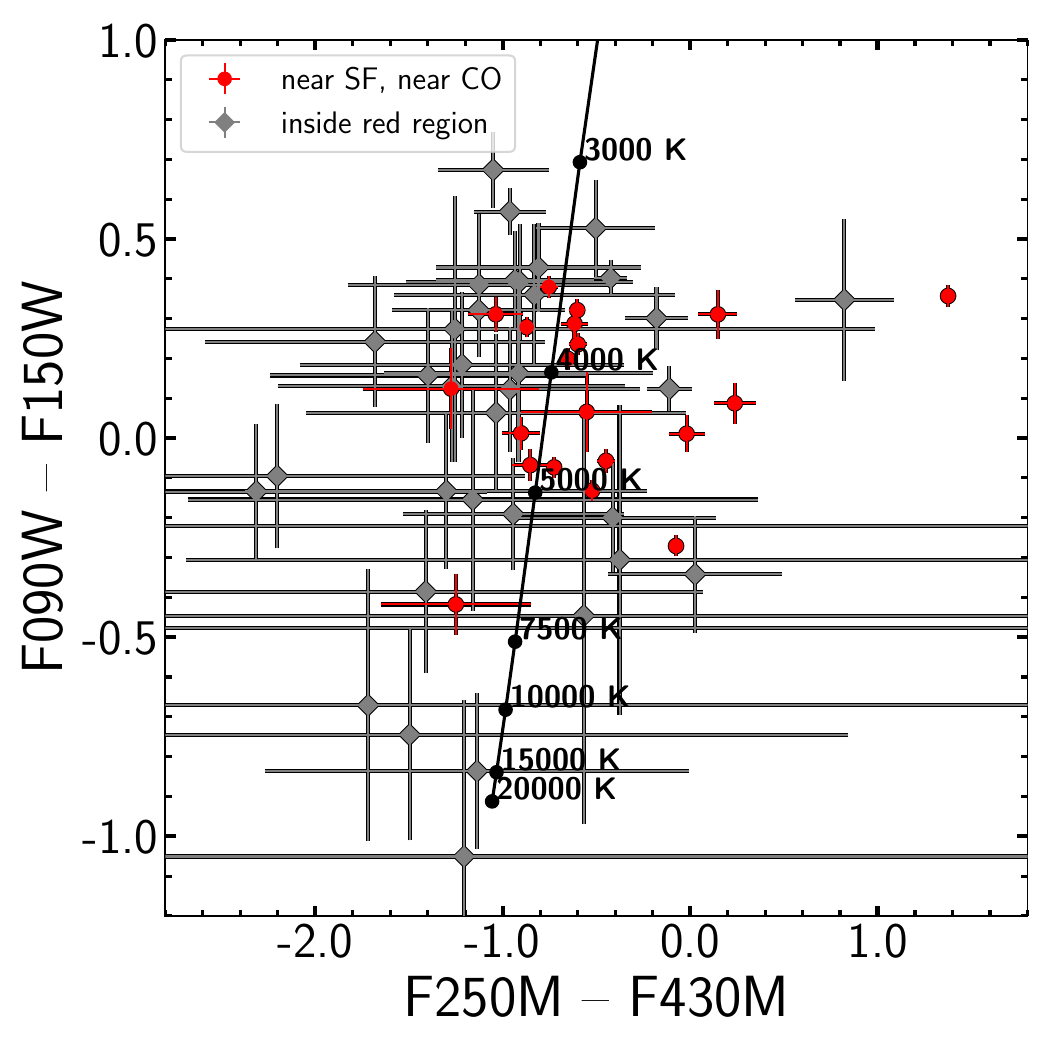}
\caption{Color-color diagram of F090W--F150W versus F250M--F430M for type 1 objects (red circles) and resolved NIR sources within type 1 objects (gray diamonds). The black line corresponds to the effective temperature of a blackbody.
\label{fig:ccd_small}}
\end{figure}
\begin{figure}[tbh!]
\epsscale{1.145}\plotone{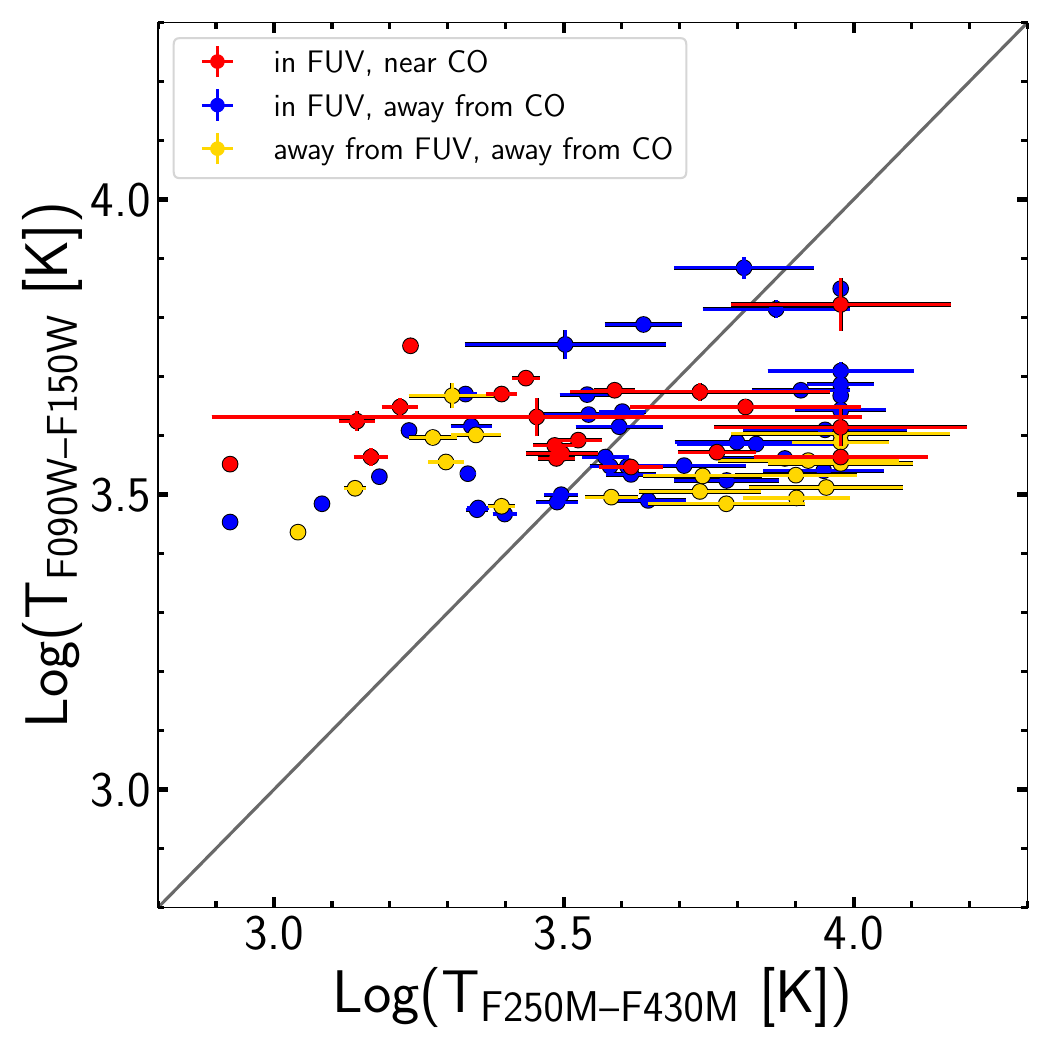}
\caption{Log(temperature)-log(temperature) plot of F090W--F150W versus F250M--F430M for all objects (red: type 1 objects, blue: type 2 objects, and yellow: type 3 objects). The gray line shows where temperatures for both colors are the same to indicate how well the objects are represented by a blackbody. If our objects are embedded clusters, the infrared colors are affected by extinction and infrared excess of young stars and less sensitive to effective temperature, which may explain the larger F250M--F430M uncertainties at higher temperatures.
\label{fig:TT}}
\end{figure}
\begin{figure*}[hbt!]
\captionsetup[subfloat]{farskip=-5pt}
\centering 
\subfloat{\includegraphics[width=0.33\linewidth]{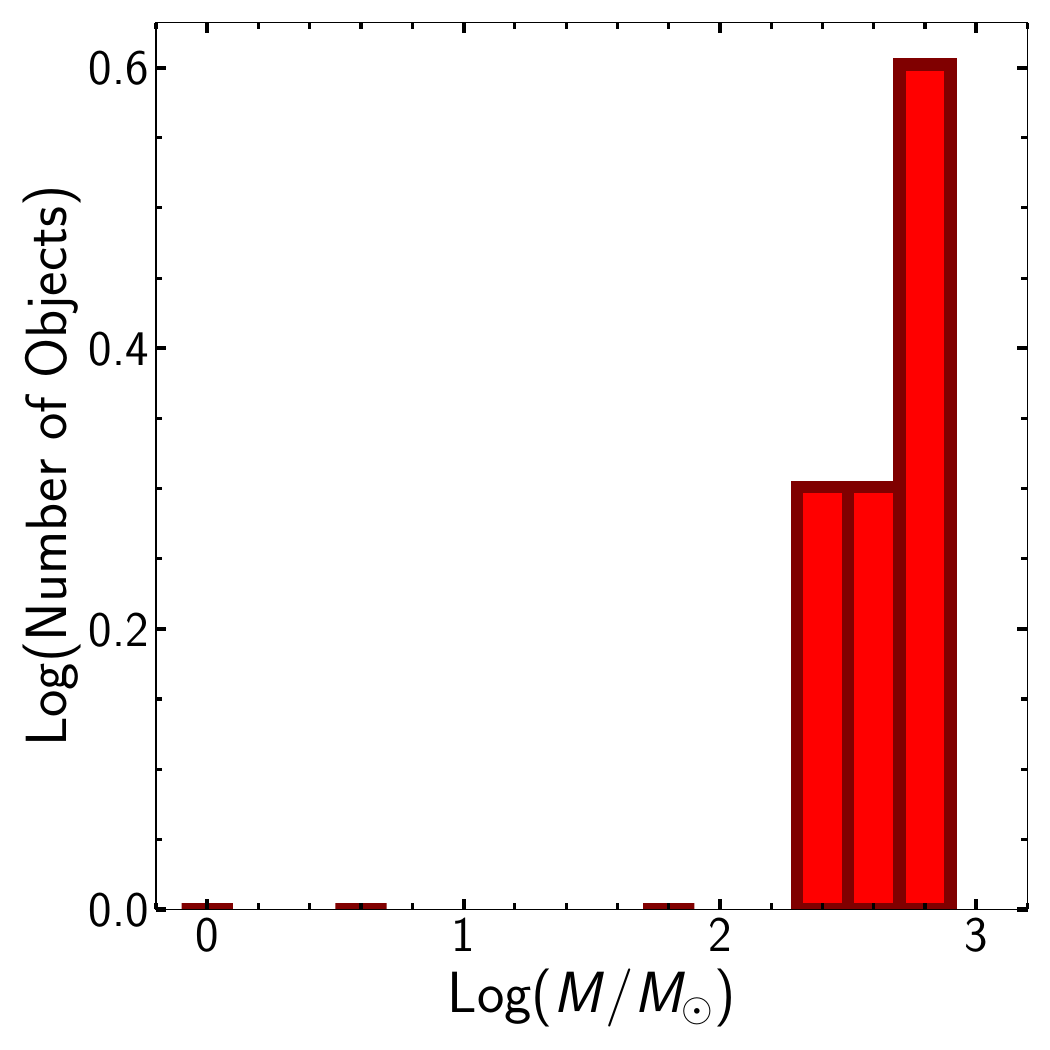}}\hspace{-0.69em}
\subfloat{\includegraphics[width=0.33\linewidth]{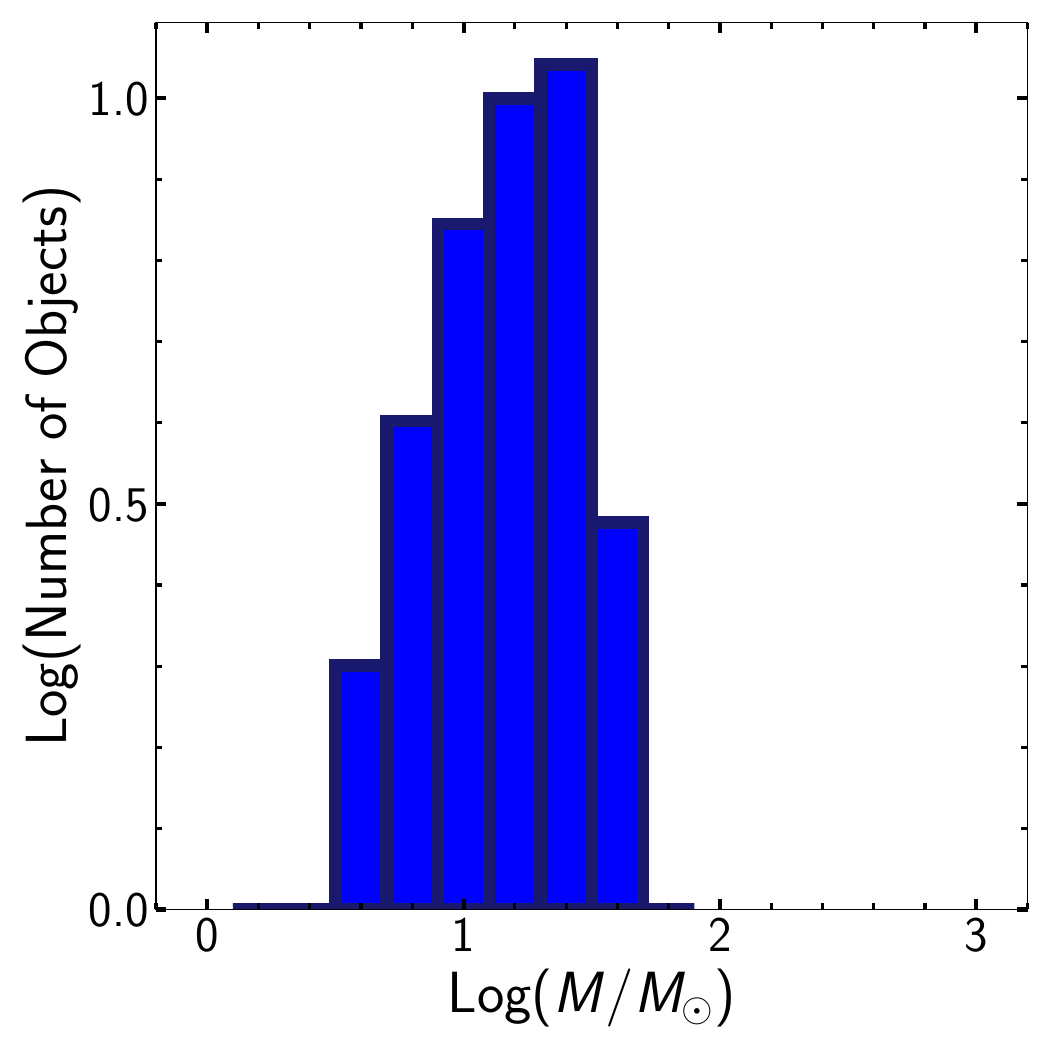}}\hspace{-0.67em}
\subfloat{\includegraphics[width=0.33\linewidth]{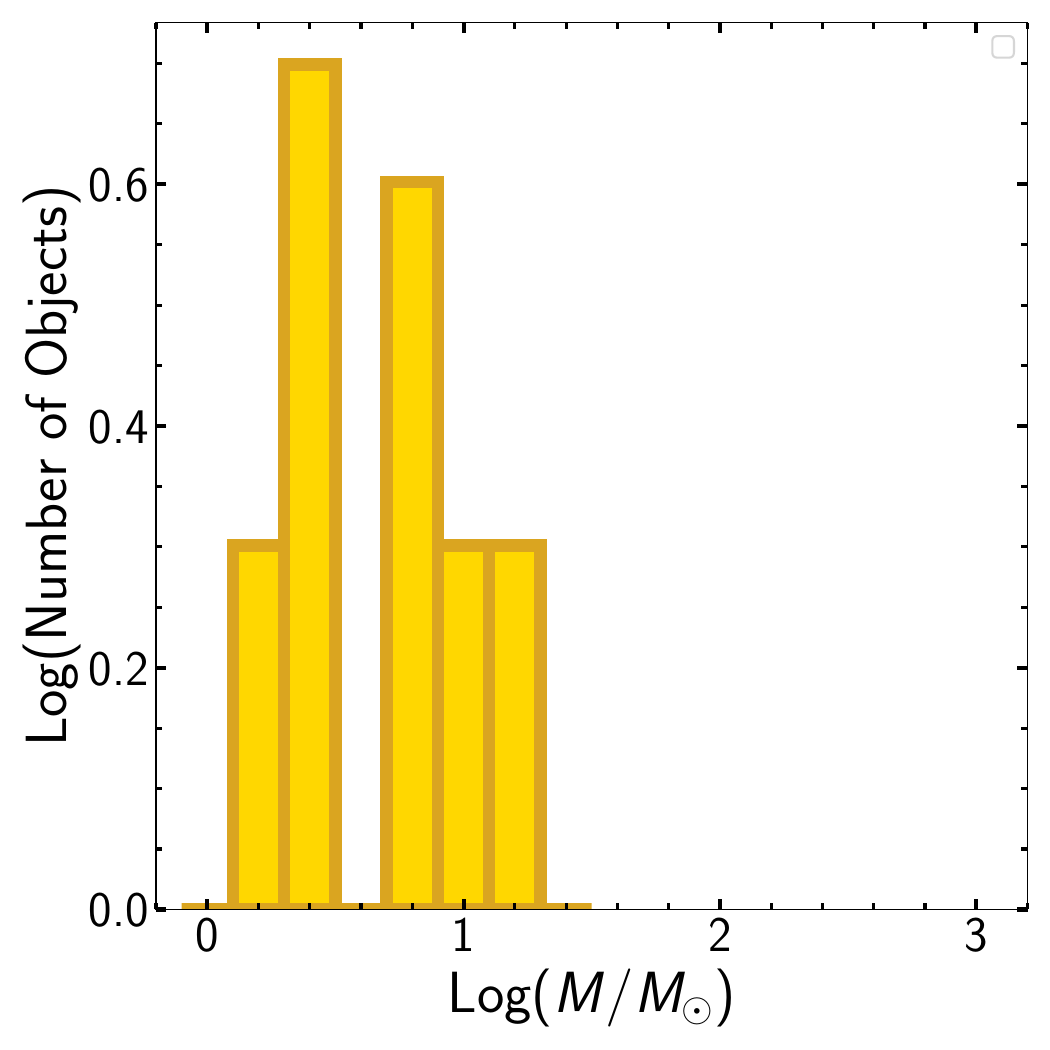}}

\subfloat{\includegraphics[width=0.33\linewidth]{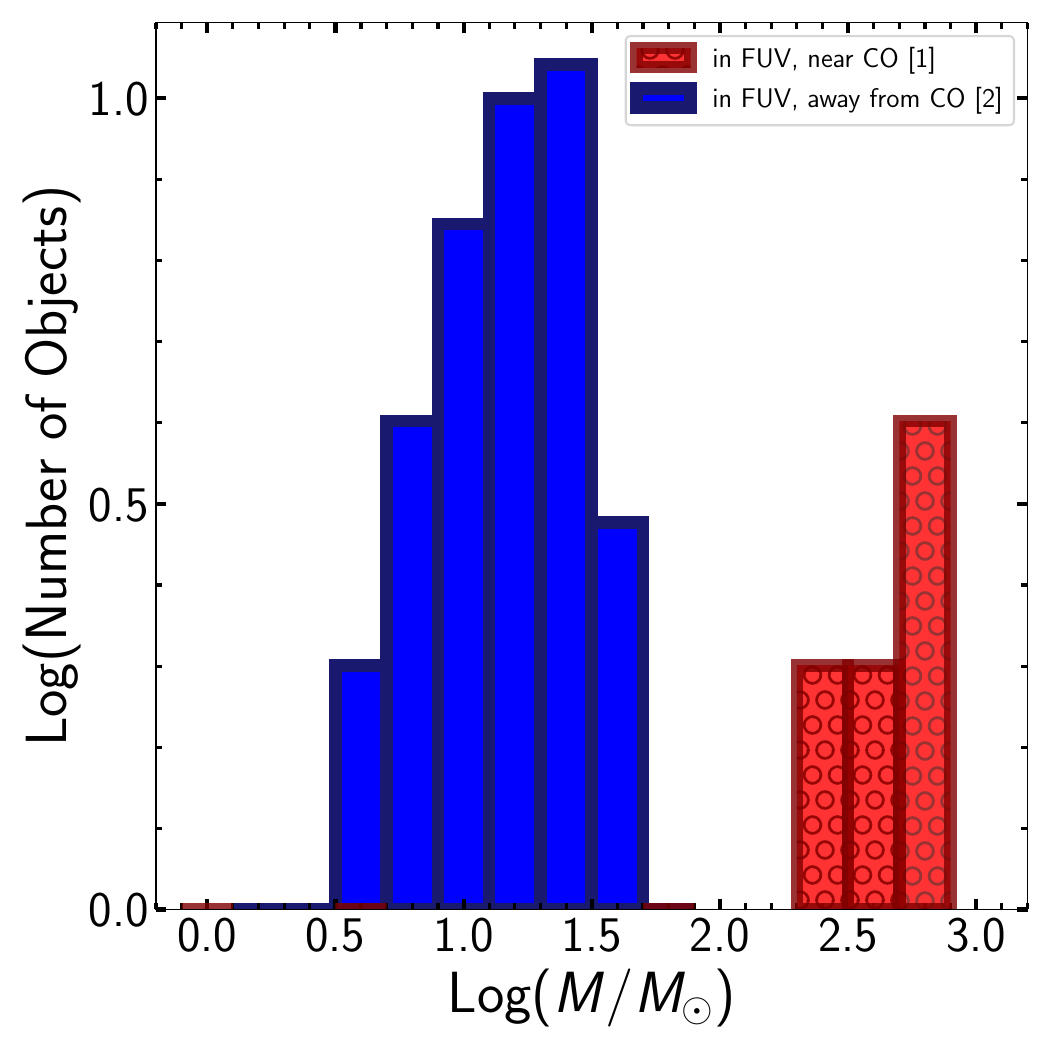}}\hspace{-0.69em}
\subfloat{\includegraphics[width=0.33\linewidth]{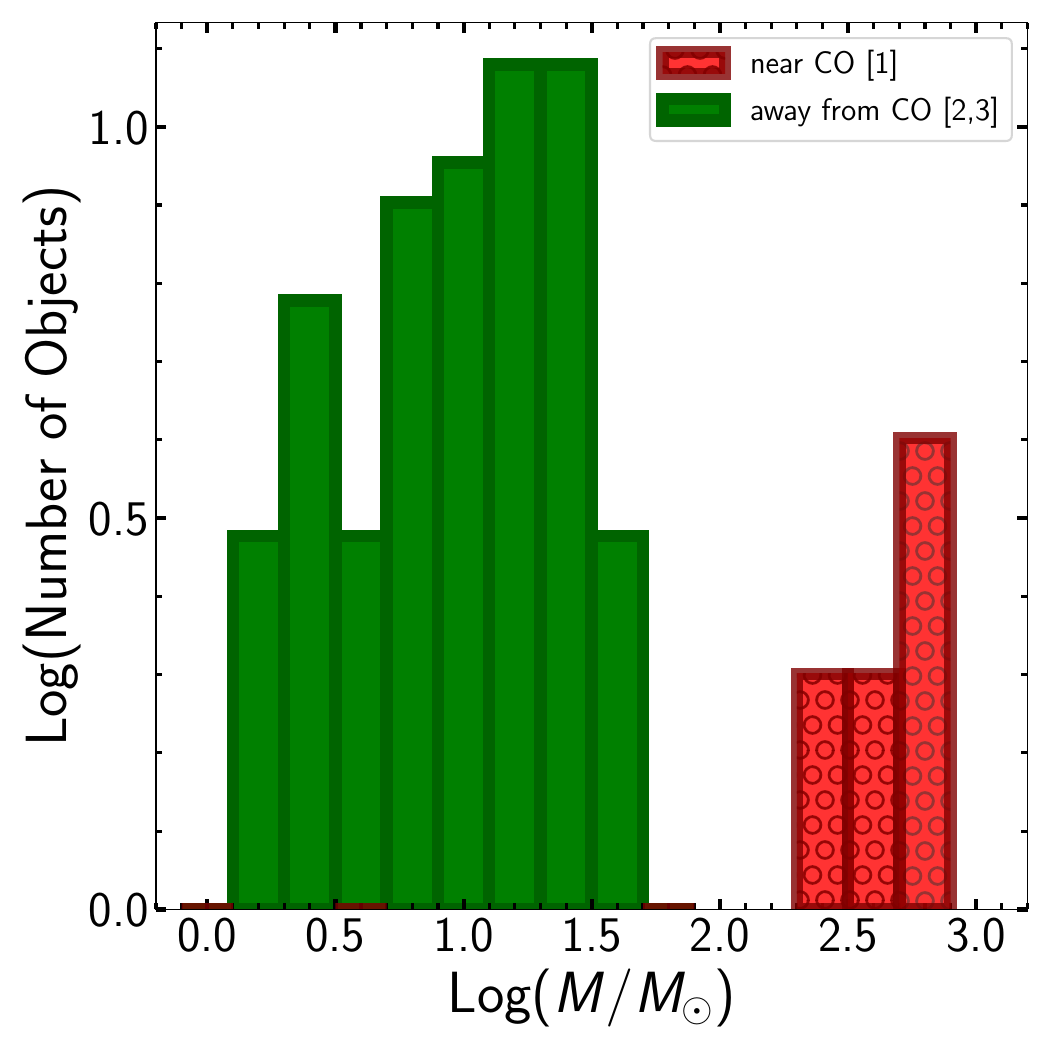}}\hspace{-0.67em}
\subfloat{\includegraphics[width=0.33\linewidth]{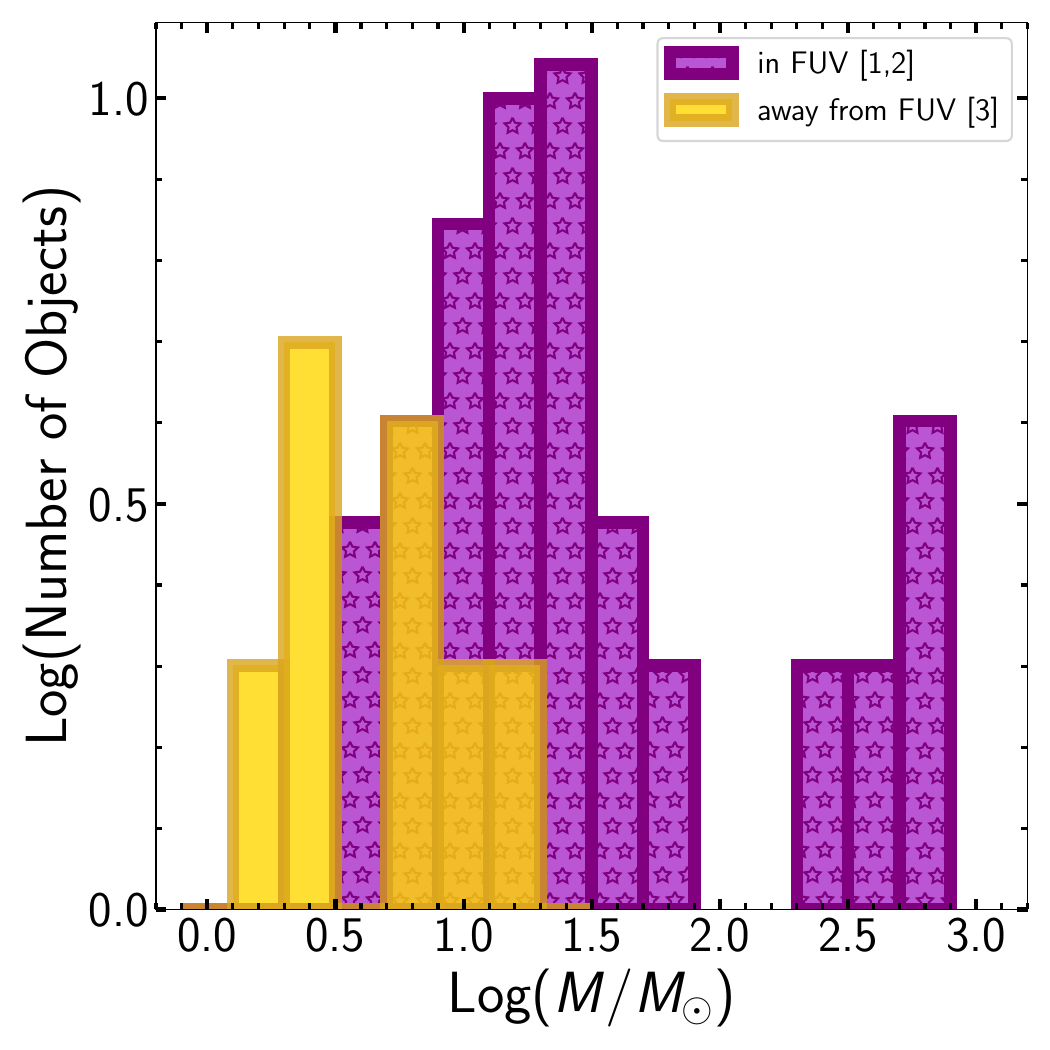}}
\caption{\textbf{Top:} \textit{Left-} Mass distribution of type 1 objects (red). \textit{Center-} Mass distribution of type 2 objects (blue). \textit{Right-} Mass distribution of type 3 objects (yellow). \textbf{Bottom:} \textit{Left-} Mass distribution comparing only objects in FUV knots -- those near CO (red with circle hatches, type 1) and those away from CO (blue, type 2). \textit{Center-} Mass distribution comparing objects near (red with circle hatches, type 1) and away from CO (green, types 2 and 3), independent of proximity to FUV. \textit{Right-} Mass distribution comparing objects near (purple with star hatches, types 1 and 2) and away from (yellow, type 3) FUV, independent of proximity to CO. We note some bins contain one object and appear only as thicker lines at log(number of objects)=0. Mass is in units of $M_\odot$.
\label{fig:mass_hist_all}}
\end{figure*}

\section{Results} \label{sec:results}
\subsection{Color-Magnitude and Color-Color Diagrams} \label{subsec:cmd_ccd}

\begin{figure}[tbh!]
\epsscale{1.14}\plotone{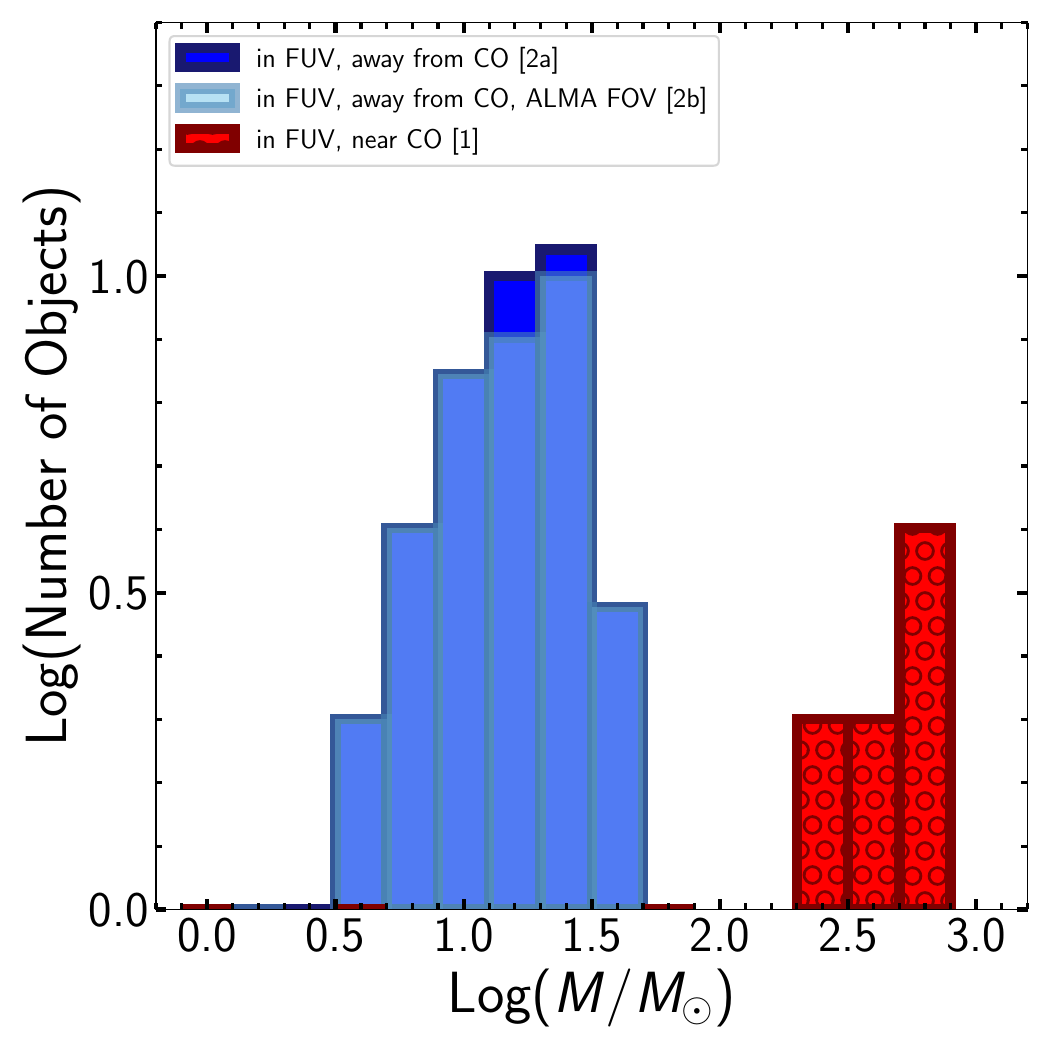}
\caption{Mass histograms near star formation: type 1 (red with circle hatches) and a comparison of all type 2 objects (dark blue, 2a) to only type 2 objects in the ALMA FOV where CO was observered (light blue, 2b). Mass is in units of $M_\odot.$
\label{fig:mass_ALMA}}
\end{figure}
Figure \ref{fig:CMDs} shows color-magnitude diagrams (CMDs) of the F090W versus F090W--F150W filters and F250M versus F250M--F430M filters for all objects. Many uncertainties are smaller than the size of the plot markers for both colors ($<$0.05), but we see that uncertainties in the color increase as objects get fainter (up to 0.10 for F090W--F150W and 0.47 for F250M--F430M). We include the blackbody temperatures corresponding to the color for the given filters on the opposing x-axis, with the temperature range from the F090W versus F090W--F150W CMD shown in gray on the F250M versus F250M--F430M CMD. Objects of all three types appear randomly distributed in the F090W versus F090W--F150W CMD, with the data centered around a color of 0.3. The data in the F250M versus F250M--F430M CMD appear clustered along a ridge from 23 to 18.5 in magnitude and -1.4 to 0 in color for objects of all three types, although several objects fall redward of this ridge.

Figure \ref{fig:ccd} shows a color-color diagram of F090W--F150W versus F250M--F430M for all objects, where many uncertainties are again smaller than the size of the plot markers. A line indicating the effective temperature of a blackbody is overlaid on the plot. We find that most objects, regardless of their proximity to FUV or CO, are concentrated around the effective temperature line with blackbody temperatures between approximately 3000 to 7500 K, while some objects appear scattered to the right of the effective temperature line. We also include seven NIR sources that were obvious, resolved background galaxies (light purple stars in plot) to see if they displayed a preferential color, and find that they occupy a F090W--F150W color range from 0.2 to 1.0 and a F250M--F430M color range from -0.8 to 0.8. In Figure \ref{fig:ccd_small} we show a color-color diagram of F090W--F150W versus F250M--F430M for individual resolved NIR sources (gray) that comprise the type 1 objects (red) to compare the colors of the individual sources and larger clumps. We find that the larger uncertainties for the smaller sources place them in the same color ranges as the inclusive clumps.

We also show a log(temperature)-log(temperature) plot in Figure \ref{fig:TT} of the effective temperatures corresponding to the F090W--F150W colors versus the effective temperatures corresponding to the F250M--F430M colors for all objects. If our objects were well represented by a black body, the best-match effective temperature would be the same for all colors, and the objects would fall along the gray diagonal line.
We see that objects cover a much larger F250M--F430M effective temperature range compared to the F090W--F150W effective temperatures for all object types. 

Table \ref{tab:type1} contains the F090W and F250M AB magnitudes, along with the F090W--F150W and F250M--F430M colors and corresponding effective temperatures of type 1 objects. The same AB magnitudes, colors, and effective temperatures for objects of all three types are available in a table provided in the online materials. The RA, Dec, radius, F090W AB magnitude, F090W--F150W color, F250M AB magnitude, and F250M--F430M color of the background galaxies are listed in Table \ref{tab:bg_gals}. Likewise, the RA, Dec, radius, F090W AB magnitude, F090W--F150W color, F250M AB magnitude, and F250M--F430M color for some of the individual sources inside type 1 objects we measured are listed in Table \ref{tab:ind_source}, with the full table of individual sources used in our sample included in a table in the online materials.

\begin{figure}[hbt!]
\captionsetup[subfloat]{farskip=-4pt}
\centering 
\subfloat{\includegraphics[width=\columnwidth]{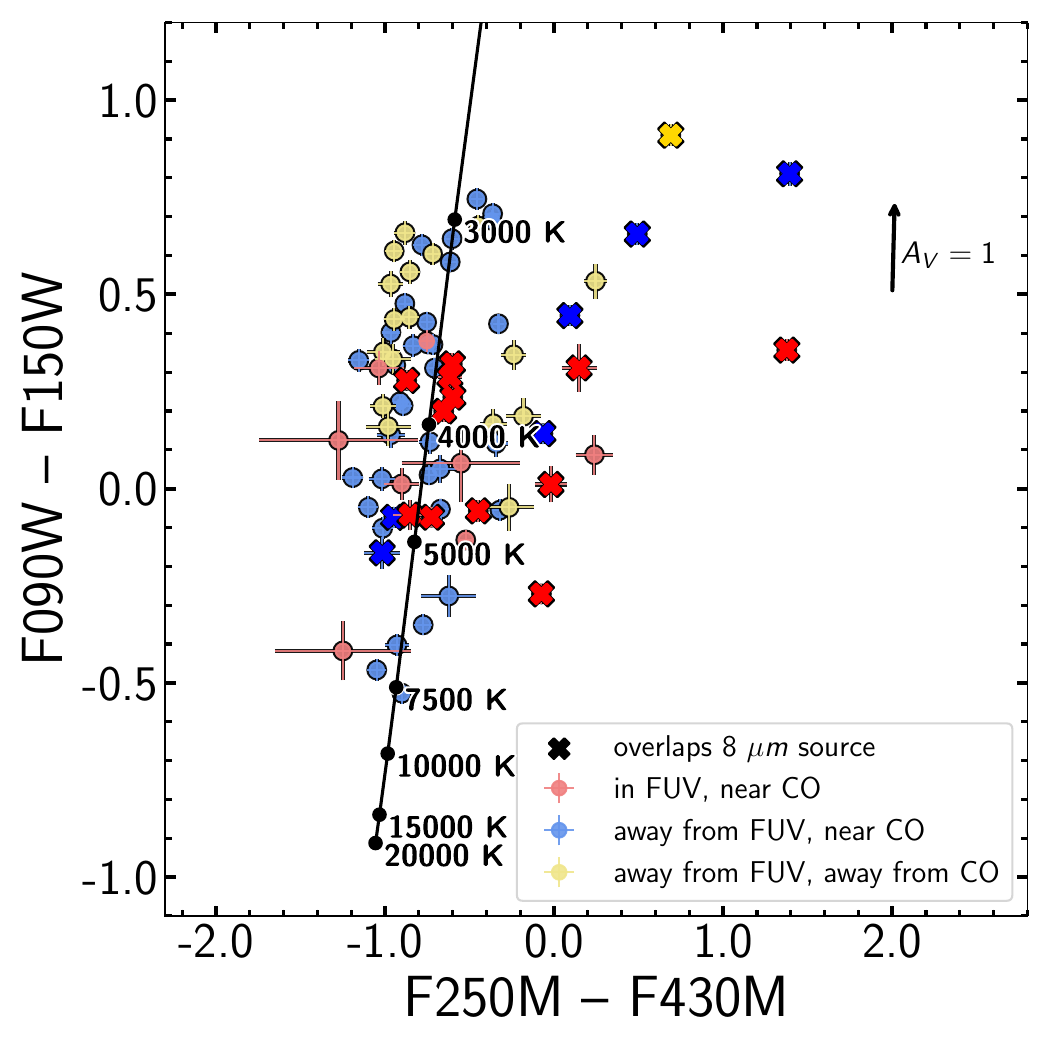}}

\subfloat{\includegraphics[width=\columnwidth]{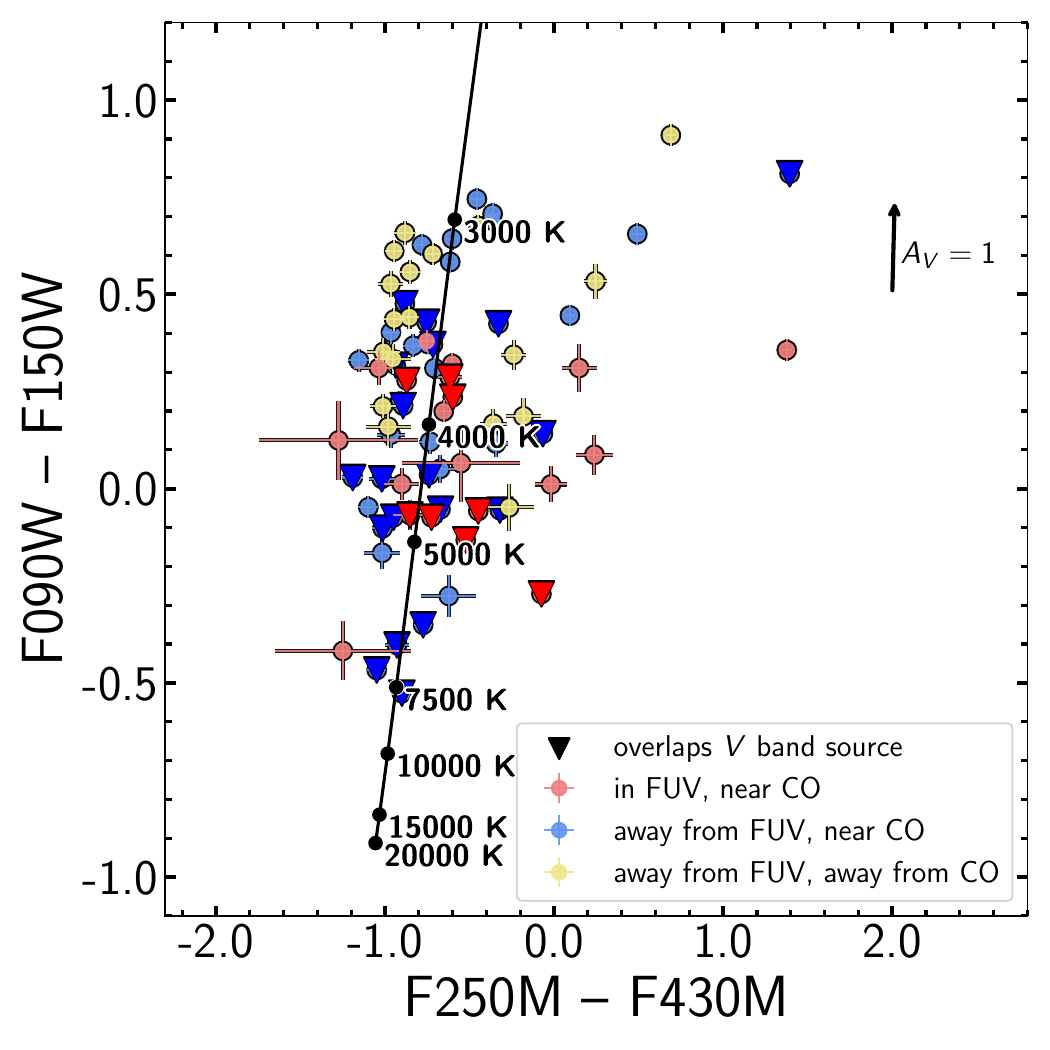}}
\caption{Color-color diagrams of F090W--F150W versus F250M--F430M for all objects (red circles: in FUV knots and near CO, blue circles: in FUV knots and away from CO, and yellow circles: away from FUV knots and away from CO). Objects that overlap 8 $\mu$m sources are marked with a $\pmb{\times}$ instead of a circle (top), and objects that overlap $V$ band sources are marked with a $\pmb{\nabla}$ instead of a circle (bottom). A reddening vector of $A_V=1$ assuming SMC-like extinction is included in the upper left of the plot, and the black line shows the effective temperature of a blackbody.
\label{fig:ccd-8V}}
\end{figure}

\subsection{Mass} \label{subsec:mass}
The top panel of Figure \ref{fig:mass_hist_all} shows histograms of the mass for each of the three object types. Type 1 objects range from approximately $1$ to $5930\ M_\odot$ and span the largest range of masses of the three categories, with eight objects between $100\ M_\odot$ and $1000\ M_\odot$ and nine objects larger than $1000\ M_\odot$. Some of the mass bins contain one object and appear only as a thick line on the bottom x-axis at log(number of objects)=0. Type 2 objects comprise the largest number of objects and cover a range of masses approximately $1$ to $66\ M_\odot$. Type 3 objects span the lowest range of masses from approximately $1$ to $33\ M_\odot$. The masses of type 1 objects are listed in Table \ref{tab:type1}, and the masses of all objects are available in the table provided in the online materials.

The bottom panel of Figure \ref{fig:mass_hist_all} groups the objects differently to compare the mass. The histogram on the left compares only objects in FUV knots: type 1 (in FUV knots and near CO, red) and type 2 (in FUV knots and away from CO, blue). We see that objects near CO and FUV have higher masses than those in FUV and away from CO. The histogram in the center compares objects near and away from CO, independent of proximity to FUV. Object types 2 (blue) and 3 (yellow) are away from CO and combined into one category (green), while type 1 objects (red) are near CO and in their own category. We again see objects near CO have higher masses than those away from CO. The histogram on the right compares objects near and away from FUV knots, independent of proximity to CO. Here we combine objects 1 (red) and 2 (blue) into one category (purple) since they are both in FUV knots, while type 3 objects (yellow) are their own category. Objects away from FUV knots cover a smaller range of masses than those in FUV knots, but still appear to fall at the lower end of the mass range that objects in FUV knots span. We note that when the mass is measured using \textsc{prospector}, the masses of all three types of objects cover the same mass range.

We also show histograms of type 1 objects, all type 2 objects, and only type 2 objects within the same area that CO was observed in Figure \ref{fig:mass_ALMA} to compare how the mass distributions are affected by the addition of type 2 objects that were not in the ALMA FOV and could contain CO that was not observed. We see that masses of type 2 objects appear to have similar distributions independent of whether the entire set or the ALMA FOV subset of type 2 objects are included.

\section{Discussion} \label{sec:discussion}
The effective temperatures from the F090W versus F090W--F150W CMD in Figure \ref{fig:CMDs} match that of the color-color diagram in Figure \ref{fig:ccd}, with most objects having effective temperatures ranging from $\sim$2600 K to 7400 K. However, the effective temperatures in the F250M versus F250M--F430M CMD cover a much wider range of $<$1000 K to 20000 K. We see in Figure \ref{fig:TT} that the temperatures of our objects fall near the grey line delineating a perfect blackbody, albeit with large F250M--F430M uncertainties at temperatures above 3000 K. In the case that our objects are embedded clusters, the infrared colors of stars are considerably influenced by extinction and infrared excess associated with young stars, while not being inherently sensitive to effective temperature \citep{Lada_2003}.
The small errors in the F250M--F430M color indicate that uncertainties in the color are not responsible for scatter of these objects, suggesting the scatter may be due to infrared excess. While we did not measure the extinction of our objects, we include reddening vectors assuming SMC-like extinction in the CMDs and color-color diagram (Figures \ref{fig:CMDs} and \ref{fig:ccd}) to indicate how our objects could be affected by extinction.

We examined our objects in a Spitzer Infrared Array Camera (IRAC) 8 $\mu$m image of WLM and find several coincident with sources bright at 8 $\mu$m, suggesting the regions are dominated by hot dust. We also examined our objects in a $V$ band image of WLM from observations taken with the Lowell Observatory Hall 1.07-m Telescope \citep{Hunter_2006} and also find several coincident with sources bright in $V$. We cannot reliably measure the brightness of the objects in either the 8 $\mu$m or $V$ images due to the difference in angular resolution of 1\farcs2 in 8 $\mu$m and 1\farcs1 in $V$ compared to 0\farcs063 in F250M. However, we show in Figure \ref{fig:ccd-8V} the F090W--F150W versus F250M--F430M color-color diagrams for the objects of all three types, this time noting which objects overlap the 8 $\mu$m sources (top, marked with a $\pmb{\times}$) and $V$ sources (bottom, marked with a $\pmb{\nabla}$). We see most objects that overlap 8 $\mu$m sources are also redward of the effective temperature line in F250M--F430M, which we would expect for clusters embedded in hot dust. Similarly, most objects that overlap $V$ sources are along the effective temperature line, which we would expect from the photospheres of stars. For type 1 objects, 60\% overlap 8 $\mu$m sources and 40\% overlap $V$ sources, with 39\% of the objects overlapping both. For type 2 objects, 15\% overlap 8 $\mu$m sources and 48\% overlap $V$ sources, with 11\% of the objects overlapping both. For type 3 objects, 5\% overlap 8 $\mu$m sources and 0\% overlap $V$ sources, thus 0\% of the objects overlapping both. The higher percentage of type 1 and 2 objects overlapping $V$ sources than type 3 objects is expected due to type 1 and 2 objects being near FUV sources. 

\begin{figure}[bth!]
\captionsetup[subfloat]{farskip=-2pt}
\centering 
\subfloat{\includegraphics[width=\columnwidth]{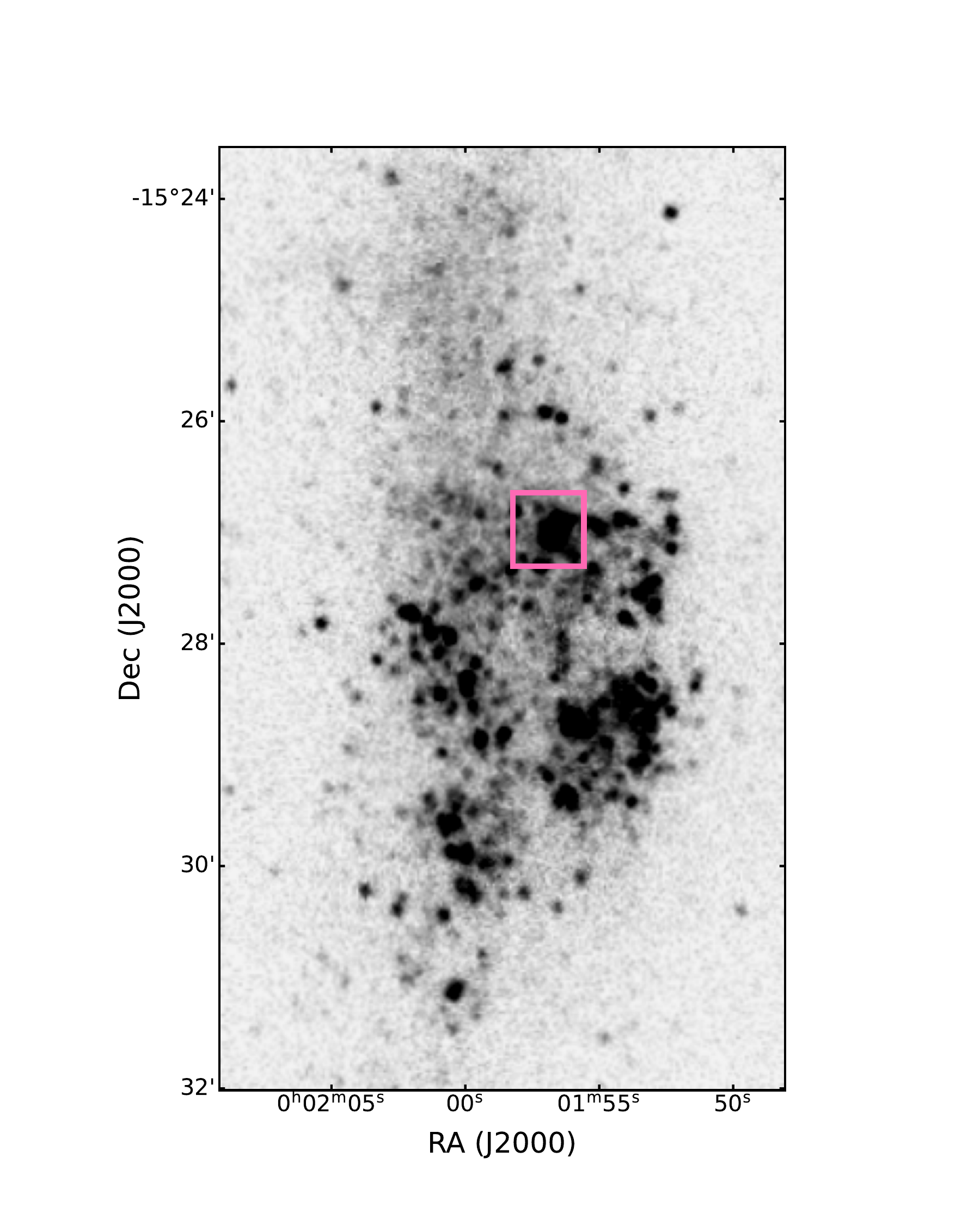}}

\subfloat{\includegraphics[width=\columnwidth]{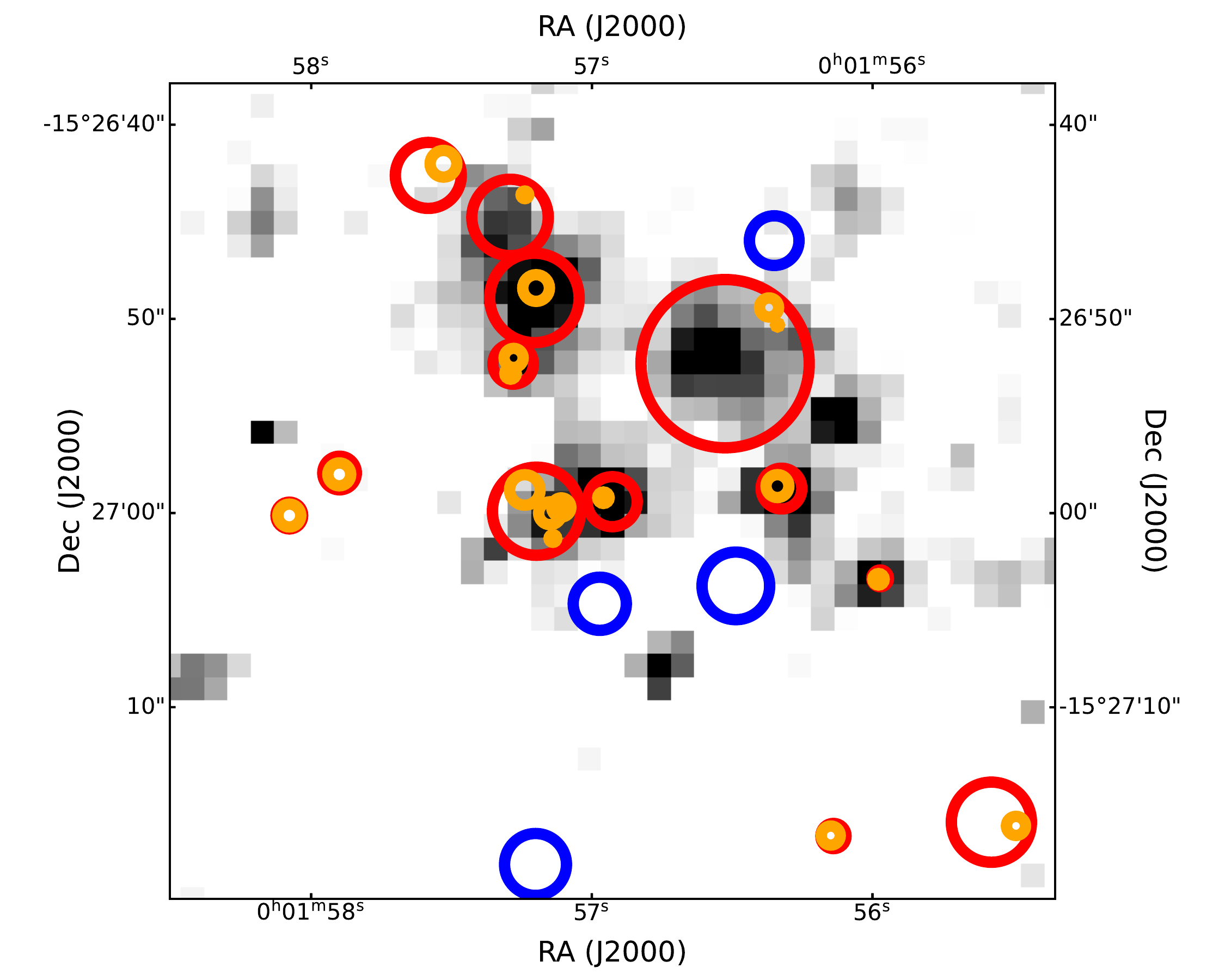}}
\caption{Top: \textit{GALEX} FUV image of WLM showing FUV-bright star-forming region (pink box). Bottom: Close-up of the FUV-bright star-forming region in Spitzer 8 $\mu$m showing that type 1 objects (red circles) are typically associated with 8 $\mu$m sources while most type 2 objects (blue circles) do not overlap 8 $\mu$m sources. The CO cores within the type 1 objects are also included (orange circles).
\label{fig:reg2}}
\end{figure}

The difference in objects overlapping 8 $\mu$m sources between types 1 and 2 may be the result of differing stages of star formation. We focused on a prominent FUV-bright region which we show in the top panel Figure \ref{fig:reg2}. \citet{archer_2022} also examined this region and estimated the age to be approximately 7 Myr. \citet{Jones_2023} found hundreds of young stellar objects (YSOs) and pre-main sequence stars in NGC 346, a young star cluster in the SMC with an age of $\sim$3 Myr, so we do not expect the age of the FUV region to preclude younger embedded star formation.
The Spitzer 8 $\mu$m image of this region is shown in the bottom panel of Figure \ref{fig:reg2} with the CO cores, type 1 objects, and type 2 objects overlaid. We see that most type 1 objects appear to be associated with sources bright at 8 $\mu$m, while no type 2 sources appear coincident with the 8 $\mu$m sources. A possible explanation for this is that the type 1 objects are still embedded, while type 2 objects have already cleared their dust envelopes. This could also explain the comparable masses of type 1 objects between the \textsc{prospector} and TIR luminosity methods when the TIR luminosity method assumes an age of 1 Myr, while the type 2 and type 3 objects have comparable masses between the two methods when the TIR luminosity method assumes an age of 10 Myr (Figure \ref{fig:mass_comp}). However, we cannot be certain without accurate age estimates for the objects.

\begin{figure}[thb!]
\captionsetup[subfloat]{farskip=-2pt}
\centering 
\subfloat{\includegraphics[width=\columnwidth]{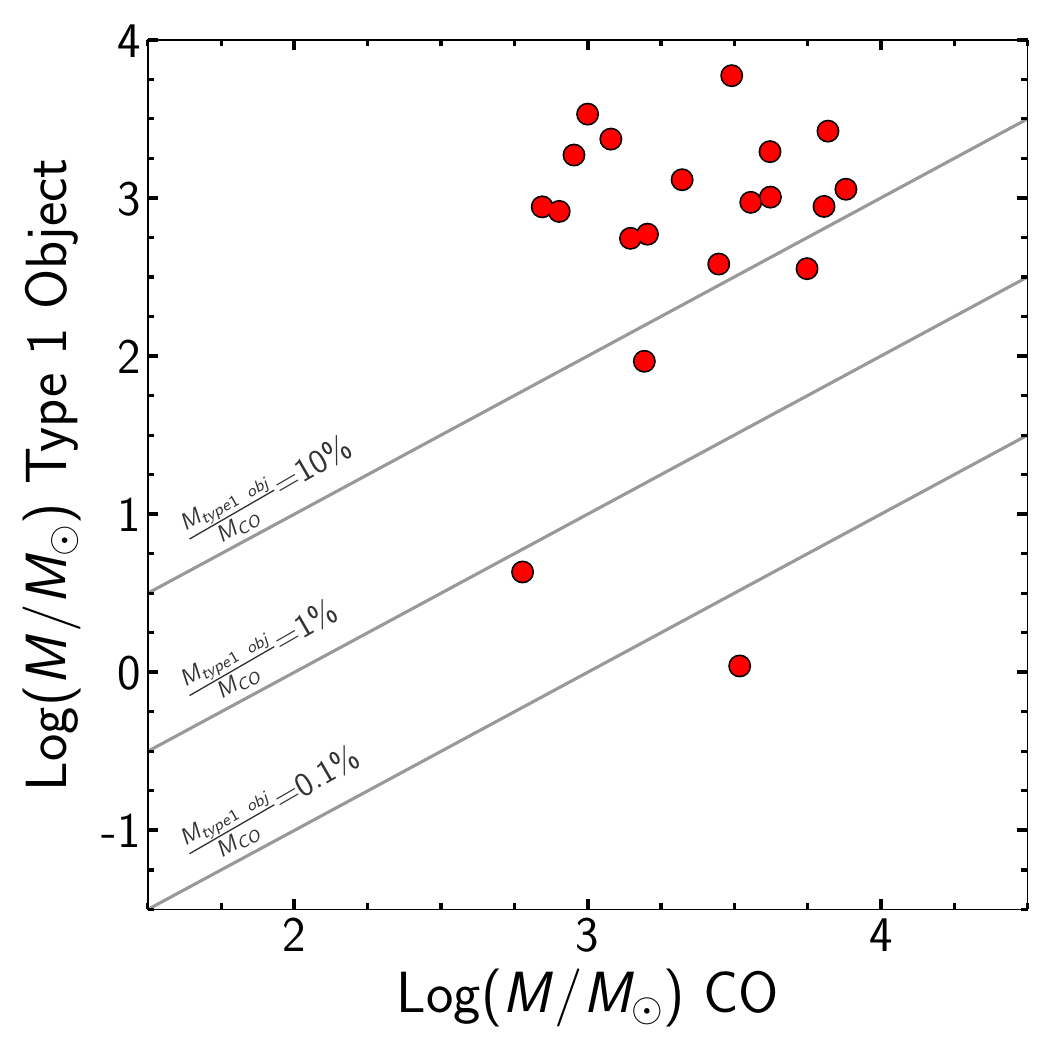}}

\subfloat{\includegraphics[width=\columnwidth]{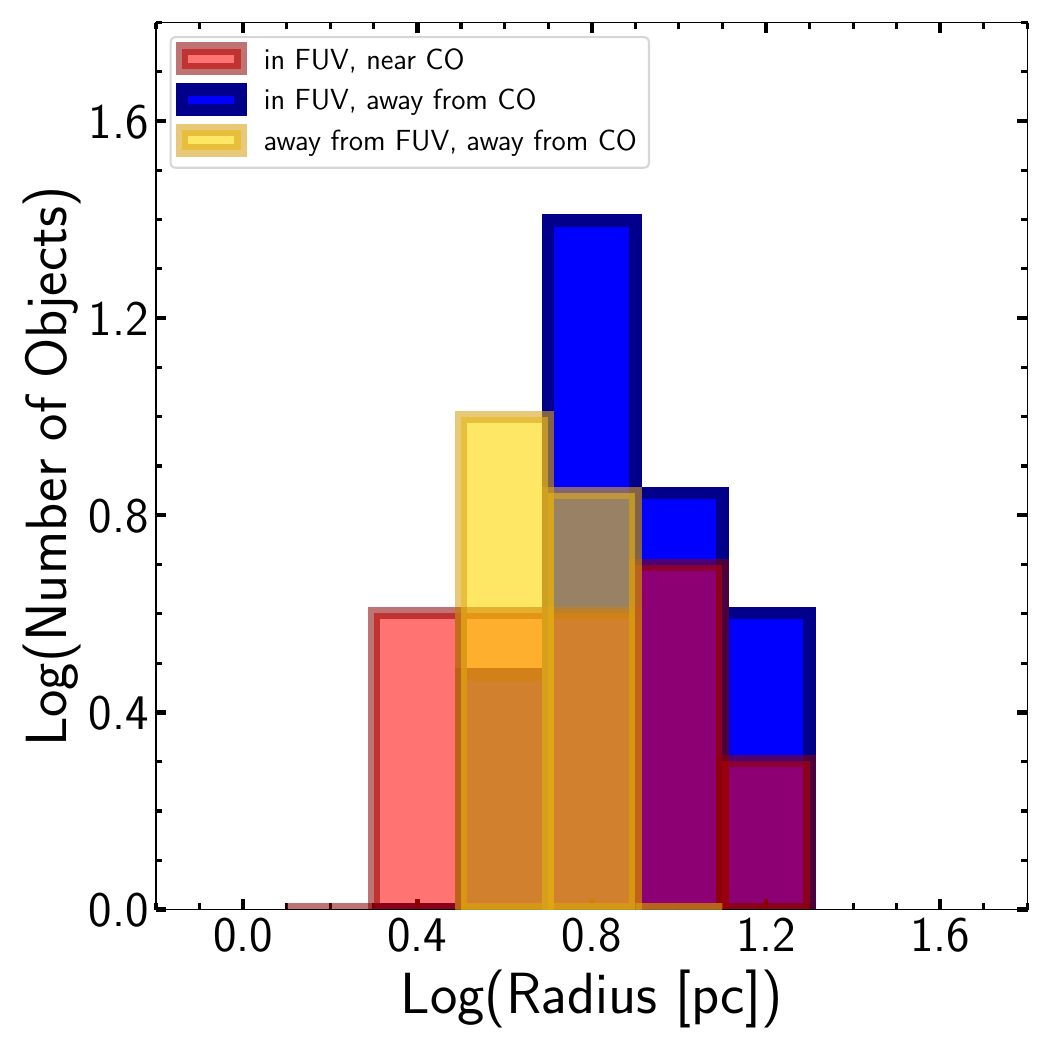}}
\caption{Top: Mass of type 1 objects versus the mass of the CO core that the object is associated with. Bottom: Histogram of object radii in pc for each of the three objects types.
\label{fig:mass-rad}}
\end{figure}

Embedded star clusters are associated with the most massive molecular cores in GMCs, with CO clump masses ranging from approximately $10^2$ to $10^4$ $M_\odot$ in the Milky Way and Large Magellanic Cloud (LMC) \citep{Carpenter_1996,Phelps_1997,Nayak_2016,Nayak_2018} and $10^2$ to $10^5$ in the SMC \citep{Saldano_2023}. A catalog of embedded star clusters within 2 kpc of the Sun compiled by \citet{Lada_2003} finds that the cluster masses range from approximately $10^1$ to $10^4$ $M_\odot$, and \citet{Elmegreen_2019} find 8 $\mu$m cores in dusty spirals on the order of $10^2$ to $10^4$ $M_\odot$. In the top panel of Figure \ref{fig:mass-rad}, we find that the masses of our CO cores are comparable to the masses of CO clumps where embedded star clusters are found in the Milky Way, SMC, and LMC. Our objects associated with CO (type 1) have masses up to $>$5000 $M_\odot$, while embedded star clusters from \citet{Lada_2003} and \citet{Elmegreen_2019} have higher upper mass limits. Figure \ref{fig:mass_hist_all} shows that type 1 objects do not appear to be sampled to the upper mass limit, so we cannot say whether the lower masses of our objects are the result of some physical constraint. However, the size of our objects, which have radii between 2 and 21 pc (Figure \ref{fig:mass-rad}, bottom panel), are larger than the $<$2 pc radii embedded clusters from \citet{Lada_2003} and smaller on average than the $\sim$25 pc radii 8 $\mu$m cores in local spirals from \citet{Elmegreen_2018}. This may be a result of our grouping individual IR sources into larger objects compared to those in \citet{Lada_2003} while the resolution of the NIRCam images in this work are much higher than the IRAC images used in \citet{Elmegreen_2018}.

In the top panel of Figure \ref{fig:mass-rad}, we also include lines delineating where the ratio of the type 1 object mass to CO core mass is 0.1\%, 1\%, and 10\%. One possibility for objects in the 10\% range is that the molecular gas has already been destroyed and the objects left behind are bound clusters. However, H$_2$ can be self-shielded from the photodissociation that results in the smaller CO cores seen in low-metallicity galaxies, leaving the cores surrounded by a potentially large reservoir of CO-dark H$_2$ gas \citep[e.g.][]{Wolfire_2010,planck_2011,Pineda_2014,Cormier_2017,Madden_2020}. If there is CO-dark H$_2$ gas present, then the ratios of type 1 object mass to molecular gas mass would be lower.

\section{Summary and Conclusions} \label{sec:summary}
In this work we present an analysis of the relationship between CO and early star-forming regions seen as embedded IR sources in the low metallicity dIrr galaxy WLM. We use ALMA-detected CO data from \citet{Rubio_2015} and Rubio et al.\ (in preparation), \textit{GALEX} FUV data from \citet{Zhang_2012}, and \textit{JWST} NIRCam data from \citet{Weisz_2023}. The IR sensitivity and high angular and spatial resolution of \textit{JWST} allow us to probe embedded, early star-forming regions in WLM at an unprecedented depth and in regions of higher extinction.

We compare and contrast objects that are: near FUV knots and near CO (\textbf{type 1}), in FUV knots and away from CO (\textbf{type 2}), and away from FUV knots and away from CO (\textbf{type 3}) to identify whether the early star-forming regions have similar physical properties in different local environments and analyze the relationship between early star formation and CO in low metallicity environments. The objects in this work are clumps of NIR sources. A test analysis of the individual NIR sources inside a subset of our type 1 objects finds that the colors of the smaller individual sources and larger objects are consistent within the uncertainties.

We find that objects of all three types have comparable NIR colors and magnitudes. Likewise, the three object types also cover the same effective temperature range, although the infrared colors of young star-forming regions do not exhibit high sensitivity to effective temperature and are notably impacted by extinction and infrared excess. Type 1 objects appear to have higher masses than the other two object types when computed using the TIR luminosity. However, when the masses are measured using \textsc{prospector}, all object types cover the same mass range.

We find IR sources associated with CO cores throughout WLM (type 1 objects). The masses of our objects associated with the CO, along with the masses of their corresponding CO cores, are in agreement with the masses of embedded clusters and their CO cores found elsewhere in the SMC, LMC, Milky Way, and other spirals. However, the sizes of our objects are larger than embedded clusters from \citet{Lada_2003} and smaller than those from \citet{Elmegreen_2018}. A possible explanation for this is that our objects are clumps of IR sources compared to individual sources in \citet{Lada_2003}, while our \textit{JWST} NIRCam images have a much higher resolution than the Spitzer IRAC images used in \citet{Elmegreen_2018}. Our type 1 objects also appear to coincide with bright 8 $\mu$m sources at a higher frequency than the other two object types, which suggests young embedded star formation in these CO cores. These type 1 objects may be at an earlier evolutionary stage than the other object types, but we cannot confirm this without accurate age estimates.

\section{Acknowledgements}
This material is based upon work supported by the National Science Foundation under Grant AST-1907492 to D.A.H. 

M.R.\ wishes to acknowledge support from ANID(CHILE) through Basal FB210003 and FONDECYT grant No1190684.

This paper makes use of the following ALMA data: ADS/JAO.ALMA\texttt{\#}2012.1.00208.S and ADS/JAO.ALMA\texttt{\#}2018.1.00337.S. ALMA is a partnership of ESO (representing its member states), NSF (USA) and NINS (Japan), together with NRC (Canada), MOST and ASIAA (Taiwan), and KASI (Republic of Korea), in cooperation with the Republic of Chile. The Joint ALMA Observatory is operated by ESO, AUI/NRAO and NAOJ. The National Radio Astronomy Observatory is a facility of the National Science Foundation operated under cooperative agreement by Associated Universities, Inc.

We would like to thank the referee for carefully reading our manuscript and for providing constructive comments which greatly enhanced the quality and clarity of this manuscript.

Lowell Observatory sits at the base of mountains sacred to tribes throughout the region. We honor their past, present, and future generations, who have lived here for millennia and will forever call this place home.

%

\vspace{5mm}
\facilities{ALMA, JWST(NIRCam)}


\software{IRAF \citep{Tody_1986}, SEDpy \citep{sedpy}, prospector \citep{Leja_2017,Johnson_2021,python-fsps}, dynesty \citep{Conroy_2009,Conroy_2010,Speagle_2020,dynesty}}


\appendix
Postage stamps showing the \textit{JWST} three-color image of each individual object. Type 1 objects and their CO cores are shown in Figure \ref{fig:type1}, type 2 objects in Figure \ref{fig:type2}, type 3 objects in Figure \ref{fig:type3}, and select resolved background galaxies used in this work in Figure \ref{fig:bgGal}.

\begin{figure*}[phtb!]
\epsscale{1.1}\plotone{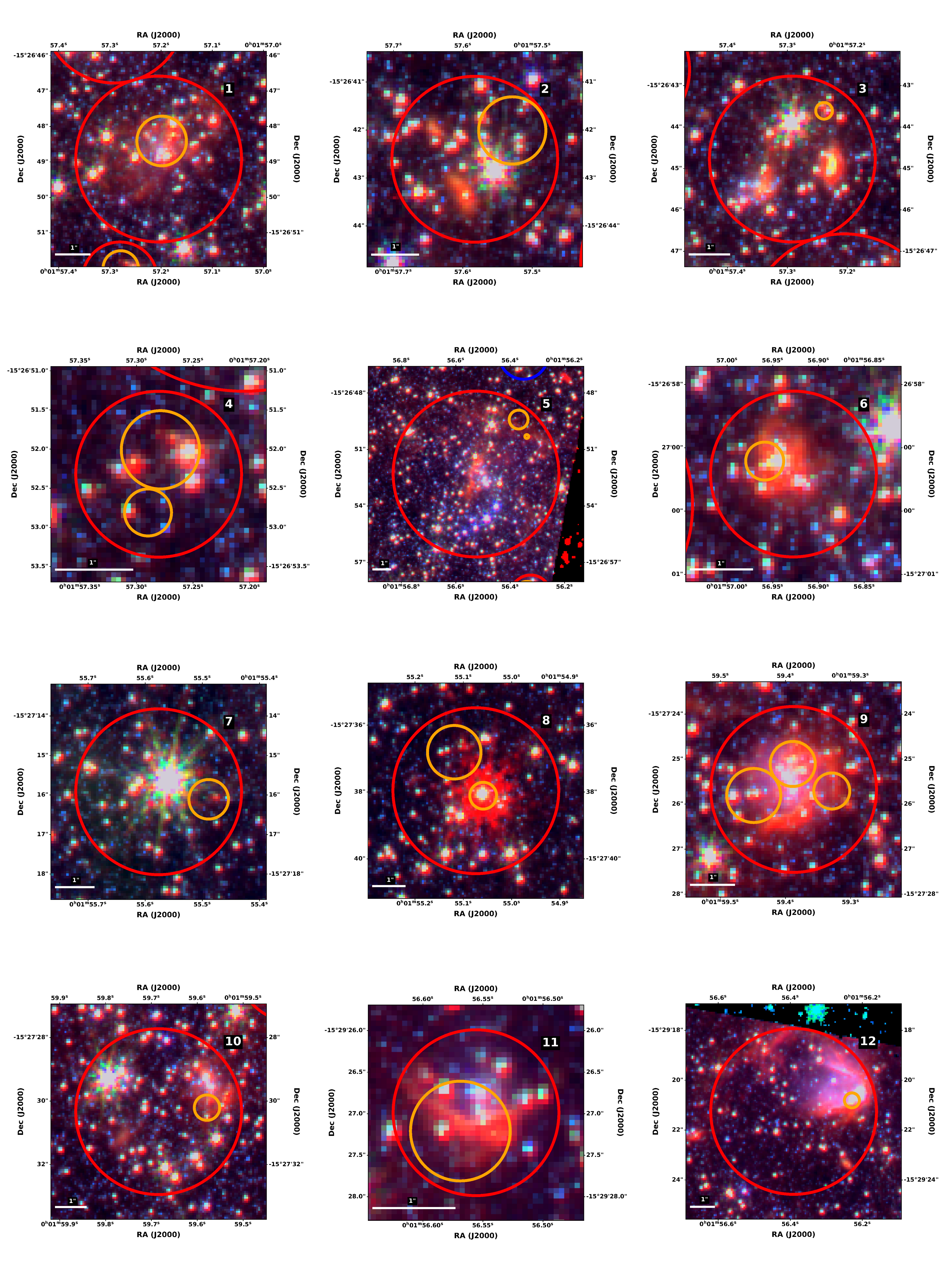}
\caption{F250M (red), F150W (green), F090W (blue) three-color image of each type 1 object (in FUV and near CO, red circles) and the CO cores (orange circles) within them. The orientation of the image is such that North is up and East is to the left. The object number corresponding to the number in Table \ref{tab:type1} is given in the upper right of the image.
\label{fig:type1}}
\end{figure*}

\begin{figure*}[phtb!]
\ContinuedFloat
\epsscale{1.1}\plotone{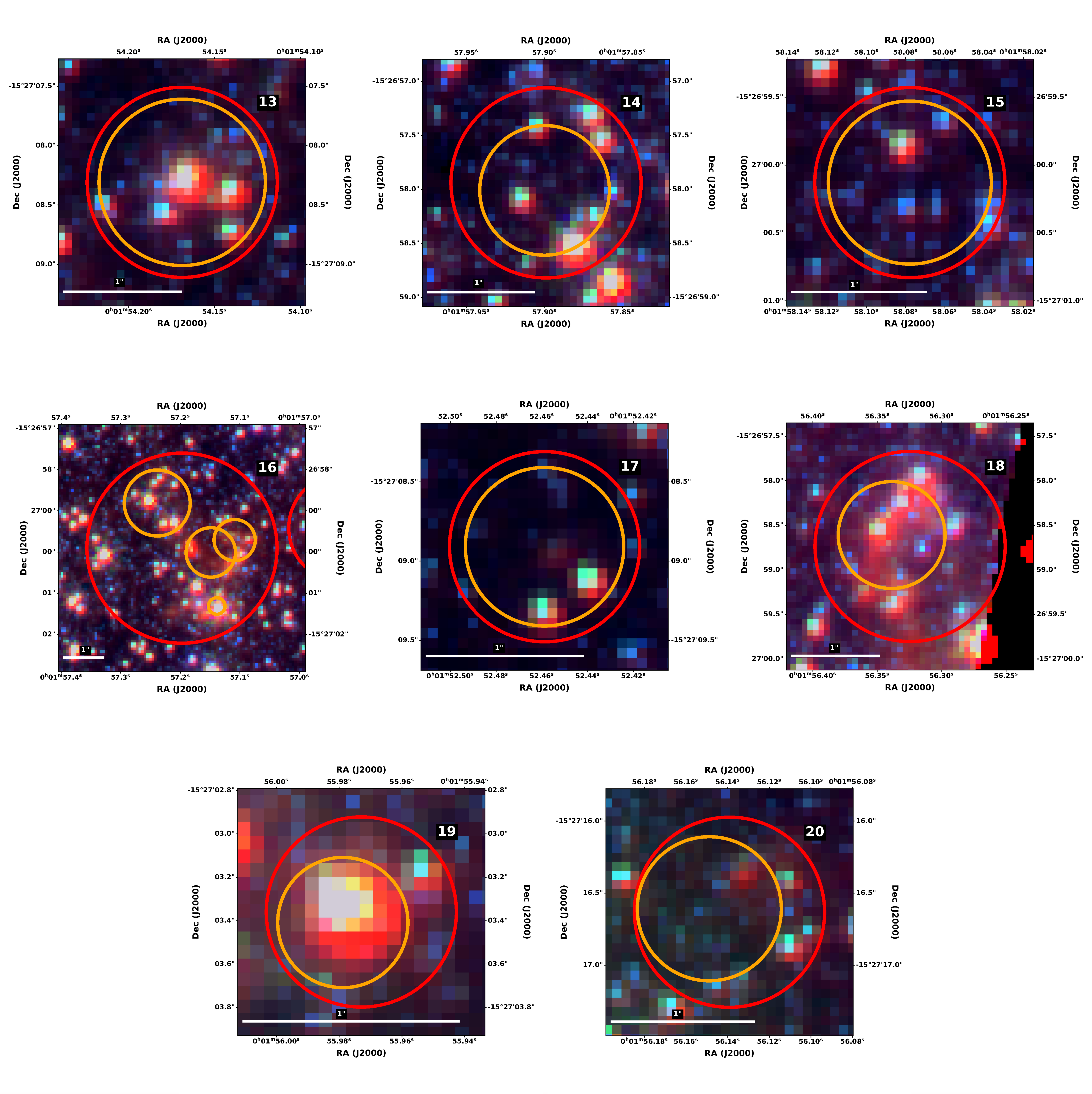}
\caption{\textbf{cont.} F250M (red), F150W (green), F090W (blue) three-color image of each type 1 object (in FUV and near CO, red circles) and the CO cores (orange circles) within them. The orientation of the image is such that North is up and East is to the left. The object number corresponding to the number in Table \ref{tab:type1} is given in the upper right of the image.}
\end{figure*}

\begin{figure*}[phtb!]
\epsscale{1.1}\plotone{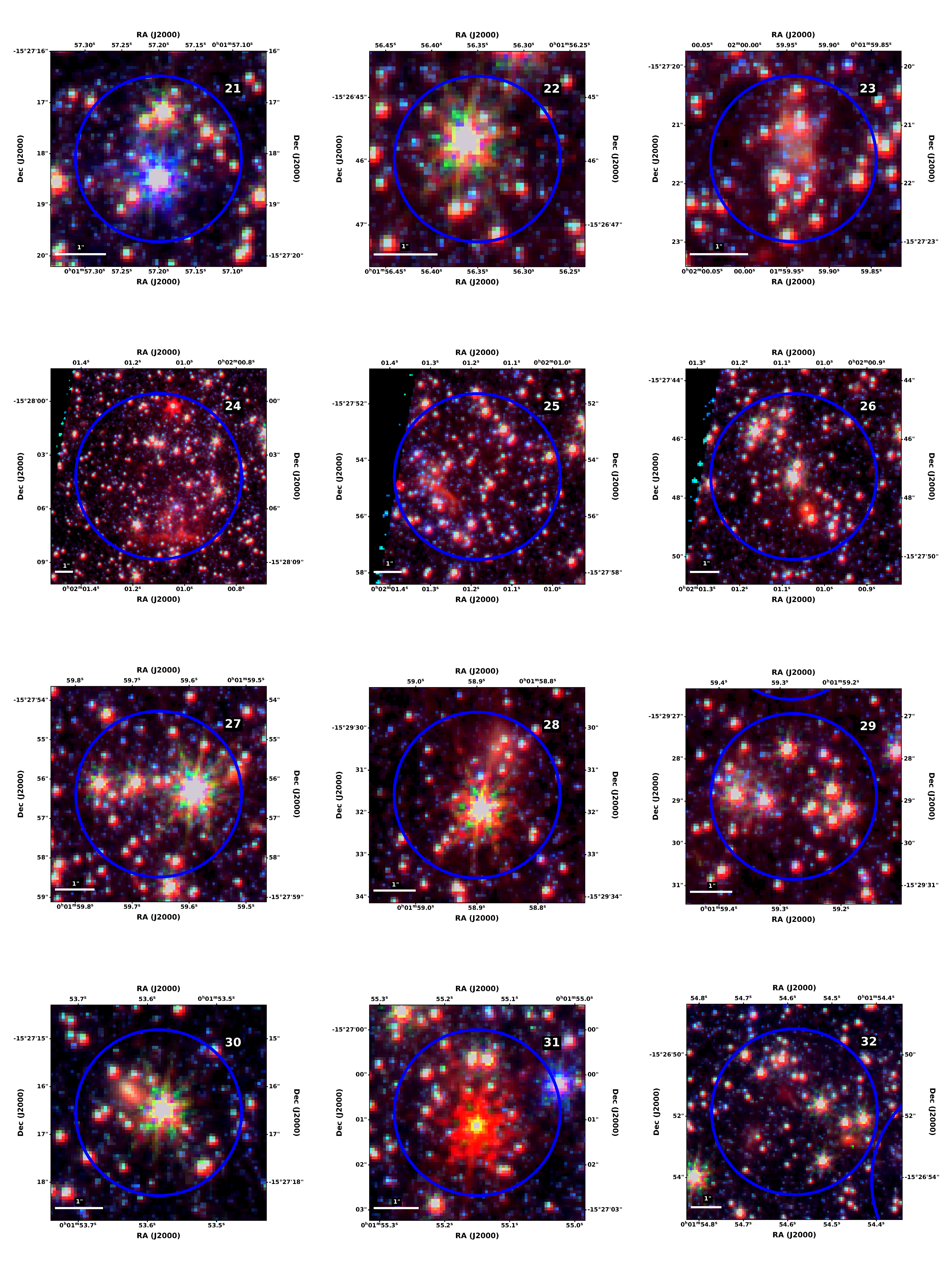}
\caption{F250M (red), F150W (green), F090W (blue) three-color image of each type 2 object (in FUV and away from CO, blue circles). The orientation of the image is such that North is up and East is to the left. The object number corresponding to the number in the extended version of Table \ref{tab:type1} provided in the online materials is given in the upper right of the image.
\label{fig:type2}}
\end{figure*}

\begin{figure*}[phtb!]
\ContinuedFloat
\epsscale{1.1}\plotone{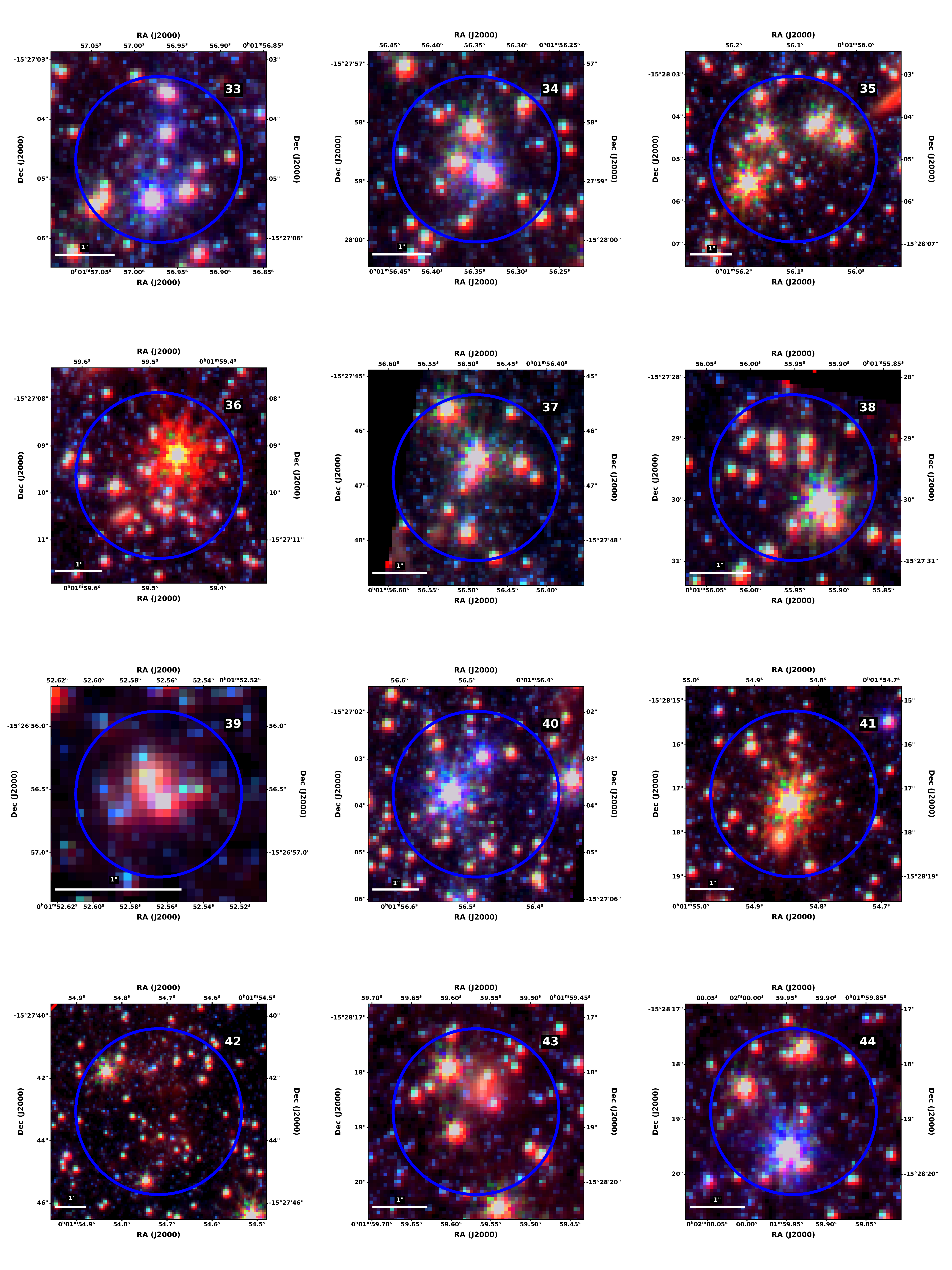}
\caption{\textbf{cont.} F250M (red), F150W (green), F090W (blue) three-color image of each type 2 object (in FUV and away from CO, blue circles). The orientation of the image is such that North is up and East is to the left. The object number corresponding to the number in the extended version of Table \ref{tab:type1} provided in the online materials is given in the upper right of the image.}
\end{figure*}

\begin{figure*}[phtb!]
\ContinuedFloat
\epsscale{1.1}\plotone{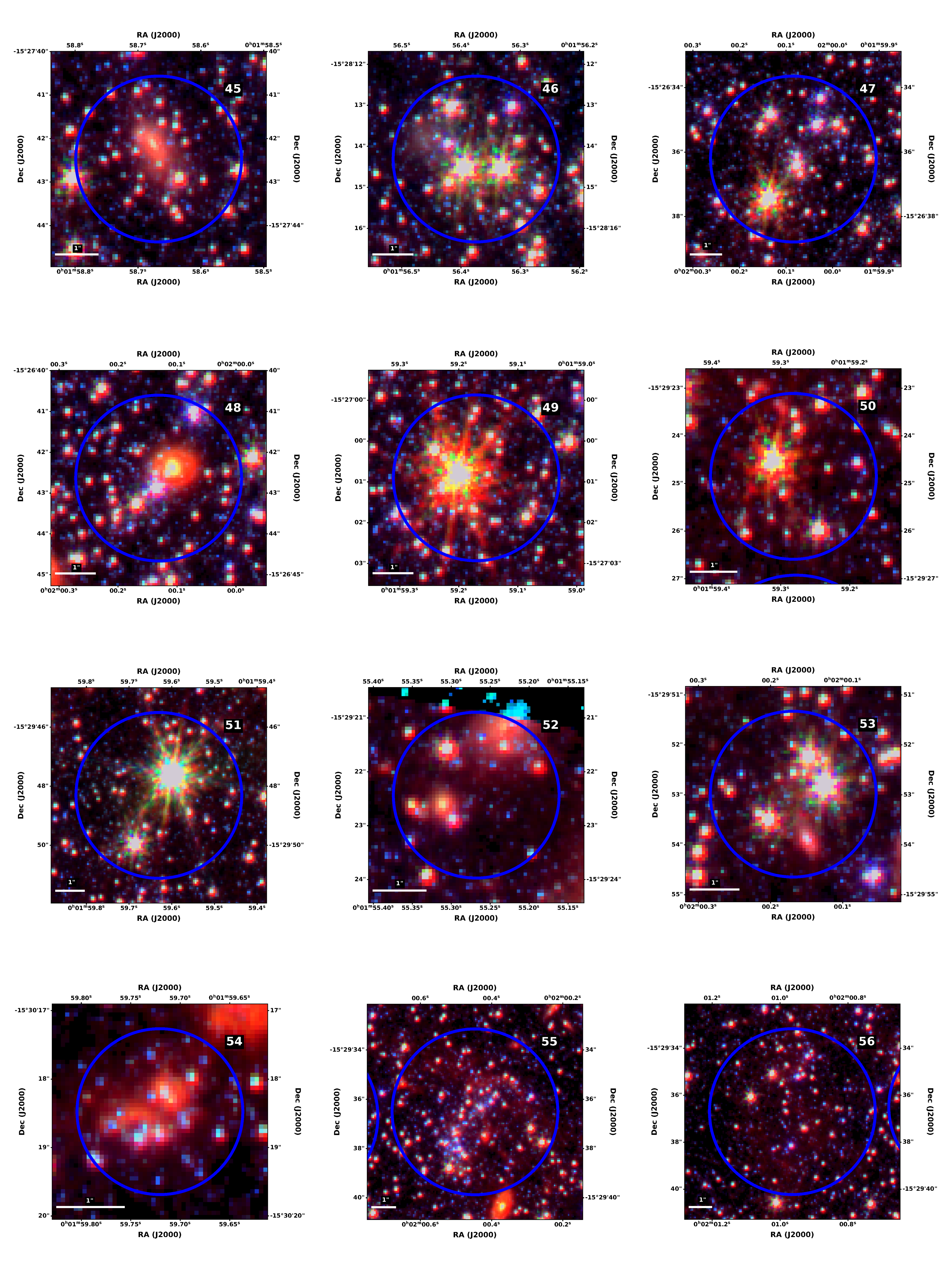}
\caption{\textbf{cont.} F250M (red), F150W (green), F090W (blue) three-color image of each type 2 object (in FUV and away from CO, blue circles). The orientation of the image is such that North is up and East is to the left. The object number corresponding to the number in the extended version of Table \ref{tab:type1} provided in the online materials is given in the upper right of the image.}
\end{figure*}

\begin{figure*}[phtb!]
\ContinuedFloat
\epsscale{1.1}\plotone{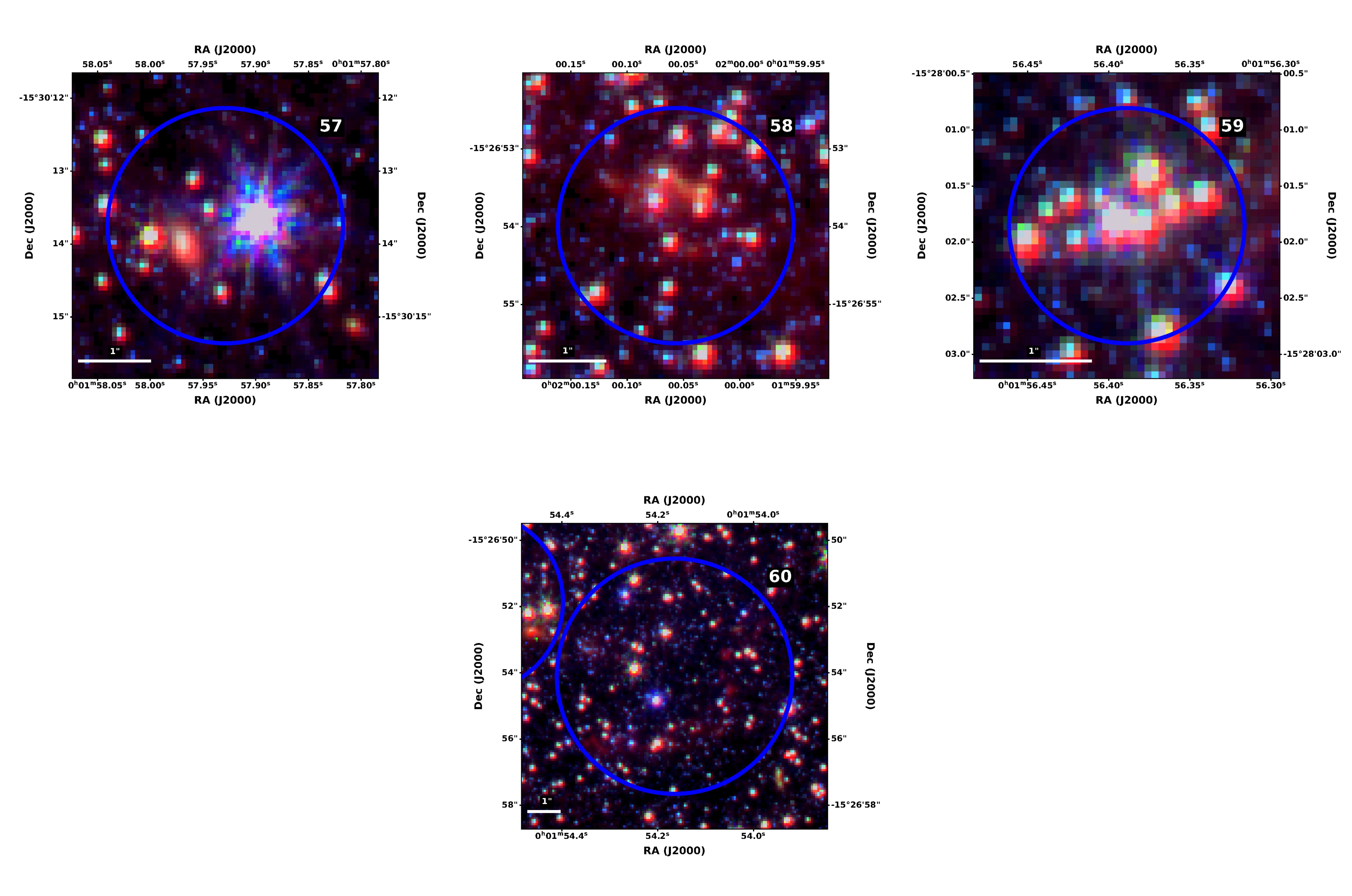}
\caption{\textbf{cont.} F250M (red), F150W (green), F090W (blue) three-color image of each type 2 object (in FUV and away from CO, blue circles). The orientation of the image is such that North is up and East is to the left. The object number corresponding to the number in the extended version of Table \ref{tab:type1} provided in the online materials is given in the upper right of the image.}
\end{figure*}

\begin{figure*}[phtb!]
\epsscale{1.1}\plotone{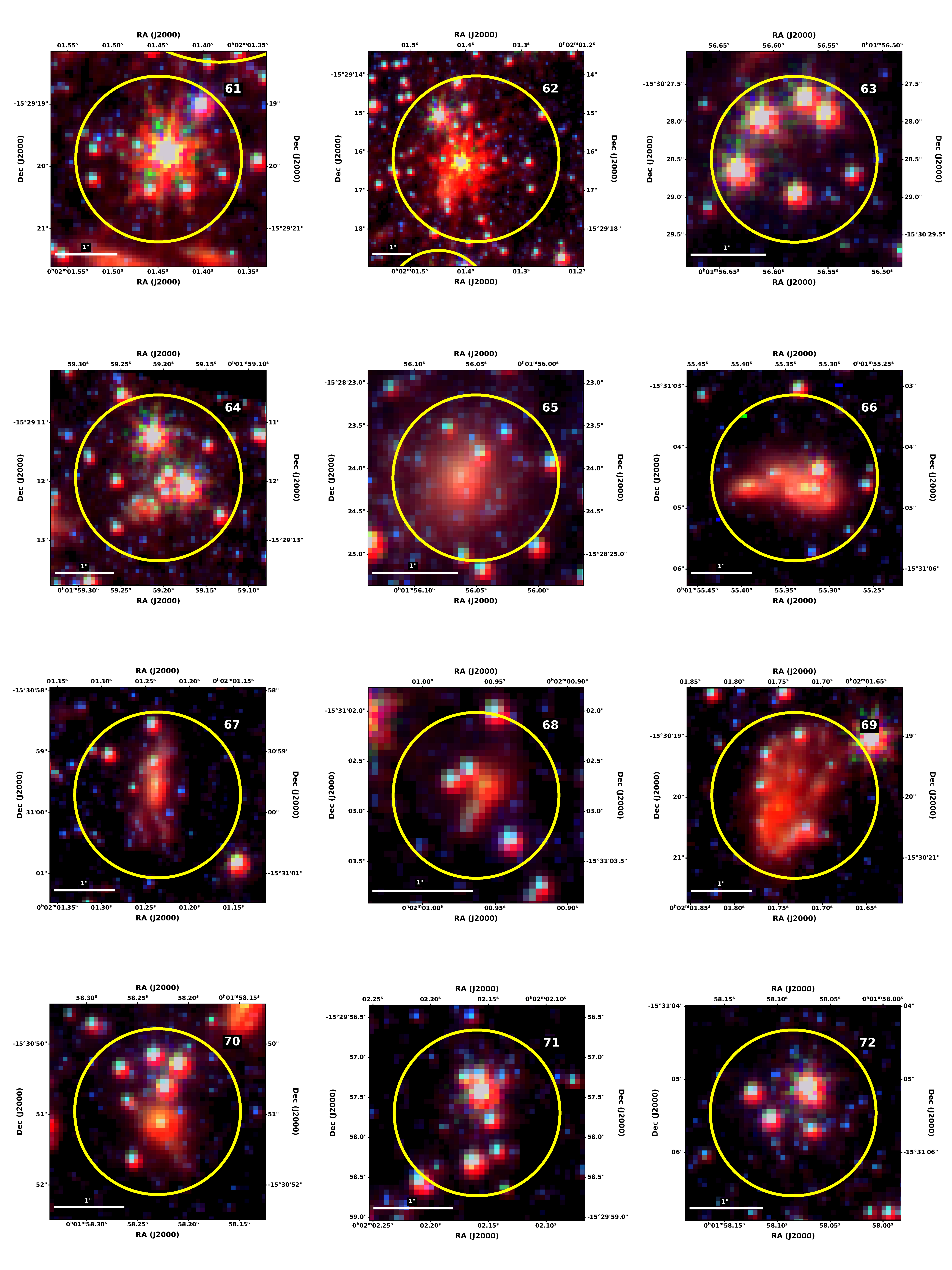}
\caption{F250M (red), F150W (green), F090W (blue) three-color image of each type 3 object (away from FUV and away from CO, yellow circles). The orientation of the image is such that North is up and East is to the left. The object number corresponding to the number in the extended version of Table \ref{tab:type1} provided in the online materials is given in the upper right of the image.
\label{fig:type3}}
\end{figure*}

\begin{figure*}[phtb!]
\ContinuedFloat
\epsscale{1.1}\plotone{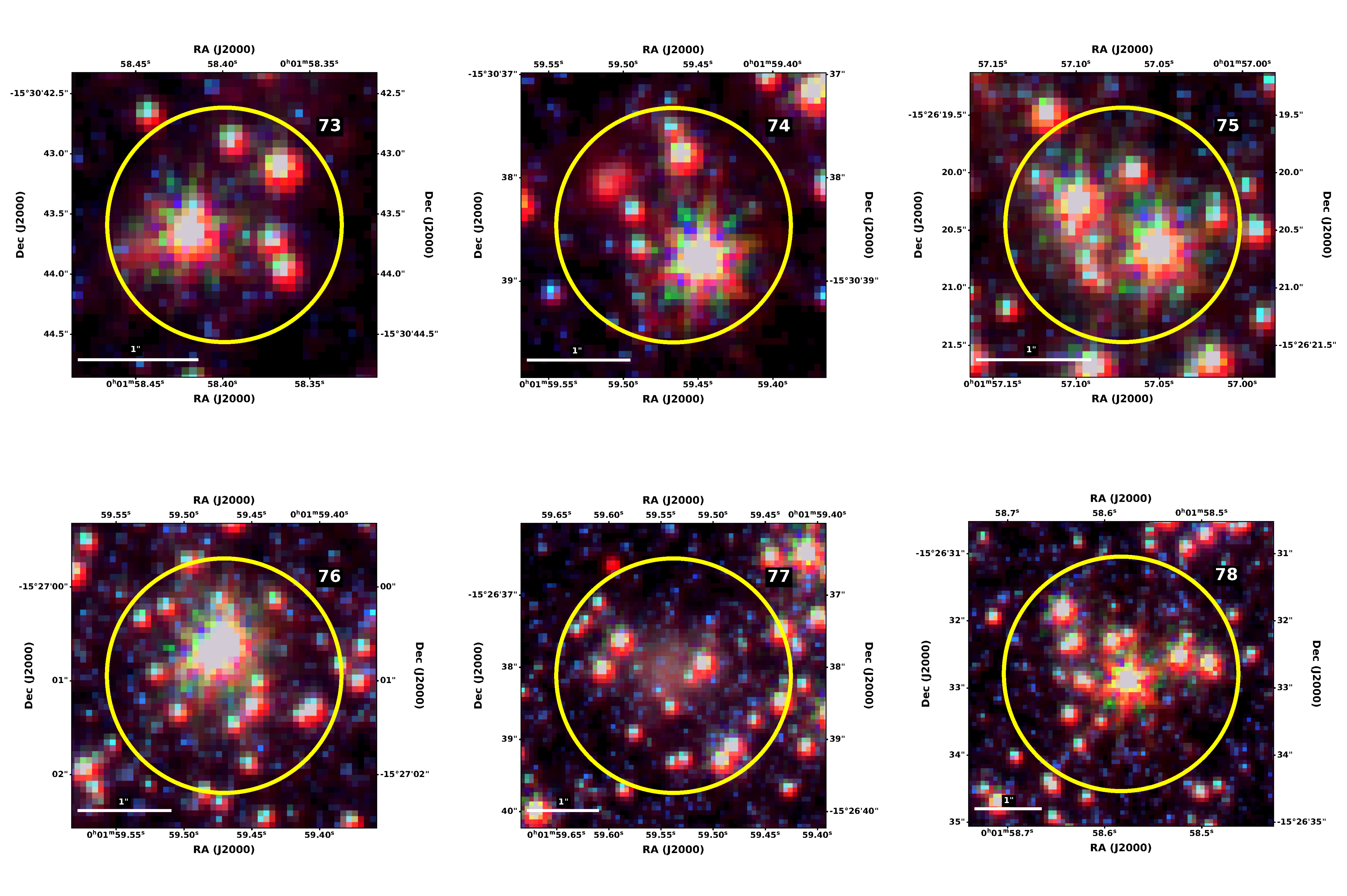}
\caption{\textbf{cont.} F250M (red), F150W (green), F090W (blue) three-color image of each type 3 object (away from FUV and away from CO, yellow circles). The orientation of the image is such that North is up and East is to the left. The object number corresponding to the number in the extended version of Table \ref{tab:type1} provided in the online materials is given in the upper right of the image.}
\end{figure*}

\begin{figure*}[phtb!]
\epsscale{1.1}\plotone{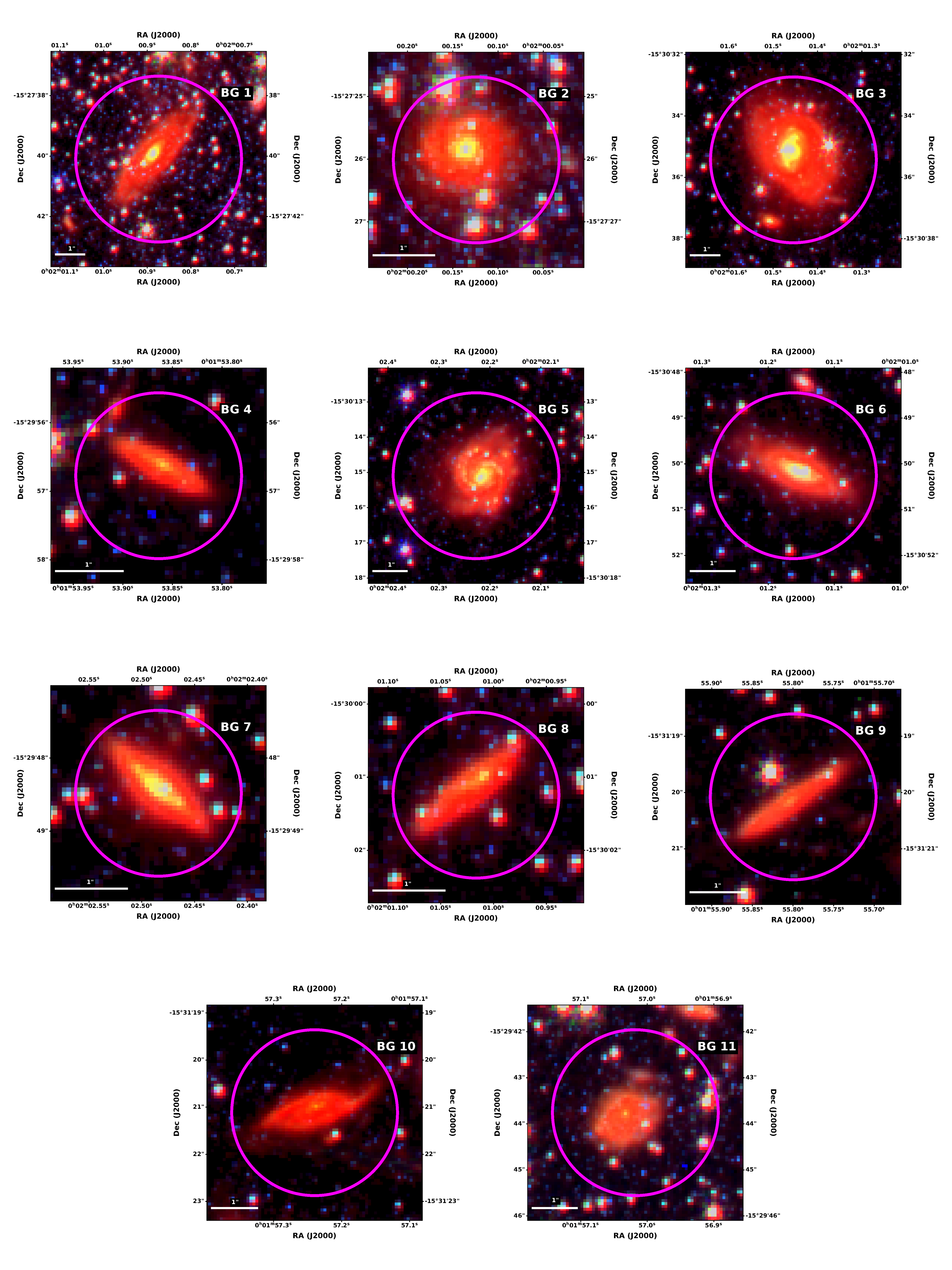}
\caption{F250M (red), F150W (green), F090W (blue) three-color image of select obvious background galaxies (magenta circles). The orientation of the image is such that North is up and East is to the left. The object number corresponding to the number in the extended version of Table \ref{tab:type1} provided in the online materials is given in the upper right of the image.
\label{fig:bgGal}}
\end{figure*}


\bibliography{wlm_jwst}{}

\begin{thebibliography}{}
\expandafter\ifx\csname natexlab\endcsname\relax\def\natexlab#1{#1}\fi
\providecommand{\url}[1]{\href{#1}{#1}}
\providecommand{\dodoi}[1]{doi:~\href{http://doi.org/#1}{\nolinkurl{#1}}}
\providecommand{\doeprint}[1]{\href{http://ascl.net/#1}{\nolinkurl{http://ascl.net/#1}}}
\providecommand{\doarXiv}[1]{\href{https://arxiv.org/abs/#1}{\nolinkurl{https://arxiv.org/abs/#1}}}

\bibitem[{{Albers} {et~al.}(2019){Albers}, {Weisz}, {Cole}, {Dolphin}, {Skillman}, {Williams}, {Boylan-Kolchin}, {Bullock}, {Dalcanton}, {Hopkins}, {Leaman}, {McConnachie}, {Vogelsberger}, \& {Wetzel}}]{Albers_2019}
{Albers}, S.~M., {Weisz}, D.~R., {Cole}, A.~A., {et~al.} 2019, \mnras, 490, 5538, \dodoi{10.1093/mnras/stz2903}

\bibitem[{{Archer} {et~al.}(2022){Archer}, {Hunter}, {Elmegreen}, {Cigan}, {Jansen}, {Windhorst}, {Hunt}, \& {Rubio}}]{archer_2022}
{Archer}, H.~N., {Hunter}, D.~A., {Elmegreen}, B.~G., {et~al.} 2022, \aj, 163, 141, \dodoi{10.3847/1538-3881/ac4e88}

\bibitem[{{Bastian} \& {Goodwin}(2006)}]{Bastian_2006}
{Bastian}, N., \& {Goodwin}, S.~P. 2006, \mnras, 369, L9, \dodoi{10.1111/j.1745-3933.2006.00162.x}

\bibitem[{{Bigiel} {et~al.}(2011){Bigiel}, {Leroy}, {Walter}, {Brinks}, {de Blok}, {Kramer}, {Rix}, {Schruba}, {Schuster}, {Usero}, \& {Wiesemeyer}}]{Bigiel_2011}
{Bigiel}, F., {Leroy}, A.~K., {Walter}, F., {et~al.} 2011, \apjl, 730, L13, \dodoi{10.1088/2041-8205/730/2/L13}

\bibitem[{{Bolatto} {et~al.}(1999){Bolatto}, {Jackson}, \& {Ingalls}}]{Bolatto_1999}
{Bolatto}, A.~D., {Jackson}, J.~M., \& {Ingalls}, J.~G. 1999, \apj, 513, 275, \dodoi{10.1086/306849}

\bibitem[{{Bolatto} {et~al.}(2013){Bolatto}, {Wolfire}, \& {Leroy}}]{Bolatto_2013}
{Bolatto}, A.~D., {Wolfire}, M., \& {Leroy}, A.~K. 2013, \araa, 51, 207, \dodoi{10.1146/annurev-astro-082812-140944}

\bibitem[{{Brosch} {et~al.}(1998){Brosch}, {Heller}, \& {Almoznino}}]{Brosch_1998}
{Brosch}, N., {Heller}, A., \& {Almoznino}, E. 1998, \apj, 504, 720, \dodoi{10.1086/306127}

\bibitem[{{Bruzual} \& {Charlot}(2003)}]{BruzualCharlot_2003}
{Bruzual}, G., \& {Charlot}, S. 2003, \mnras, 344, 1000, \dodoi{10.1046/j.1365-8711.2003.06897.x}

\bibitem[{{Calzetti}(2013)}]{Calzetti_2013}
{Calzetti}, D. 2013, in Secular Evolution of Galaxies, ed. J.~{Falc{\'o}n-Barroso} \& J.~H. {Knapen} (Cambridge University Press), 419, \dodoi{10.48550/arXiv.1208.2997}

\bibitem[{{Carpenter} {et~al.}(1995){Carpenter}, {Snell}, \& {Schloerb}}]{Carpenter_1996}
{Carpenter}, J.~M., {Snell}, R.~L., \& {Schloerb}, F.~P. 1995, \apj, 450, 201, \dodoi{10.1086/176132}

\bibitem[{{Chevance} {et~al.}(2020){Chevance}, {Madden}, {Fischer}, {Vacca}, {Lebouteiller}, {Fadda}, {Galliano}, {Indebetouw}, {Kruijssen}, {Lee}, {Poglitsch}, {Polles}, {Cormier}, {Hony}, {Iserlohe}, {Krabbe}, {Meixner}, {Sabbi}, \& {Zinnecker}}]{Chevance_2020}
{Chevance}, M., {Madden}, S.~C., {Fischer}, C., {et~al.} 2020, \mnras, 494, 5279, \dodoi{10.1093/mnras/staa1106}

\bibitem[{{Conroy} \& {Gunn}(2010)}]{Conroy_2010}
{Conroy}, C., \& {Gunn}, J.~E. 2010, \apj, 712, 833, \dodoi{10.1088/0004-637X/712/2/833}

\bibitem[{{Conroy} {et~al.}(2009){Conroy}, {Gunn}, \& {White}}]{Conroy_2009}
{Conroy}, C., {Gunn}, J.~E., \& {White}, M. 2009, \apj, 699, 486, \dodoi{10.1088/0004-637X/699/1/486}

\bibitem[{{Cormier} {et~al.}(2017){Cormier}, {Bendo}, {Hony}, {Lebouteiller}, {Madden}, {Galliano}, {Glover}, {Klessen}, {Abel}, {Bigiel}, \& {Clark}}]{Cormier_2017}
{Cormier}, D., {Bendo}, G.~J., {Hony}, S., {et~al.} 2017, \mnras, 468, L87, \dodoi{10.1093/mnrasl/slx034}

\bibitem[{{Dale} {et~al.}(2015){Dale}, {Ercolano}, \& {Bonnell}}]{Dale_2015}
{Dale}, J.~E., {Ercolano}, B., \& {Bonnell}, I.~A. 2015, \mnras, 451, 987, \dodoi{10.1093/mnras/stv913}

\bibitem[{{Draine} \& {Li}(2007)}]{Draine_2007}
{Draine}, B.~T., \& {Li}, A. 2007, \apj, 657, 810, \dodoi{10.1086/511055}

\bibitem[{{Elmegreen}(1989)}]{Elmegreen_1989}
{Elmegreen}, B.~G. 1989, \apj, 338, 178, \dodoi{10.1086/167192}

\bibitem[{{Elmegreen} \& {Elmegreen}(2019)}]{Elmegreen_2019}
{Elmegreen}, B.~G., \& {Elmegreen}, D.~M. 2019, \apjs, 245, 14, \dodoi{10.3847/1538-4365/ab4903}

\bibitem[{{Elmegreen} \& {Elmegreen}(2020)}]{Elmegreen_2020}
---. 2020, \apj, 895, 71, \dodoi{10.3847/1538-4357/ab8d20}

\bibitem[{{Elmegreen} {et~al.}(2018){Elmegreen}, {Elmegreen}, \& {Efremov}}]{Elmegreen_2018}
{Elmegreen}, B.~G., {Elmegreen}, D.~M., \& {Efremov}, Y.~N. 2018, \apj, 863, 59, \dodoi{10.3847/1538-4357/aacf9a}

\bibitem[{{Elmegreen} {et~al.}(1980){Elmegreen}, {Morris}, \& {Elmegreen}}]{Elmegreen_1980}
{Elmegreen}, B.~G., {Morris}, M., \& {Elmegreen}, D.~M. 1980, \apj, 240, 455, \dodoi{10.1086/158251}

\bibitem[{{Elmegreen} {et~al.}(2013){Elmegreen}, {Rubio}, {Hunter}, {Verdugo}, {Brinks}, \& {Schruba}}]{Elmegreen_2013}
{Elmegreen}, B.~G., {Rubio}, M., {Hunter}, D.~A., {et~al.} 2013, \nat, 495, 487, \dodoi{10.1038/nature11933}

\bibitem[{{Fukui} \& {Kawamura}(2010)}]{Fukui_2010}
{Fukui}, Y., \& {Kawamura}, A. 2010, \araa, 48, 547, \dodoi{10.1146/annurev-astro-081309-130854}

\bibitem[{{Glover} \& {Mac Low}(2011)}]{Glover_2011}
{Glover}, S.~C.~O., \& {Mac Low}, M.~M. 2011, \mnras, 412, 337, \dodoi{10.1111/j.1365-2966.2010.17907.x}

\bibitem[{{Hunter} \& {Elmegreen}(2006)}]{Hunter_2006}
{Hunter}, D.~A., \& {Elmegreen}, B.~G. 2006, \apjs, 162, 49, \dodoi{10.1086/498096}

\bibitem[{{Hunter} {et~al.}(1998){Hunter}, {Elmegreen}, \& {Baker}}]{Hunter_1998}
{Hunter}, D.~A., {Elmegreen}, B.~G., \& {Baker}, A.~L. 1998, \apj, 493, 595, \dodoi{10.1086/305158}

\bibitem[{{Israel} {et~al.}(1986){Israel}, {de Graauw}, {van de Stadt}, \& {de Vries}}]{Isreal_1986}
{Israel}, F.~P., {de Graauw}, T., {van de Stadt}, H., \& {de Vries}, C.~P. 1986, \apj, 303, 186, \dodoi{10.1086/164065}

\bibitem[{{Johnson} {et~al.}(2022){Johnson}, {Foreman-Mackey}, {Sick}, {Leja}, {Byler}, {Walmsley}, {Tollerud}, {Leung}, \& {Scott}}]{python-fsps}
{Johnson}, B., {Foreman-Mackey}, D., {Sick}, J., {et~al.} 2022, {dfm/python-fsps: python-fsps v0.4.2rc1}, v0.4.2rc1,  Zenodo, \dodoi{10.5281/zenodo.7113363}

\bibitem[{{Johnson}(2019)}]{sedpy}
{Johnson}, B.~D. 2019, {SEDPY: Modules for storing and operating on astronomical source spectral energy distribution}, Astrophysics Source Code Library, record ascl:1905.026.
\newblock \doeprint{1905.026}

\bibitem[{{Johnson} {et~al.}(2021){Johnson}, {Leja}, {Conroy}, \& {Speagle}}]{Johnson_2021}
{Johnson}, B.~D., {Leja}, J., {Conroy}, C., \& {Speagle}, J.~S. 2021, \apjs, 254, 22, \dodoi{10.3847/1538-4365/abef67}

\bibitem[{{Jones} {et~al.}(2023){Jones}, {Nally}, {Habel}, {Lenki{\'c}}, {Fahrion}, {Hirschauer}, {Chu}, {Meixner}, {De Marchi}, {Nayak}, {Robberto}, {Sabbi}, {Zeidler}, {Alves de Oliveira}, {Beck}, {Biazzo}, {Brandl}, {Giardino}, {Jerabkova}, {Keyes}, {Muzerolle}, {Panagia}, {Pontoppidan}, {Rogers}, {Sargent}, \& {Soderblom}}]{Jones_2023}
{Jones}, O.~C., {Nally}, C., {Habel}, N., {et~al.} 2023, Nature Astronomy, 7, 694, \dodoi{10.1038/s41550-023-01945-7}

\bibitem[{{Kennicutt}(1998)}]{Kennicutt_1998}
{Kennicutt}, Robert~C., J. 1998, \apj, 498, 541, \dodoi{10.1086/305588}

\bibitem[{{Koposov} {et~al.}(2022){Koposov}, {Speagle}, {Barbary}, {Ashton}, {Bennett}, {Buchner}, {Scheffler}, {Cook}, {Talbot}, {Guillochon}, {Cubillos}, {Asensio Ramos}, {Johnson}, {Lang}, {Ilya}, {Dartiailh}, {Nitz}, {McCluskey}, {Archibald}, {Deil}, {Foreman-Mackey}, {Goldstein}, {Tollerud}, {Leja}, {Kirk}, {Pitkin}, {Sheehan}, {Cargile}, {Ruskin23}, \& {Angus}}]{dynesty}
{Koposov}, S., {Speagle}, J., {Barbary}, K., {et~al.} 2022, {joshspeagle/dynesty: v2.0.3}, v2.0.3,  Zenodo, \dodoi{10.5281/zenodo.7388523}

\bibitem[{{Lada} \& {Lada}(2003)}]{Lada_2003}
{Lada}, C.~J., \& {Lada}, E.~A. 2003, \araa, 41, 57, \dodoi{10.1146/annurev.astro.41.011802.094844}

\bibitem[{{Leaman} {et~al.}(2012){Leaman}, {Venn}, {Brooks}, {Battaglia}, {Cole}, {Ibata}, {Irwin}, {McConnachie}, {Mendel}, \& {Tolstoy}}]{Leaman_2012}
{Leaman}, R., {Venn}, K.~A., {Brooks}, A.~M., {et~al.} 2012, \apj, 750, 33, \dodoi{10.1088/0004-637X/750/1/33}

\bibitem[{{Lee} {et~al.}(2021){Lee}, {Freedman}, {Madore}, {Owens}, {Monson}, \& {Hoyt}}]{Lee_2021}
{Lee}, A.~J., {Freedman}, W.~L., {Madore}, B.~F., {et~al.} 2021, \apj, 907, 112, \dodoi{10.3847/1538-4357/abd253}

\bibitem[{{Lee} {et~al.}(2005){Lee}, {Skillman}, \& {Venn}}]{Lee_2005}
{Lee}, H., {Skillman}, E.~D., \& {Venn}, K.~A. 2005, \apj, 620, 223, \dodoi{10.1086/427019}

\bibitem[{{Leja} {et~al.}(2017){Leja}, {Johnson}, {Conroy}, {van Dokkum}, \& {Byler}}]{Leja_2017}
{Leja}, J., {Johnson}, B.~D., {Conroy}, C., {van Dokkum}, P.~G., \& {Byler}, N. 2017, \apj, 837, 170, \dodoi{10.3847/1538-4357/aa5ffe}

\bibitem[{{Leroy} {et~al.}(2008){Leroy}, {Walter}, {Brinks}, {Bigiel}, {de Blok}, {Madore}, \& {Thornley}}]{Leroy_2008}
{Leroy}, A.~K., {Walter}, F., {Brinks}, E., {et~al.} 2008, \aj, 136, 2782, \dodoi{10.1088/0004-6256/136/6/2782}

\bibitem[{{Leroy} {et~al.}(2009){Leroy}, {Bolatto}, {Bot}, {Engelbracht}, {Gordon}, {Israel}, {Rubio}, {Sandstrom}, \& {Stanimirovi{\'c}}}]{Leroy_2009}
{Leroy}, A.~K., {Bolatto}, A., {Bot}, C., {et~al.} 2009, \apj, 702, 352, \dodoi{10.1088/0004-637X/702/1/352}

\bibitem[{{Madden} {et~al.}(2020){Madden}, {Cormier}, {Hony}, {Lebouteiller}, {Abel}, {Galametz}, {De Looze}, {Chevance}, {Polles}, {Lee}, {Galliano}, {Lambert-Huyghe}, {Hu}, \& {Ramambason}}]{Madden_2020}
{Madden}, S.~C., {Cormier}, D., {Hony}, S., {et~al.} 2020, \aap, 643, A141, \dodoi{10.1051/0004-6361/202038860}

\bibitem[{{Martin} {et~al.}(2005){Martin}, {Fanson}, {Schiminovich}, {Morrissey}, {Friedman}, {Barlow}, {Conrow}, {Grange}, {Jelinsky}, {Milliard}, {Siegmund}, {Bianchi}, {Byun}, {Donas}, {Forster}, {Heckman}, {Lee}, {Madore}, {Malina}, {Neff}, {Rich}, {Small}, {Surber}, {Szalay}, {Welsh}, \& {Wyder}}]{Martin_2005}
{Martin}, D.~C., {Fanson}, J., {Schiminovich}, D., {et~al.} 2005, \apjl, 619, L1, \dodoi{10.1086/426387}

\bibitem[{{Nayak} {et~al.}(2018){Nayak}, {Meixner}, {Fukui}, {Tachihara}, {Onishi}, {Saigo}, {Tokuda}, \& {Harada}}]{Nayak_2018}
{Nayak}, O., {Meixner}, M., {Fukui}, Y., {et~al.} 2018, \apj, 854, 154, \dodoi{10.3847/1538-4357/aaab5f}

\bibitem[{{Nayak} {et~al.}(2016){Nayak}, {Meixner}, {Indebetouw}, {De Marchi}, {Koekemoer}, {Panagia}, \& {Sabbi}}]{Nayak_2016}
{Nayak}, O., {Meixner}, M., {Indebetouw}, R., {et~al.} 2016, \apj, 831, 32, \dodoi{10.3847/0004-637X/831/1/32}

\bibitem[{{Osman} {et~al.}(2020){Osman}, {Bekki}, \& {Cortese}}]{Osman_2020}
{Osman}, O., {Bekki}, K., \& {Cortese}, L. 2020, \mnras, 497, 2002, \dodoi{10.1093/mnras/staa1554}

\bibitem[{{Phelps} \& {Lada}(1997)}]{Phelps_1997}
{Phelps}, R.~L., \& {Lada}, E.~A. 1997, \apj, 477, 176, \dodoi{10.1086/303713}

\bibitem[{{Pineda} {et~al.}(2014){Pineda}, {Langer}, \& {Goldsmith}}]{Pineda_2014}
{Pineda}, J.~L., {Langer}, W.~D., \& {Goldsmith}, P.~F. 2014, \aap, 570, A121, \dodoi{10.1051/0004-6361/201424054}

\bibitem[{{Planck Collaboration} {et~al.}(2011){Planck Collaboration}, {Ade}, {Aghanim}, {Arnaud}, {Ashdown}, {Aumont}, {Baccigalupi}, {Balbi}, {Banday}, {Barreiro}, {Bartlett}, {Battaner}, {Benabed}, {Beno{\^\i}t}, {Bernard}, {Bersanelli}, {Bhatia}, {Bock}, {Bonaldi}, {Bond}, {Borrill}, {Bouchet}, {Boulanger}, {Bucher}, {Burigana}, {Cabella}, {Cardoso}, {Catalano}, {Cay{\'o}n}, {Challinor}, {Chamballu}, {Chiang}, {Chiang}, {Christensen}, {Clements}, {Colombi}, {Couchot}, {Coulais}, {Crill}, {Cuttaia}, {Dame}, {Danese}, {Davies}, {Davis}, {de Bernardis}, {de Gasperis}, {de Rosa}, {de Zotti}, {Delabrouille}, {Delouis}, {D{\'e}sert}, {Dickinson}, {Dobashi}, {Donzelli}, {Dor{\'e}}, {D{\"o}rl}, {Douspis}, {Dupac}, {Efstathiou}, {En{\ss}lin}, {Eriksen}, {Falgarone}, {Finelli}, {Forni}, {Fosalba}, {Frailis}, {Franceschi}, {Fukui}, {Galeotta}, {Ganga}, {Giard}, {Giardino}, {Giraud-H{\'e}raud}, {Gonz{\'a}lez-Nuevo}, {G{\'o}rski}, {Gratton}, {Gregorio}, {Grenier}, {Gruppuso}, {Hansen}, {Harrison}, {Helou},
  {Henrot-Versill{\'e}}, {Herranz}, {Hildebrandt}, {Hivon}, {Hobson}, {Holmes}, {Hovest}, {Hoyland}, {Huffenberger}, {Jaffe}, {Jones}, {Juvela}, {Kawamura}, {Keih{\"a}nen}, {Keskitalo}, {Kisner}, {Kneissl}, {Knox}, {Kurki-Suonio}, {Lagache}, {Lamarre}, {Lasenby}, {Laureijs}, {Lawrence}, {Leach}, {Leonardi}, {Leroy}, {Lilje}, {Linden-V{\o}rnle}, {L{\'o}pez-Caniego}, {Lubin}, {Mac{\'\i}as-P{\'e}rez}, {MacTavish}, {Maffei}, {Maino}, {Mandolesi}, {Mann}, {Maris}, {Martin}, {Mart{\'\i}nez-Gonz{\'a}lez}, {Masi}, {Matarrese}, {Matthai}, {Mazzotta}, {McGehee}, {Meinhold}, {Melchiorri}, {Mendes}, {Mennella}, {Miville-Desch{\^e}nes}, {Moneti}, {Montier}, {Morgante}, {Mortlock}, {Munshi}, {Murphy}, {Naselsky}, {Natoli}, {Netterfield}, {N{\o}rgaard-Nielsen}, {Noviello}, {Novikov}, {Novikov}, {O'Dwyer}, {Onishi}, {Osborne}, {Pajot}, {Paladini}, {Paradis}, {Pasian}, {Patanchon}, {Perdereau}, {Perotto}, {Perrotta}, {Piacentini}, {Piat}, {Plaszczynski}, {Pointecouteau}, {Polenta}, {Ponthieu}, {Poutanen}, {Pr{\'e}zeau},
  {Prunet}, {Puget}, {Reach}, {Reinecke}, {Renault}, {Ricciardi}, {Riller}, {Ristorcelli}, {Rocha}, {Rosset}, {Rowan-Robinson}, {Rubi{\~n}o-Mart{\'\i}n}, {Rusholme}, {Sandri}, {Santos}, {Savini}, {Scott}, {Seiffert}, {Shellard}, {Smoot}, {Starck}, {Stivoli}, {Stolyarov}, {Stompor}, {Sudiwala}, {Sygnet}, {Tauber}, {Terenzi}, {Toffolatti}, {Tomasi}, {Torre}, {Tristram}, {Tuovinen}, {Umana}, {Valenziano}, {Vielva}, {Villa}, {Vittorio}, {Wade}, {Wandelt}, {Wilkinson}, {Yvon}, {Zacchei}, \& {Zonca}}]{planck_2011}
{Planck Collaboration}, {Ade}, P.~A.~R., {Aghanim}, N., {et~al.} 2011, \aap, 536, A19, \dodoi{10.1051/0004-6361/201116479}

\bibitem[{{Rosolowsky} \& {Leroy}(2006)}]{Rosolowsky_2006}
{Rosolowsky}, E., \& {Leroy}, A. 2006, \pasp, 118, 590, \dodoi{10.1086/502982}

\bibitem[{{Rubio} {et~al.}(2015){Rubio}, {Elmegreen}, {Hunter}, {Brinks}, {Cort{\'e}s}, \& {Cigan}}]{Rubio_2015}
{Rubio}, M., {Elmegreen}, B.~G., {Hunter}, D.~A., {et~al.} 2015, \nat, 525, 218, \dodoi{10.1038/nature14901}

\bibitem[{{Salda{\~n}o} {et~al.}(2023){Salda{\~n}o}, {Rubio}, {Bolatto}, {Verdugo}, {Jameson}, \& {Leroy}}]{Saldano_2023}
{Salda{\~n}o}, H.~P., {Rubio}, M., {Bolatto}, A.~D., {et~al.} 2023, \aap, 672, A153, \dodoi{10.1051/0004-6361/202142217}

\bibitem[{{Schruba} {et~al.}(2012){Schruba}, {Leroy}, {Walter}, {Bigiel}, {Brinks}, {de Blok}, {Kramer}, {Rosolowsky}, {Sandstrom}, {Schuster}, {Usero}, {Weiss}, \& {Wiesemeyer}}]{Schruba_2012}
{Schruba}, A., {Leroy}, A.~K., {Walter}, F., {et~al.} 2012, \aj, 143, 138, \dodoi{10.1088/0004-6256/143/6/138}

\bibitem[{{Solomon} {et~al.}(1987){Solomon}, {Rivolo}, {Barrett}, \& {Yahil}}]{Solomon_1987}
{Solomon}, P.~M., {Rivolo}, A.~R., {Barrett}, J., \& {Yahil}, A. 1987, \apj, 319, 730, \dodoi{10.1086/165493}

\bibitem[{{Speagle}(2020)}]{Speagle_2020}
{Speagle}, J.~S. 2020, \mnras, 493, 3132, \dodoi{10.1093/mnras/staa278}

\bibitem[{{Taylor} {et~al.}(1998){Taylor}, {Kobulnicky}, \& {Skillman}}]{Taylor_1998}
{Taylor}, C.~L., {Kobulnicky}, H.~A., \& {Skillman}, E.~D. 1998, \aj, 116, 2746, \dodoi{10.1086/300655}

\bibitem[{{Teyssier} {et~al.}(2012){Teyssier}, {Johnston}, \& {Kuhlen}}]{Teyssier_2012}
{Teyssier}, M., {Johnston}, K.~V., \& {Kuhlen}, M. 2012, \mnras, 426, 1808, \dodoi{10.1111/j.1365-2966.2012.21793.x}

\bibitem[{{Tody}(1986)}]{Tody_1986}
{Tody}, D. 1986, in Society of Photo-Optical Instrumentation Engineers (SPIE) Conference Series, Vol. 627, Instrumentation in astronomy VI, ed. D.~L. {Crawford}, 733, \dodoi{10.1117/12.968154}

\bibitem[{{Wakelam} {et~al.}(2017){Wakelam}, {Bron}, {Cazaux}, {Dulieu}, {Gry}, {Guillard}, {Habart}, {Hornek{\ae}r}, {Morisset}, {Nyman}, {Pirronello}, {Price}, {Valdivia}, {Vidali}, \& {Watanabe}}]{Wakelam_2017}
{Wakelam}, V., {Bron}, E., {Cazaux}, S., {et~al.} 2017, Molecular Astrophysics, 9, 1, \dodoi{10.1016/j.molap.2017.11.001}

\bibitem[{{Weisz} {et~al.}(2023){Weisz}, {McQuinn}, {Savino}, {Kallivayalil}, {Anderson}, {Boyer}, {Correnti}, {Geha}, {Dolphin}, {Sandstrom}, {Cole}, {Williams}, {Skillman}, {Cohen}, {Newman}, {Beaton}, {Bressan}, {Bolatto}, {Boylan-Kolchin}, {Brooks}, {Bullock}, {Conroy}, {Cooper}, {Dalcanton}, {Dotter}, {Fritz}, {Garling}, {Gennaro}, {Gilbert}, {Girardi}, {Johnson}, {Johnson}, {Kalirai}, {Kirby}, {Lang}, {Marigo}, {Richstein}, {Schlafly}, {Schmidt}, {Tollerud}, {Warfield}, \& {Wetzel}}]{Weisz_2023}
{Weisz}, D.~R., {McQuinn}, K. B.~W., {Savino}, A., {et~al.} 2023, arXiv e-prints, arXiv:2301.04659, \dodoi{10.48550/arXiv.2301.04659}

\bibitem[{{Windhorst} {et~al.}(2023){Windhorst}, {Cohen}, {Jansen}, {Summers}, {Tompkins}, {Conselice}, {Driver}, {Yan}, {Coe}, {Frye}, {Grogin}, {Koekemoer}, {Marshall}, {O'Brien}, {Pirzkal}, {Robotham}, {Ryan}, {Willmer}, {Carleton}, {Diego}, {Keel}, {Porto}, {Redshaw}, {Scheller}, {Wilkins}, {Willner}, {Zitrin}, {Adams}, {Austin}, {Arendt}, {Beacom}, {Bhatawdekar}, {Bradley}, {Broadhurst}, {Cheng}, {Civano}, {Dai}, {Dole}, {D'Silva}, {Duncan}, {Fazio}, {Ferrami}, {Ferreira}, {Finkelstein}, {Furtak}, {Gim}, {Griffiths}, {Hammel}, {Harrington}, {Hathi}, {Holwerda}, {Honor}, {Huang}, {Hyun}, {Im}, {Joshi}, {Kamieneski}, {Kelly}, {Larson}, {Li}, {Lim}, {Ma}, {Maksym}, {Manzoni}, {Meena}, {Milam}, {Nonino}, {Pascale}, {Petric}, {Pierel}, {del Carmen Polletta}, {R{\"o}ttgering}, {Rutkowski}, {Smail}, {Straughn}, {Strolger}, {Swirbul}, {Trussler}, {Wang}, {Welch}, {B. Wyithe}, {Yun}, {Zackrisson}, {Zhang}, \& {Zhao}}]{Windhorst_2023}
{Windhorst}, R.~A., {Cohen}, S.~H., {Jansen}, R.~A., {et~al.} 2023, \aj, 165, 13, \dodoi{10.3847/1538-3881/aca163}

\bibitem[{{Wolfire} {et~al.}(2010){Wolfire}, {Hollenbach}, \& {McKee}}]{Wolfire_2010}
{Wolfire}, M.~G., {Hollenbach}, D., \& {McKee}, C.~F. 2010, \apj, 716, 1191, \dodoi{10.1088/0004-637X/716/2/1191}

\bibitem[{{Xu} {et~al.}(2001){Xu}, {Lonsdale}, {Shupe}, {O'Linger}, \& {Masci}}]{Xu_2001}
{Xu}, C., {Lonsdale}, C.~J., {Shupe}, D.~L., {O'Linger}, J., \& {Masci}, F. 2001, \apj, 562, 179, \dodoi{10.1086/323430}

\bibitem[{{Yang} {et~al.}(2022){Yang}, {Ianjamasimanana}, {Hammer}, {Higgs}, {Namumba}, {Carignan}, {J{\'o}zsa}, \& {McConnachie}}]{Yang_2022}
{Yang}, Y., {Ianjamasimanana}, R., {Hammer}, F., {et~al.} 2022, \aap, 660, L11, \dodoi{10.1051/0004-6361/202243307}

\bibitem[{{Zhang} {et~al.}(2012){Zhang}, {Hunter}, {Elmegreen}, {Gao}, \& {Schruba}}]{Zhang_2012}
{Zhang}, H.-X., {Hunter}, D.~A., {Elmegreen}, B.~G., {Gao}, Y., \& {Schruba}, A. 2012, \aj, 143, 47, \dodoi{10.1088/0004-6256/143/2/47}

\end{thebibliography}
\bibliographystyle{aasjournal}



\end{document}